\documentclass[a4paper,11pt]{article}
\usepackage{jcappub}

\usepackage{amssymb}
\usepackage{graphicx,bm,color}
\graphicspath{{./fig/}{./png/}}

\newcommand{\BoldVec}[1]{\mathchoice%
  {\mbox{\boldmath $\displaystyle     #1$}}%
  {\mbox{\boldmath $\textstyle        #1$}}%
  {\mbox{\boldmath $\scriptstyle      #1$}}%
  {\mbox{\boldmath $\scriptscriptstyle#1$}}%
}
\newcommand{\EQ}{\begin{equation}}
\newcommand{\EN}{\end{equation}}
\newcommand{\EQA}{\begin{eqnarray}}
\newcommand{\ENA}{\end{eqnarray}}

\newcommand{\EEq}[1]{Equation~(\ref{#1})}
\newcommand{\Eq}[1]{equation~(\ref{#1})}
\newcommand{\Eqs}[2]{equations~(\ref{#1}) and~(\ref{#2})}
\newcommand{\EEqs}[2]{Equations~(\ref{#1}) and~(\ref{#2})}

\newcommand{\Eqss}[2]{equations~(\ref{#1})--(\ref{#2})}
\newcommand{\Sec}[1]{section~\ref{#1}}
\newcommand{\Secs}[2]{sections~\ref{#1} and~\ref{#2}}
\newcommand{\App}[1]{Appendix~\ref{#1}}
\newcommand{\Apps}[2]{Appendices~\ref{#1} and~\ref{#2}}
\newcommand{\Fig}[1]{figure~\ref{#1}}
\newcommand{\FFig}[1]{Figure~\ref{#1}}
\newcommand{\Tab}[1]{table~\ref{#1}}
\newcommand{\Figs}[2]{figures~\ref{#1} and \ref{#2}}

\newcommand{\bra}[1]{\langle #1\rangle}

{}
{}
{}
\newcommand{\nullvector}{{\bf0}}

\newcommand{\xx}{\BoldVec{x}{}}

\newcommand{\uu}{\BoldVec{u} {}}
\newcommand{\UU}{\BoldVec{U} {}}
\newcommand{\vv}{\BoldVec{v} {}}

\renewcommand{\UU}{\BoldVec{U} {}}

\newcommand{\BB}{\BoldVec{B} {}}

\newcommand{\AAA}{\BoldVec{A} {}}

\newcommand{\JJ}{\BoldVec{J} {}}
\newcommand{\nn}{\BoldVec{n} {}}
\newcommand{\ee}{\BoldVec{e} {}}

\newcommand{\tilff}{\BoldVec{\tilde f} {}}
\newcommand{\tilBB}{\BoldVec{\tilde B} {}}
\newcommand{\tilAA}{\BoldVec{\tilde A} {}}

\newcommand{\kk}{\BoldVec{k} {}}

\newcommand{\nab}{\BoldVec{\nabla} {}}

\newcommand{\oo}{\BoldVec{\omega} {}}

\newcommand{\SSSS}{\bm{\mathsf{S}}}

\newcommand{\FFF}{\mbox{\boldmath ${\cal F}$} {}}

\newcommand{\dd}{{\rm d} {}}

\def\EEM{{\cal E}_{\rm M}}
\def\EEMmax{{\cal E}_{\rm M}^{\rm max}}
\def\EEGW{{\cal E}_{\rm GW}}
\def\OmGW{{\Omega}_{\rm GW}}

\def\OmGWsat{{\Omega}_{\rm GW}^{\rm sat}}
\def\XiGWsat{{\Xi}_{\rm GW}^{\rm sat}}
\def\PPM{{\cal P}_{\rm M}}
\def\sigM{\sigma_{\rm M}}
\def\PPGW{{\cal P}_{\rm GW}}
\def\PPh{{\cal P}_h}
\def\XiGW{{\Xi}_{\rm GW}}

\def\EEcrit{{\cal E}_{\rm crit}}

\def\HHM{{\cal H}_{\rm M}}
\def\EGW{E_{\rm GW}}
\def\HGW{H_{\rm GW}}

\def\EM{E_{\rm M}}
\def\HM{H_{\rm M}}
\def\FM{F_{\rm M}}
\def\GM{G_{\rm M}}
\def\hc{h_{\rm c}}
\def\Sh{S_h}
\def\Shd{S_{\dot{h}}}
\def\Ah{A_h}
\def\Ahd{A_{\dot{h}}}

\def\kf{k_*}
\def\kB{k_{\rm B}}
\def\kGW{k_{\rm GW}}
\def\fH{f_{\rm H}}

\newcommand{\ea}{{\em et al.}}

\def\half{{\textstyle{1\over2}}}
\def\threehalf{{\textstyle{3\over2}}}

\def\threeeigth{{\textstyle{3\over8}}}
\def\onethird{{\textstyle{1\over3}}}
\def\twothird{{\textstyle{2\over3}}}
\def\fourthird{{\textstyle{4\over3}}}
\def\quarter{{\textstyle{1\over4}}}

\newcommand{\Hz}{\,{\rm Hz}}
\newcommand{\mHz}{\,{\rm mHz}}

\newcommand{\K}{\,{\rm K}}

\newcommand{\s}{\,{\rm s}}

\newcommand{\picom}{{\rm \, pm}}
\newcommand{\fm}{{\rm \, fm}}
\newcommand{\km}{\,{\rm km}}

\newcommand{\Mpc}{\,{\rm Mpc}}
\newcommand{\yr}{\,{\rm yr}}

\newcommand{\GeV}{\,{\rm GeV}}
\newcommand{\MeV}{\,{\rm MeV}}
\newcommand{\Mon}{{\cal M}_{OO'}^\lambda}
\newcommand{\Don}{{\cal D}_{OO'}^\lambda}
\newcommand{\Monpm}{{\cal M}_{OO'}^{\pm}}
\newcommand{\Donpm}{{\cal D}_{OO'}^{\pm}}
\newcommand{\Qab}{{\cal Q}_{O}^{ab}}
\newcommand{\Qcd}{{\cal Q}_{O'}^{cd}}
\newcommand{\Qi}{{\cal Q}_{i}^{ab}}
\newcommand{\Qone}{{\cal Q}_{1}^{ab}}
\newcommand{\Qtwo}{{\cal Q}_{2}^{ab}}
\newcommand{\Qthree}{{\cal Q}_{3}^{ab}}
\newcommand{\SNR}{{\rm SNR}}

\def\tmax{t_{\rm max}}
\newcommand{\yana}[4]{, {\em #4}, {Astron. Astrophys.} {\bf #2}, #3 (#1).}

\newcommand{\ysci}[4]{, {\em #4}, Science {\bf #2}, #3 (#1).}

\newcommand{\yjetp}[4]{, {\em #4}, Sov. Phys. JETP {\bf #2}, #3 (#1).}

\newcommand{\yprd}[4]{, {\em #4}, {Phys.\ Rev. D} {\bf #2}, #3 (#1).}

\newcommand{\yprl}[4]{, {\em #4}, Phys.\ Rev.\ Lett.\ {\bf #2}, #3 (#1).}

\newcommand{\yapj}[4]{, {\em #4}, Astrophys.\ J. {\bf #2}, #3 (#1).}

\newcommand{\yanar}[4]{, {\em #4}, Astron. Astrophys. Rev. {\bf #2}, #3 (#1).}

\newcommand{\ypf}[3]{, {\em #3}, Phys. Fluids {\bf #2} (#1).}

\newcommand{\ygafd}[4]{, {\em #4}, Geophys. Astrophys. Fluid Dynam. {\bf #2}, #3 (#1).}

\newcommand{\yjour}[5]{, {\em #5}, #2 {\bf #3}, #4 (#1).}

\title{Polarization of gravitational waves from helical MHD turbulent sources}

\author[a,b]{Alberto~Roper~Pol,}
\author[c,b]{Sayan~Mandal,}
\author[d,e,f,b]{Axel Brandenburg,}
\author[f,b,g]{Tina Kahniashvili}

\affiliation[a]{
Laboratoire Astroparticule et Cosmologie, CNRS UMR 7164,
Universit\'e de Paris,
10 Rue Alice Domon et L\'eonie Duquet, Paris, F-75013, France}
\affiliation[b]{School of Natural Sciences and Medicine,
Ilia State University,
3-5 Cholokashvili Ave, Tbilisi, GE-0194, Georgia}
\affiliation[c]{Physics and Astronomy Department,
Stony Brook University, Stony Brook,
NY, 11794, USA}
\affiliation[d]{Nordita, KTH Royal Institute of Technology and
Stockholm University, Hannes Alfv\'ens v\"ag 12,
Stockholm, SE-10691, Sweden}
\affiliation[e]{The Oskar Klein Centre, Department of Astronomy,
Stockholm University, AlbaNova, Stockholm, SE-10691, Sweden}
\affiliation[f]{McWilliams Center for Cosmology and Department of 
Physics, Carnegie Mellon University, 5000 Forbes Ave,
Pittsburgh, PA, 15213, USA}
\affiliation[g]{Abastumani Astrophysical Observatory,
Tbilisi, GE-0179, Georgia}

\emailAdd{roperpol@apc.in2p3.fr}
\emailAdd{sayan.mandal@stonybrook.edu}
\emailAdd{brandenb@nordita.org}
\emailAdd{tinatin@andrew.cmu.edu}
\emailAdd{\today}

\abstract{
We use direct numerical simulations of decaying primordial hydromagnetic
turbulence with helicity to compute the resulting gravitational wave (GW)
production and its degree of circular polarization.
The turbulence is sourced by magnetic fields that
are either initially present or
driven by an electromotive force applied for a short duration,
given as a fraction of one Hubble time.
In both types of simulations, we find a clear dependence of the
polarization of the resulting GWs on the fractional
helicity of the turbulent source.
We find a low frequency tail below the spectral peak
shallower than the $f^3$ scaling expected at super-horizon scales,
in agreement with similar recent numerical simulations.
This type of spectrum facilitates its observational detection with
the planned Laser Interferometer Space Antenna (LISA).
We show that driven magnetic fields produce
GWs more efficiently than magnetic fields that are initially present, leading
to larger spectral amplitudes, and to modifications of the spectral shape.
In particular, we observe a sharp drop of GW energy above the spectral peak
that is in agreement with the previously obtained results.
The helicity does not have a huge impact on the maximum spectral amplitude
in any of the two types of turbulence considered.
However, the GW spectrum at wave numbers away from the peak becomes smaller
for larger values of the magnetic fractional helicity.
Such variations of the spectrum are most noticeable when magnetic fields are driven.
The degree of circular polarization approaches zero at frequencies below the peak,
and reaches its maximum at the peak.
At higher frequencies, it stays finite if the magnetic field is
initially present, and it approaches zero if it is driven.
We predict that the spectral peak of the GW signal can be detected
by LISA if the turbulent energy density is at least $\sim\!3\%$ of the
radiation energy density,
and the characteristic scale is a hundredth of the horizon at the electroweak
scale.
We show that the resulting GW polarization is unlikely to be
detectable by the anisotropies induced by our proper motion in the dipole
response function of LISA.
Such signals can, however, be detectable
by cross-correlating data from the LISA--Taiji network
for turbulent energy densities of $\sim\!5\%$,
and fractional helicity
of 0.5 to 1.
Second-generation space-based GW detectors, such as the Big Bang Observer (BBO)
and  the DECi-hertz Interferometer Gravitational wave Observatory (DECIGO), would
allow for the detection of a larger range of the GW spectrum and smaller
amplitudes of the magnetic field.
}

\keywords{
cosmological phase transitions, gravitational waves/sources, magnetohydrodynamics,
primordial magnetic fields
}

\arxivnumber{2107.05356}

\renewcommand\afterProceedingsSpace{\vskip 3pt}
\renewcommand\afterTitleSpace{\vskip 25pt}

\begin{document}
\maketitle
\flushbottom

\section{Introduction}

Primordial turbulent magnetic fields produced and/or present during phase transitions
in the early universe generate a stochastic
background of gravitational waves (GWs) 
\cite{Deryagin:1986qq}; see refs.~\cite{Widrow:2002ud,Grasso:2000wj} for reviews.
Assuming the standard energy scales of cosmological phase transitions 
($T_* \sim 100 \GeV$ for the electroweak, and $T_* \sim 100
\MeV$ for the QCD phase transition),
and accounting that the characteristic
scale of the magnetic field at the moment of its
generation is limited by the Hubble horizon scale and it is taken to be a 
fraction of it (this fraction is determined by the size of 
the magnetic field eddies, which is related to the phase transition bubble size 
in the case of a first order phase transition, and/or by the 
energy containing wave number $\kf$ of the turbulent motions in other scenarios),
the characteristic typical frequencies of the 
GW spectrum range from nHz to Hz
\cite{Hogan:1986qda,Krauss:1991qu,Kosowsky:1992rz}.
Future space-based GW detectors such as the Laser Interferometer Space Antenna
(LISA) \cite{Audley:2017drz}, planned to be launched in 2034, as well as
TianQin \cite{Luo:2015ght} and Taiji \cite{Guo:2018npi}, will be sensitive to
GWs in frequencies ranging from 10 $\mu$Hz to a few Hz, with a peak
sensitivity around 1 mHz (LISA and Taiji) and a few mHz (TianQin).
Actually, this is a typical Hubble 
frequency range for the electroweak phase transition (EWPT) if occurring around 10 TeV.
In this range of frequencies, we expect magnetic fields and turbulence 
yielding GW signals generated at the EWPT;
see ref.~\cite{Kamionkowski:1993fg} for pioneering work and refs.~\cite{Kosowsky:2001xp,Apreda:2001us,Dolgov:2002ra}
for subsequent studies.
Importantly, the main parameters of the turbulence (and, correspondingly,
the characteristics of the phase transitions) are imprinted on the
GW signal shape, amplitude, and polarization 
\cite{Grojean:2006bp,Kahniashvili:2008pf}.

Additional sources from a first-order phase transition producing
GW radiation in this range of frequencies
include the collision of scalar field shells, sound waves induced into the
surrounding plasma, and subsequent turbulent motions;
see ref.~\cite{Hindmarsh:2013xza} for a pioneering work and
refs.~\cite{Weir:2017wfa,Caprini:2019egz} for recent reviews, and
references therein.
In addition, the next generation of space-based GW detectors is planned
to improve the sensitivity to GW signals and to cover the range from
mHz to 10 Hz (which lies in between the sensitive frequencies of
space-based GW detectors such as LISA and ground-based GW detectors such as
the LIGO-Virgo-Kagra network),
e.g., the DECi-hertz Interferometer Gravitational wave
Observatory (DECIGO) \cite{Seto:2001qf},
and the Big Bang Observer (BBO) \cite{BBO_NASA,Crowder:2005nr}.
In the lower regime of frequencies, measuring the time of arrival by a
pulsar timing array (PTA) allows one to detect GW signals in the range from
$10^{-9}$--$10^{-7}$ Hz, which corresponds to the GW signals generated during
a phase transition with typical energy scales from a few MeV to one GeV,
such as the QCD phase transition, with a typical scale of about $100 \MeV$; see
refs.~\cite{Witten:1984rs,Hogan:1986qda,Signore:1989,Thorsett:1996dr} and
ref.~\cite{Caprini:2018mtu} for a review, and references therein.
This frequency range is also typical for the blue tilted GW spectrum 
originated from the inflationary epoch; see ref.~\cite{Caprini:2018mtu} for
a review and references therein.
Even smaller frequencies can be probed by the indirect detection of
$B$-modes in the cosmic microwave background (CMB) polarization
\cite{Kamionkowski:1996zd,Seljak:1996gy}, which, along with temperature 
and $E$-mode polarization
anisotropies, can be produced by inflation-generated GW signals \cite{Polnarev:1985,Starobinsky:1985ww}.
The treatment of the early-universe generated GW energy density
spectrum allows one to constrain different scenarios by PTA measurements,
laser interferometer experiments, and big bang nucleosynthesis (BBN)
bounds \cite{Boyle:2007zx}.
In addition, large surveys of stars like Gaia \cite{Brown:2018dum}
or the proposed Theia \cite{Boehm:2017wie}
have been recently proposed to detect GWs in the range of frequencies
around the QCD scale \cite{Moore:2017ity,Garcia-Bellido:2021zgu}.

There is various kind of evidence for magnetic fields in the largest
scales of the universe \cite{Widrow:2002ud}, which can have their origin
in astrophysical or cosmological seed fields.
In particular, primordial magnetic fields are motivated by the
lower limits on the strength of extragalactic magnetic fields inferred
by observations of blazar spectra by the Fermi Gamma-ray Observatory;
see ref.~\cite{Neronov:1900zz} for a pioneering work and ref.~\cite{Vachaspati:2020blt}
for a recent review, and references therein.
Such fields are strongly coupled to the primordial plasma due to the high
conductivity of the early universe, inevitably leading to
magnetohydrodynamic (MHD) turbulence; see refs.~\cite{Ahonen:1996nq,
Brandenburg:1996fc} for pioneering work
and ref.~\cite{Brandenburg:2017neh} for a recent study.
In addition, any primordial turbulent process during the early universe
can also reinforce the magnetic field; see ref.~\cite{Brandenburg:2017rnt} 
for a discussion of the dynamo mechanism in decaying turbulence.

The cosmological evolution of the magnetic field strongly depends on
helicity \cite{BM99,Christensson:2000sp},
yielding magnetic fields with larger coherent scales
and favoring the constraints from Fermi observations
(see ref.~\cite{Vachaspati:2020blt} for a review and references therein). 
Parity-violating processes at the EWPT
leading to the generation of helical magnetic fields have been proposed.
Some examples are: via sphaleron decay
(see ref.~\cite{Vachaspati:1991nm} for a non-helical case and
ref.~\cite{Vachaspati:2001nb} for a helical case),
due to the generation of Chern-Simons number through $B+L$ anomalies
\cite{Cornwall:1997ms}, 
and due to inhomogeneities in the Higgs field in low-scale electroweak hybrid inflation
\cite{Joyce:1997uy,Garcia-Bellido:1999xos,Cornwall:2000eu,GarciaBellido:2002aj,Garcia-Bellido:2003wva,Diaz-Gil:2007fch}.
The presence of a cosmic axion field also leads to the generation of
helicity in existing primordial magnetic fields \cite{Forbes:2000gr,Campanelli:2005ye}.
Magnetic fields can also be produced during inflation; see
refs.~\cite{Turner:1987bw,Vachaspati:1991nm,Garretson:1992vt,
Ratra:1991bn,Dolgov:1993vg,Gasperini:1995dh}
for pioneering work
and the reviews~\cite{Durrer:2013pga,Subramanian:2015lua,Vachaspati:2020blt},
and references therein.
Some mechanisms have been proposed to add helicity into the inflationary
magnetogenesis models; see ref.~\cite{Giovannini:1998kg}
for pioneering work and 
refs.~\cite{Subramanian:2015lua, Vachaspati:2020blt} for reviews, and
references therein.

As expected, primordial helical magnetic fields
produce circularly polarized GWs \cite{Kahniashvili:2005qi,
Kisslinger:2015hua, Kahniashvili:2020jgm,Ellis:2020uid}.
In particular, the detection of circularly polarized GWs,
proposed in refs.~\cite{Seto:2006hf,Seto:2006dz},
will shed light on phenomena of fundamental symmetry breaking in the early 
universe, such as parity violation, and potentially can serve as an explanation
of the 
lepto- and baryogenesis asymmetry problem; see refs.~\cite{Sakharov:1967dj,Kuzmin:1985mm,
Shaposhnikov:1986jp,Cohen:1990py,Cohen:1990it} for pioneering work, 
ref.~\cite{Morrissey:2012db} for a review, 
and refs.~\cite{Fujita:2016igl,Kamada:2016eeb,Kamada:2016cnb}
for recent work.
The dependence of the degree of polarization of GWs on the
helicity of the source
has been a matter of uncertainty owing to the approximations made in the analytical
calculations available to date.
In particular, previous works (see
refs.~\cite{Kahniashvili:2005qi,Kisslinger:2015hua}) 
showed that the maximum
circular polarization depends on the relation between the magnetic energy and
the magnetic helicity spectra.
Assuming Kolmogorov-type turbulence, with spectral index of $-5/3$ for
the magnetic energy density and $-8/3$ for the helicity,\footnote{%
The spectral indices in refs.~\cite{Kahniashvili:2005qi,Kisslinger:2015hua}
refer to spectra defined with a $1/k^2$ factor with respect to those used
in the present work (defined in \Sec{spectra_def}),
where a scale-invariant spectrum is $\propto k^{-1}$.
Hence, the spectral indices $-5/3$ and $-8/3$ correspond to $-11/3$ and
$-14/3$ in their works.}
following phenomenological modeling of ref.~\cite{Kraichnan:1965zz},
the circular polarization of GWs would be at most about 80\% for a maximally
helical magnetic field \cite{Kahniashvili:2005qi,Kisslinger:2015hua}.
Following refs.~\cite{Kahniashvili:2005qi,Kisslinger:2015hua},
we call this {\em helical Kolmogorov} (HK) turbulence.
It can actually reach nearly 100\% in the case
when the spectral indices are equal, which we call a Moiseev-Chkhetiani 
type spectrum; see ref.~\cite{MC96}, or {\it helical transfer} (HT)  turbulence,
following refs.~\cite{Kahniashvili:2005qi,Kisslinger:2015hua}. 
On small scales, the HT turbulence is dominated by helicity dissipation,
and hence, the transfer of helicity is effective.
Thus, the resulting degree of polarization stays constant at large wave
numbers, and approximately equal to the fractional magnetic helicity.
On the other hand, the HK turbulence is dominated by energy dissipation,
such that the polarization decays to zero due to the vanishing
helicity \cite{Kahniashvili:2005qi}.
In both cases, the spectral peak of the polarization spectrum is at
the same scale as the GW spectral peak, which
is at a wave number approximately
twice the wave number of the magnetic spectral
peak.\footnote{The GW energy density is
sourced by the stress tensor, computed from the convolution of
the magnetic field in Fourier space.
This causes stress spectrum to peak at a wave number twice that of the magnetic peak.}
For low values of the fractional magnetic helicity, the maximum
degree of polarization also diminishes.
More recently, the two types of spectra have been studied 
to model the turbulence produced in a first-order EWPT
in ref.~\cite{Ellis:2020uid}, in the context of detection prospects with LISA 
by using the dipole
modulation induced by the proper motion of the solar system,
as proposed in refs.~\cite{Seto:2006hf,Seto:2006dz}, and recently applied to
LISA in ref.~\cite{Domcke:2019zls}.
The potential detection of polarization can be improved by cross-correlating
two space-based GW detectors
as, for example, LISA and Taiji \cite{Seto:2020zxw,Orlando:2020oko}.
We show that the circular degree of polarization computed from direct
numerical simulations follows the HT turbulence model of previous analytical
works if the magnetic field is assumed to be present at the initial time
of generation. 
This scenario neglects the production of GWs that occurs while the magnetic field
is generated by any of the described magnetogenesis mechanisms or via MHD dynamo.
When we consider a magnetic field that is initially zero and it builds up
during the simulation, following ref.~\cite{Pol:2019yex},
both the HK and the HT models of turbulence fail to predict the
spectrum of the degree of circular polarization, and numerical simulations are required
\cite{Kahniashvili:2020jgm}.

Recently, in refs.~\cite{Pol:2018pao,Pol:2019yex},
the authors have described the
implementation of a GW solver into the {\sc Pencil Code} \cite{PC} and
have presented direct numerical simulations for modeling development and dynamics of
primordial hydrodynamic and
hydromagnetic turbulence from phase transitions,
and subsequent generation of a stochastic GW
background, also computed numerically.
Their simulations included fully helical sources, but
the estimation of the GW polarization degree spectrum,
as well as the polarization detection prospects,
were not the focus of their studies.
We present here two types of simulations, similar to the two types of
hydromagnetic simulations presented in ref.~\cite{Pol:2019yex}: one where
a primordial magnetic field is assumed to be given as the initial condition
and one where a magnetic field is generated by an electromotive force
$\FFF(\xx,t)$ that depends on time $t$ and 
position $\xx$.
In particular, regarding the first type,
we study the cases with an initial stochastic
magnetic field with different values of the fractional helicity,
from non-helical up to the fully helical case.
In ref.~\cite{Kahniashvili:2020jgm}, the degree of circular polarization for
kinetically and magnetically forced turbulence was presented,
which is similar to our second type of simulations.
We complement their analysis in the present work by studying the variation
of the polarization degree 
in the different scenarios of the magnetic field generation.
Furthermore, we study cases where turbulence is driven for times
significantly shorter than what was considered in
ref.~\cite{Kahniashvili:2020jgm}.
The driving is then applied during a short time interval
(around a 10\% of the Hubble time),
and then switched off such that
turbulence decays for later times.
By using suitably scaled variables and conformal time, the governing
equations describing the evolution of GWs and turbulent magnetic fields
in an expanding universe in the radiation era can be brought into a
form that is best suited for numerical simulations \cite{Pol:2018pao}.
We explore the detectability of the generated GW signal and its
polarization with planned space-based GW detectors.

We begin by summarizing our approach and the equations
solved in \Sec{model}.
We then present the magnetic and GW energy spectra obtained from the numerical simulations in
\Sec{numerical_sec}.
In particular, the degree of circular polarization is shown
in \Sec{spec_pol_section} and compared with previous analytical models in \Sec{an_spec_pol_section}.
We explore the potential detectability of the
GW background amplitude and polarization by space-based GW detectors and, in particular, by combining LISA and Taiji,
in \Sec{detectability_sec}, and we conclude in \Sec{conclusions}.

Throughout this work, electromagnetic quantities are expressed in
Lorentz--Heaviside units where the vacuum permeability is unity.
Einstein index notation is used so summation is assumed over repeated
indices.
Latin indices $i$ and $j$ refer to spatial coordinates 1 to 3.
The Kronecker delta is indicated by $\delta_{ij}$, the Levi-Civita tensor by
$\varepsilon_{ijk}$, the Dirac delta function by $\delta (x)$, and the
Heaviside step function by $\Theta(x)$.

\section{The model}
\label{model}

\subsection{Gravitational signal from MHD turbulence}

We perform direct numerical simulations of the MHD turbulence starting
at the time of generation, which belongs to the radiation-dominated era
and can be appropriately scaled to, e.g., the EWPT.
At every time step of the MHD simulation, we compute the
contributions from velocity and magnetic fields to the stress tensor
$T_{ij}$.
Then, we solve the GW equation to compute the strains $h_{ij}$,
sourced by the traceless and transverse projection of the stress tensor. 
The details of the numerical setup and application to the electroweak scale
are described in refs.~\cite{Pol:2018pao,Pol:2019yex}.

We use the linear polarization modes $+$ and $\times$ to describe
the two gauge-independent components of the tensor mode perturbations,\footnote{%
When the GWs are unpolarized, the amplitudes of the $+$ and $\times$ modes
are the same, and we only have one gauge-independent component.}
such that $\tilde{h}_{ij}(\kk) = \tilde{h}_+(\kk) e_{ij}^+(\hat{\kk}) + 
\tilde{h}_\times(\kk) e_{ij}^\times (\hat{\kk})$ \cite{Varshalovich:1988ye},
where the tilde indicates that this decomposition is performed
in Fourier space,\footnote{We use the Fourier convention,
$\tilde h (\kk) = \int h(\xx) e^{-i \kk \cdot \xx} \dd^3 \xx$, such that
the inverse Fourier transform is $h (\xx) = (2 \pi)^{-3}
\int \tilde h (\kk) e^{i \kk \cdot \xx} \dd^3 \kk$.}
and $\hat \kk=\kk/|\kk|$.
The linear polarization basis tensors are
\begin{equation}
e_{ij}^+ (\hat \kk) = e_i^1 e_j^1 - e_i^2 e_j^2,
\quad
e_{ij}^\times (\hat \kk) = e_i^1 e_j^2 + e_i^2 e_j^1,
\label{basis_pol}
\end{equation} 
where $\ee^1$ and $\ee^2$ form a basis with the unit vector $\hat \kk$
\cite{Hu:1997hp}. 
We solve the non-dimensional
GW equation in the radiation era for the scaled strains
$\tilde{h}_{+,\times} (\kk, t)$ \cite{Deryagin:1986qq}, using conformal time,
normalized to unity at the initial time of magnetic field generation
$t_\ast$, and comoving wave vector, normalized by
$1/(c t_\ast)$, as described in refs.~\cite{Pol:2018pao,Pol:2019yex},
\begin{equation}
\left( \partial^2_t + \kk^2\right)
\tilde{h}_{+, \times} (\kk, t)\, =\, {6 \over t}
\tilde{T}_{+, \times}^{\rm TT} (\kk, t),
\label{GW}
\end{equation}
where $\tilde{T}_{+, \times}^{\rm TT}(\kk, t)$ is the comoving stress tensor, projected
into the traceless and transverse (TT) gauge, described by
the linear polarization modes $+$ and $\times$, and normalized by the
energy density at $t_\ast$.
The scaled strains are tensor mode perturbations over
the Friedmann-Lema\^itre-Robertson-Walker metric
tensor, such that the line element is
$\dd s^2 = a^2(-\dd t^2 + [\delta_{ij} 
+ h_{ij}/a]\, \dd x_i \, \dd x_j)$.
During the radiation-dominated epoch, the equation of state is
$p = \onethird \rho$, where $\rho$ is the energy density and
$p$ the pressure.
This leads to a linear evolution of
the scale factor $a$ with $t$, and allows one to get rid of the 
damping term \cite{Gris74}, which should be included otherwise in
\Eq{GW}.\footnote{%
We take the scale factor $a$ to be unity at the time
of generation, which allows one to simply write $a = t$, and $t_*=H_*^{-1}$,
where $H_*$ is the Hubble rate at the time of generation.
More generally}, one should substitute $t$ by $a$ in 
the denominator of the sourcing term of \Eq{GW}.
The stress is composed of magnetic and kinetic contributions
and computed in physical space as
\begin{equation}
T_{ij} (\xx) =\frac{4}{3}\frac{\rho u_i u_j}{1 - \uu^2} - B_i B_j +
\left(\frac{\rho}{3} + \frac{\BB^2}{2}\right)
\delta_{ij},
\label{Tij}
\end{equation}
where $\uu$ is the plasma velocity and
$\BB$ is the magnetic field.
The total enthalpy is $w = p + \rho = \fourthird \rho$.
Since $T_{ij}$ refers to comoving and normalized stress tensor, the
MHD fields ($\rho$, $\uu$, and $\BB$) are accordingly normalized and
comoving.

The non-dimensional and comoving MHD equations for an ultrarelativistic
gas in a flat expanding universe in the radiation-dominated 
era after the EWPT are given by \cite{Brandenburg:1996fc}
\begin{eqnarray}
{\partial\ln\rho\over\partial t}
&=&-\frac{4}{3}\left(\nab\cdot\uu+\uu\cdot\nab\ln\rho\right)
+{1\over\rho}\left[\uu\cdot(\JJ\times\BB)+\eta \JJ^2\right], \label{dlnrhodt} \\
{\partial\uu\over\partial t}&=&-\uu\cdot\nab\uu
+{\uu\over3}\left(\nab\cdot\uu+\uu\cdot\nab\ln\rho\right)
+{2\over\rho}\nab\cdot\left(\rho\nu\SSSS\right) \nonumber \\
&&-{1\over4}\nab\ln\rho -{\uu\over\rho}\left[\uu\cdot(\JJ\times\BB)+\eta\JJ^2\right] +{3\over4\rho}\JJ\times\BB, \\
{\partial\BB\over\partial t}&=&\nab\times(\uu\times\BB-\eta\JJ+\FFF),\quad
\JJ=\nab\times\BB, \label{dBdt}
\end{eqnarray}
where ${\sf S}_{ij}=\half(\partial_j u_i +\partial_i u_j)-\onethird\delta_{ij}\nab\cdot\uu$
are the components of the rate-of-strain tensor,
$\JJ$ is the current density,
$\nu$ is the kinematic viscosity, and $\eta$ is the
magnetic diffusivity.
The electromotive force $\FFF$ is used to model the generation of magnetic
fields.

\subsection{Magnetic fields present at the initial time}

In the first type of runs, we consider the magnetic field to be present
at the initial time of the simulation, so
we set the electromotive force term $\FFF$ to be zero at all times.
We generate a random three-dimensional vector field in Fourier space,
\begin{equation}
\tilde B_i({\kk})=B_0\left(P_{ij}(\hat \kk)-i \sigma\epsilon_{ijl}
\hat k_l\right) \tilde g_j({\hat \kk})\, g_0(k),
\label{Bikk}
\end{equation}
where $B_0$ is the magnetic field amplitude,
$\tilde g_j(\hat \kk)$ is the Fourier transform of a $\delta$-correlated
vector field in three dimensions with Gaussian fluctuations,
i.e., $g_i(\xx) g_j(\xx') = \delta_{ij} \delta^3(\xx - \xx')$,
$\sigma$ is a parameter that allows one to control the fractional magnetic helicity,
$P_{ij} (\hat \kk) = \delta_{ij} - \hat k_i \hat k_j$ is the projection operator.
The spectral shape is determined by $g_0(k)$~\cite{Brandenburg:2017neh},
\begin{equation}
g_0(k)={k_*^{-3/2} (k/k_*)^{\alpha/2-1}
\over[1+(k/k_*)^{2(\alpha-\beta)}]^{1/4}},
\label{Sfunction}
\end{equation}
where $\kf$ sets the scale of the spectral peak, which is identified with
the initial wave number of the energy-carrying eddies.
The magnetic energy density is
$\EEM=\half \bra{\BB^2}$,\footnote{Angle brackets denote ensemble average over stochastic realizations,
which can be approximated as the average of the random field over the
physical domain for a statistically homogeneous field.\label{foot_angle_br}}
such that its initial value is $\EEMmax=\half B_0^2$.
Due to the normalization used, this value corresponds to a fraction of the
radiation energy density at the time of magnetic field generation, and since
this case corresponds to decaying turbulence, $\EEMmax$ is the maximum
value of the magnetic energy density.
The magnetic spectrum $\EM (k)$, computed such that $\EEM =
\int \EM(k) \, \dd k$,  is proportional to $k^2 g_0^2 (k)$, with
a spectral index $\alpha=4$ in the low wave number limit 
(subinertial range) for a Batchelor spectrum,\footnote{
For magnetic fields produced by causal processes, e.g., during cosmological
phase transitions, the correlation length is finite, which leads
to a Batchelor magnetic spectrum $\EM (k) \propto k^4$ in the limit $k\rightarrow 0$
\cite{MY75}.}
and Kolmogorov-type spectral slope $\beta=-5/3$ 
in the high 
wave number range.\footnote{%
The Kolmogorov-type $k^{-5/3}$ spectrum is found and well-established in purely
hydrodynamic turbulence \cite{Kol41}.
In general MHD, a $k^{-3/2}$ Iroshnikov-Kraichnan spectrum has been proposed
in refs.~\cite{Irosh64,Kraichnan:1965zz}.
However, direct numerical simulations of MHD turbulence have found
approximately Kolmogorov and steeper scalings 
\cite{Muller:2000zz,Christensson:2000sp}.}
Here, the $k_*^{-3/2}$ prefactor ensures that the resulting magnetic
energy $\EEM$ is independent of the value of $k_*$.
The exponents $\zeta=2$ and $\zeta^{-2}=1/4$ in the denominator of \Eq{Sfunction}
determine the transition smoothness
from one slope to the other around the spectral peak.
The initial fractional helicity of the magnetic field, $\PPM =
\kf \bra{\AAA \cdot \BB}/ \bra{\BB^2}$,\footnote{The general definition
of $\PPM$ uses a characteristic wave number $k$ computed from the integration
of the helical spectrum over wave numbers, which is, in general, different than the
spectral peak $\kf$.} is given by
$2\sigma/(1+\sigma^2)$, being $\AAA$ the magnetic vector potential,
such that $\BB = \nabla \times \AAA$.

\subsection{Magnetic fields forced at the initial time}

In the second type of simulations, to model the magnetic field generation with 
fractional magnetic helicity, we use the electromotive force $\FFF$,
which is non-zero for a short amount of time, and its value is given by
\begin{equation}
\FFF(\xx,t;\kf)={\rm Re}\{{\cal A}{\tilff}(\kk(t))\exp[i\kk(t)\cdot\xx+i\varphi]\},
\label{FFF_model}
\end{equation}
where the wave vector $\kk(t)$ and the phase $\varphi(t)$ change randomly
from one time step to the next.
This forcing function is therefore white noise in time and consists of plane
waves with average wave number $k_*$ such that $|\kk|$ lies in an interval
$k_*-\delta k/2\leq|\kk|<k_*+\delta k/2$ of width $\delta k$.
Here, ${\cal A}$ is the amplitude of the forcing term.
The Fourier amplitudes of the forcing are
\begin{equation}
\tilde{f}_i=\left(\delta_{ij}-i\sigma\epsilon_{ijl} \hat k_l \right)
\tilde{f}_j^{(0)}\Bigl/\sqrt{1+\sigma^2},\;
\label{forcing}
\end{equation}
where ${\tilff}^{\rm(0)}({\kk})=(\kk\times\ee)/[\kk^2-(\kk\cdot\ee)^2]^{1/2}$
is a non-helical forcing function.
Here, $\ee$ is an arbitrary unit vector that is not aligned with $\kk$.
Note that $|\tilff|^2=1$.
The parameter $\sigma \in [-1,1]$ is related to the fractional helicity of the 
forcing term, with $\sigma=0$ and $\sigma=\pm 1$ corresponding to non-helical
and maximally helical cases, respectively.
The forcing is only enabled during an arbitrarily short time interval
$1\leq t\leq \tmax$, to reproduce the more realistic scenario in which the magnetic
field does not appear abruptly, but it is built up to its
maximum value $\EEMmax$ at $\tmax$, and then it decays.
We chose $\tmax = 1.1$ in the present work,
which corresponds to a 10\% of the Hubble time.
In ref.~\cite{Kahniashvili:2020jgm}, the authors consider forcing
up to $\tmax = 3$, so the forcing is active for 2 Hubble times, although
its amplitude is considered to decrease linearly.

\subsection{Characterization of stochastic magnetic and strain fields}
\label{spectra_def}

The magnetic fields considered and the resulting velocity fields
and tensor mode perturbations, are all stochastic fields.
We present here the spectral functions that are used
to describe the statistical properties of these fields.

The autocorrelation function of the magnetic field,
assuming statistical homogeneity and isotropy, and a
Gaussian-distribution in space,\footnote{For a random field with Gaussian 
distribution, the two-point autocorrelation function is sufficient to describe
its statistical properties \cite{MY75}.} is
\begin{equation}
\bra{\tilde B_i^*(\kk, t) \tilde B_j(\kk', t)} = (2\pi)^6 \delta^3
(\kk - \kk') \biggl[P_{ij} (\hat \kk) \frac{\EM (k, t)}{4 \pi k^2} +
i \epsilon_{ijl} \hat k_l \frac{\HM (k, t)}{8 \pi k} \biggr],
\label{equaltime}
\end{equation}
where $\EM (k, t)$ and $\HM (k, t)$ are the magnetic and helicity
spectra, respectively.
We work here with spectra per linear wave number
interval.

The magnetic field is either given at the initial time; see \Eq{Bikk}, or
driven using the forcing term $\FFF$, described in \Eqs{FFF_model}{forcing},
for a short time $1 \leq t \leq \tmax$, being $\tmax=1.1$.
In both cases, we have introduced a parameter $\sigma$ that allows one
to control the fractional helicity of the initial magnetic
field or the initial forcing term.
In general, we define the fractional helicity $\PPM(t)$ as
\begin{equation}
\PPM (t) = \frac{\displaystyle\int_0^\infty k \HM (k, t) \, \dd k}
{2\displaystyle\int_0^\infty \EM(k, t) \, \dd k} = \frac{2\sigM (t)}
{1 + \sigM^2 (t)}.
\label{PPM}
\end{equation}
Initially, if the magnetic field is given, $\PPM$ is determined by the
chosen value of $\sigM=\sigma$.
On the other hand, if the magnetic field is initially driven, the fractional
helicity depends on the value of $\sigma$ used in the forcing term;
see \Eq{forcing}, but its exact value is obtained by solving the set of MHD equations
and $\sigM$ might differ from $\sigma$.
For later times, in both cases, the values of $\sigM(t)$ and $\PPM(t)$ are given by the
dynamical evolution of the MHD fields.
The magnetic polarization spectrum is directly computed from the helical and
magnetic spectra, $\PPM(k, t) = \half k \HM(k, t)/\EM(k, t)$; see \Eq{PPM}.
The realizability condition gives an upper bound to the helicity
spectrum $\HM(k, t)$~\cite{Mof78},
\begin{equation}
\left|\half k\HM(k,t)\right|\leq\EM(k,t),
\label{RealSpec}
\end{equation}
such that the magnetic polarization $\PPM (k, t)$ takes values from $-1$ to 1. 
Due to the realizability condition,
to directly compare the magnetic and the helicity spectra, the latter
is usually multiplied by $k/2$; see \Eqs{PPM}{RealSpec}.

The GW energy density is \cite{Misner:1974qy}
\begin{equation} 
\EEGW(t) = \frac{c^2}{32 \pi G} \bra{\dot{h}_{ij}^{\rm phys} (\xx, t)
\dot{h}_{ij}^{\rm phys} (\xx, t)},
\label{EEGW}
\end{equation}
where $h_{ij}^{\rm phys} = h_{ij}/a$ are the physical strains, 
the angle brackets denote space average; see footnote \ref{foot_angle_br},
and a dot represents derivative with respect to physical time
$t_{\rm phys}$.\footnote{%
The GW energy density in \Eq{EEGW} is given in non-normalized units,
being the strains $h_{ij}^{\rm phys}$ defined such that
$\dd s^2 = a^2 (-\dd t^2 +[\delta_{ij} + h_{ij}^{\rm phys}]\,\dd x_i \, \dd x_j)$,
and the physical time refers to non-normalized cosmic time, which is related to
conformal time as $\dd t_{\rm phys} = a \,\dd t$.}
In terms of the normalized and comoving units used in \Eq{GW},
the ratio of comoving GW energy density to critical energy
density $\OmGW = \EEGW/\EEcrit^0$ is
\begin{equation}
a^4 \OmGW(t) = \frac{1}{12} \biggl( \frac{H_*}{H_0} \biggr)^2
\bigl\langle \partial_t h_{ij}\, \partial_t h_{ij} +
 h_{ij} h_{ij}/t^2 - 2 h_{ij}\,
\partial_t h_{ij}/t \bigr\rangle,
\label{OmGW}
\end{equation}
where $\EEcrit^0 = 3 H_0^2 c^2/(8 \pi G)$, with
$H_0 = 100\,h_0 \km \s^{-1} \Mpc^{-1} \approx 3.241
\times 10^{-18}\,h_0 \s^{-1}$ being the Hubble rate
at the present time, and $h_0$ takes into account the
uncertainties in its exact value \cite{Maggiore:1999vm}.
The equal time correlation function for general tensor fields
$\Pi_{ij}^a$ and $\Pi_{ij}^b$ (assuming isotropic and homogeneous random fields) is 
expressed as \cite{Caprini:2003vc}
\begin{align}
\bra{\tilde \Pi^a_{ij} (\kk, t) \tilde \Pi^b_{lm} (\kk', t) } = & \, \frac{1}{4}
(2\pi)^6 \delta^3(\kk - \kk')
\biggl[{\cal M}_{ijlm} (\hat \kk) \frac{S_{ab} (k, t)}{4\pi k^2} +
i {\cal A}_{ijlm} (\hat \kk) \frac{A_{ab} (k, t)}{4\pi k^2} \biggr],
\label{etc}
\end{align}
where
\begin{align}
{\cal M}_{ijlm} (\hat \kk) = & \, P_{il} P_{jm} + P_{im} P_{jl} -
P_{ij} P_{lm},\\
{\cal A}_{ijlm} (\hat \kk) = & \, \half \hat{\kk}_q (P_{jm} \varepsilon_{ilq} +
P_{il} \varepsilon_{jmq} + P_{im} \varepsilon_{jlq} +
P_{jl} \varepsilon_{imq}).
\end{align}
To compute the spectral functions of the GW energy density;
see \Eq{OmGW}, we
use \Eq{etc} applied to the strains $h_{ij}$ and their time derivatives
$h'_{ij}=\partial_t h_{ij}$: when $\Pi^a_{ij} = \Pi^b_{ij} = h_{ij}$, we define
$S_h$ and $A_h$; when $\Pi^a_{ij} = \Pi^b_{ij} = h'_{ij}$, we define
$S_{h'}$ and $A_{h'}$, and when $\Pi^a_{ij} = h_{ij}$ and $\Pi^b_{ij} =
h'_{ij}$ (or vice versa, note that \Eq{etc} is symmetric in $ab$),
we define $S_{\rm mix}$ and $A_{\rm mix}$.
The spectra of GW energy density $\OmGW(k, t)$ and GW helicity/chirality
$\XiGW (k, t)$ are
\begin{align}
a^4 \OmGW(k, t) = & \, \frac{1}{12} \biggl( \frac{H_*}{H_0} \biggr)^2
k \biggl[S_{h'} (k, t) + \frac{1}{t^2} S_h (k, t) - \frac{2}{t}
S_{\rm mix} (k, t) \biggr] \nonumber \\
= & \, \bigl(H_*/H_0 \bigr)^2 \, k \EGW (k, t),
\label{OmGW_k} \\
a^4 \XiGW(k, t) = & \, \frac{1}{12} \biggl( \frac{H_*}{H_0} \biggr)^2
k \biggl[A_{h'} (k, t) + \frac{1}{t^2} A_h (k, t) - \frac{2}{t}
A_{\rm mix} (k, t) \biggr] \nonumber \\
= & \, \bigl(H_*/H_0 \bigr)^2 \, k \HGW (k, t),
\label{XiGW_k} 
\end{align}
such that $\OmGW(t) = \int \OmGW(k, t) \, \dd \ln k \propto \int \EGW(k, t) \, \dd k$.
Using the $+$ and $\times$ polarization basis defined in \Eq{basis_pol},
the functions $S_{ab} (k, t)$ and $A_{ab} (k, t)$, defined in \Eq{etc},
for generic tensor fields $\Pi_{ij}^a$ and $\Pi_{ij}^b$, can be 
expressed as \cite{Caprini:2003vc}
\begin{align}
2\bra{\tilde \Pi_+^a (\kk) \tilde \Pi_+^{b,*}(\kk') +
\tilde \Pi^a_\times(\kk) \tilde \Pi_\times^{b,*} (\kk')} = &
\, (2\pi)^6 \delta^3(\kk - \kk') \frac{S_{ab}(k)}{4 \pi k^2}, \label{SPiij}\\
2 \bra{\tilde \Pi^a_+(\kk) \tilde \Pi_\times^{b,*} (\kk') -
\tilde \Pi^{a,*}_+(\kk) \tilde \Pi_\times^b (\kk')} = &
\, i (2 \pi)^6 \delta^3(\kk - \kk') \frac{A_{ab}(k)}{4\pi k^2},
\label{APiij}
\end{align}
which allows one to compute the spectral functions in \Eqs{OmGW_k}{XiGW_k}
from $\tilde h_{+,\times}$ and $\partial_t h_{+,\times}$, solutions to the GW \Eq{GW},
via shell-integration.\footnote{The spectral functions $S(k)$ and $A(k)$,
defined in \Eqs{SPiij}{APiij}, and omitting the subscript $ab$,
correspond to shell-integrated functions
of the tensor fields.
For example,
$$S(k,t) = \frac{4 \pi k^2}{(2\pi)^6}  \int 2 \left(\tilde \Pi_+^2(\kk, t) +
\tilde \Pi_\times^2 (\kk, t)\right) \, \dd \Omega_{k},$$
where $\Omega_{k}$ is the solid angle of the shell of size $k$,
such that $\int S(k, t) \, \dd k = \bra{\Pi_{ij} (\xx,t) \Pi_{ij} (\xx, t)}$.}
The degree of circular polarization of the GW background is
$\PPh (k, t) = A_h(k, t)/S_h(k, t)$ \cite{Kahniashvili:2005qi}. 
We also define the polarization using the GW energy density spectral
functions: $\PPGW (k, t) = \HGW(k, t)/\EGW (k, t)$.\footnote{%
In previous works, these two definitions are used interchangeably, assuming that
$\EGW = k^2 S_h$, and $\HGW = k^2 A_h$,
which hold in the absence of sources.
However, turbulent sources can modify the dispersion relation of the strains,
and $\PPh \neq \PPGW$ in general.}
The total GW polarization can be expressed as the ratio of the integrated
spectra over wave numbers, $\PPGW (t) = \XiGW(t)/\OmGW(t)$.

\section{Numerical results}
\label{numerical_sec}

We have computed solutions for a range of values of $\sigma$, using
both given and driven initial fields; see
table~\ref{runs} for a summary of the different runs.
In general, the different GW modes grow up to $\delta t = t - \tmax \sim {\cal O}
(k^{-1})$; see figure~3 of ref.~\cite{Pol:2019yex}, when they start to oscillate,
taking $\tmax = 1$ in the runs with an initial magnetic field.
The duration $\delta t$ is the time that it takes to the GWs,
which propagate at the speed of light, to reach the scales corresponding to the wave number $k$.
We have set up all the runs to have a spectral peak $\kf \approx 600$
(see \Tab{runs}), which corresponds to, approximately, 100 times the
Hubble wave number $2 \pi/H_*$ or, equivalently, to a 100th of the Hubble scale,
and the smallest wave number of the simulations is
$k_0\approx 100$.
Hence, the GW spectrum stops growing and enters the oscillatory stage
at $\delta t \sim {\cal O} (10^{-2})$, and after that time we average the spectra
over oscillations in time to obtain the saturated GW spectra and
their integrated values over wave numbers $\OmGWsat$
and $\PPGW=\XiGWsat/\OmGWsat$, given in \Tab{runs} in units of
$a^{-4} (H_*/H_0)^2$.

In the present work, all the numerical simulations are performed using
a periodic cubic domain of size $L = 2\pi/k_0$ with a discretization of $n^3 = 1152^3$ mesh points.
To solve the GW equation, given by \Eq{GW},
we use the {\sc Pencil Code} \cite{PC}, following the
methodology described in sec.~2.6 of ref.~\cite{Pol:2018pao}, which is denoted 
there as approach II.
\EEq{GW} is sourced by the strain tensor,
which is obtained by solving the MHD
\Eqss{dlnrhodt}{dBdt}.
Following ref.~\cite{Pol:2019yex}, we fix the viscosity $\nu = \eta$
and choose it to be as small as possible (see \Tab{runs}), but still
large enough such that the inertial range of the computed
spectra is appropriately resolved \cite{Brandenburg:2017neh}.
Their physical values in the early universe are much smaller than what we can
accurately simulate and they would require much larger numerical resolution.
The inertial range of the turbulence would extend to higher frequencies.
However, those higher wave numbers are of little observational interest since
the GW amplitude at those wave numbers would be very low, as shown in the
results presented below.

\begin{table}[t!]
    \centering
    {\footnotesize
    \renewcommand{\arraystretch}{1.25}
    \begin{tabular}{l|l|c|c|l|l|c|c|c|c}
        Type & \multicolumn{1}{|c|}{$\sigM$} &
            $\EEMmax$ & $\OmGWsat$ & \multicolumn{1}{|c|}{$\PPM$} &
            \multicolumn{1}{|c|}{$\PPGW$} & $\kf$ & $n$ & $\nu$, $\eta$ & $k_\nu$\\
        \hline
ini & \ $0.1$ & $3.93 \times 10^{-3}$ & $2.53 \times 10^{-11}$ & \ $0.19$ & \ $0.18$ & 600 & 1152 & $5 \times 10^{-8}$ & $8.1 \times 10^{4}$ \\
ini & \ $0.3$ & $4.23 \times 10^{-3}$ & $3.00 \times 10^{-11}$ & \ $0.55$ & \ $0.54$ & 600 & 1152 & $5 \times 10^{-8}$ & $8.2 \times 10^{4}$ \\
ini & \ $0.5$ & $4.85 \times 10^{-3}$ & $4.09 \times 10^{-11}$ & \ $0.80$ & \ $0.77$ & 600 & 1152 & $5 \times 10^{-8}$ & $8.5 \times 10^{4}$ \\
ini & \ $0.7$ & $5.78 \times 10^{-3}$ & $5.72 \times 10^{-11}$ & \ $0.94$ & \ $0.91$ & 600 & 1152 & $5 \times 10^{-8}$ & $8.9 \times 10^{4}$ \\
ini & \ $1$ & $7.75 \times 10^{-3}$ & $1.04 \times 10^{-10}$ & \ $1.00$ & \ $0.95$ & 600 & 1152 & $5 \times 10^{-8}$ & $9.6 \times 10^{4}$ \\
forc & $\!\!\!-0.01$ & $1.42 \times 10^{-2}$ & $2.64 \times 10^{-9}$ & $\!\!\!-0.006$ & \ $0.01$ & 600 & 1152 & $5 \times 10^{-7}$ & $2.2 \times 10^{4}$ \\
forc & \ $0.01$ & $1.43 \times 10^{-2}$ & $2.65 \times 10^{-9}$ & \ $0.02$ & \ $0.06$ & 600 & 1152 & $5 \times 10^{-7}$ & $2.2 \times 10^{4}$ \\
forc & \ $0.3$ & $1.69 \times 10^{-2}$ & $3.14 \times 10^{-9}$ & \ $0.56$ & \ $0.58$ & 600 & 1152 & $5 \times 10^{-7}$ & $1.9 \times 10^{4}$ \\
forc & \ $0.5$ & $1.92 \times 10^{-2}$ & $3.54 \times 10^{-9}$ & \ $0.82$ & \ $0.78$ & 600 & 1152 & $5 \times 10^{-7}$ & $1.6 \times 10^{4}$ \\
forc & \ $0.7$ & $2.09 \times 10^{-2}$ & $3.96 \times 10^{-9}$ & \ $0.96$ & \ $0.88$ & 600 & 1152 & $5 \times 10^{-7}$ & $1.3 \times 10^{4}$ \\
forc & $\!\!\!-1$ & $2.12 \times 10^{-2}$ & $4.16 \times 10^{-9}$ & $\!\!\!-0.999$ & $\!\!-0.91$ & 600 & 1152 & $5 \times 10^{-7}$ & $1.1 \times 10^{4}$ \\
    \end{tabular}
    }
    \caption{Summary of numerical simulations and relevant parameters.}
    \label{runs}
\end{table}

\subsection{Runs with decaying magnetic field at the initial time}

\begin{figure*}[t!]\begin{center}
\includegraphics[width=.47\textwidth]{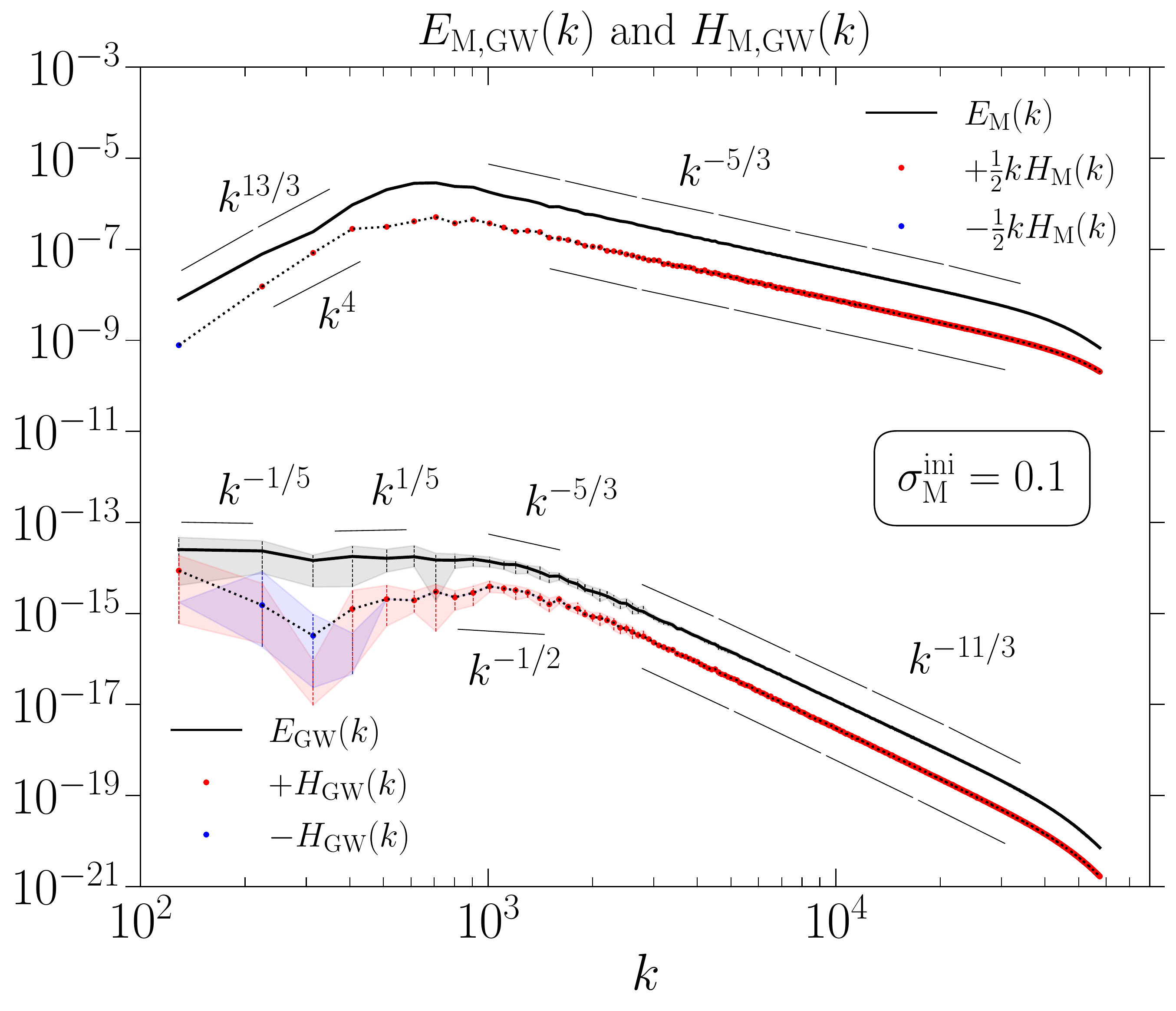}
\includegraphics[width=.47\textwidth]{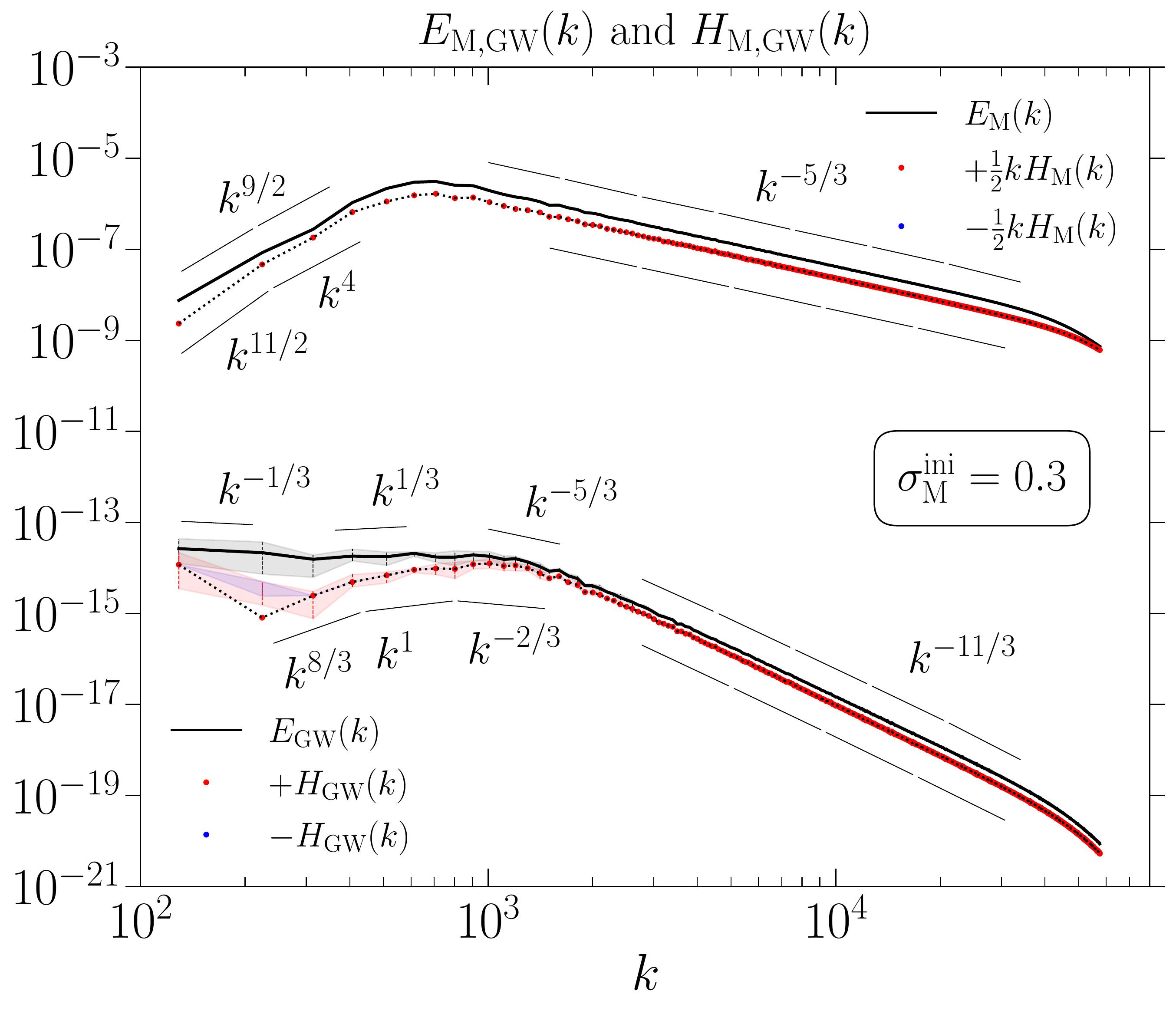}
\includegraphics[width=.47\textwidth]{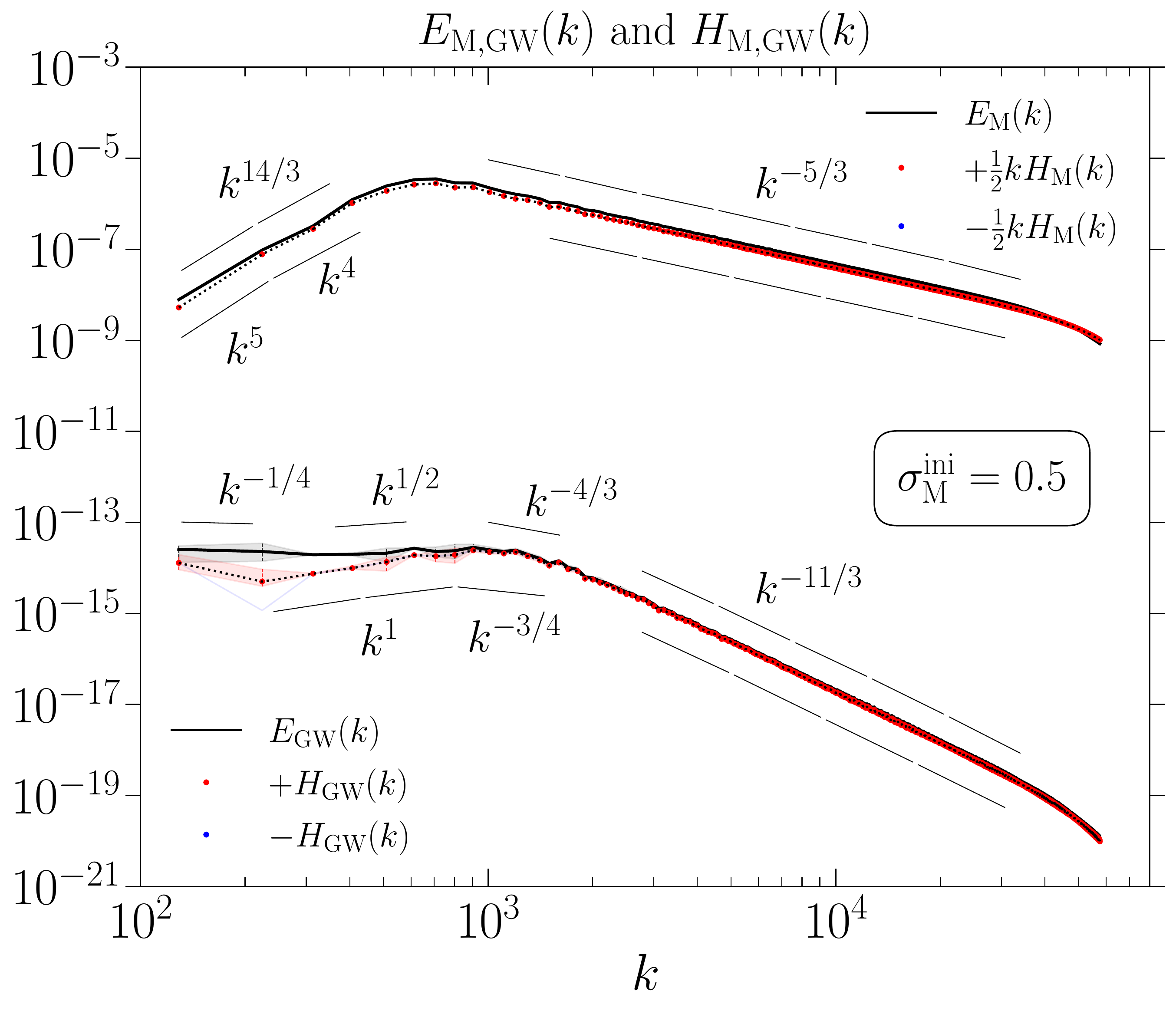}
\includegraphics[width=.47\textwidth]{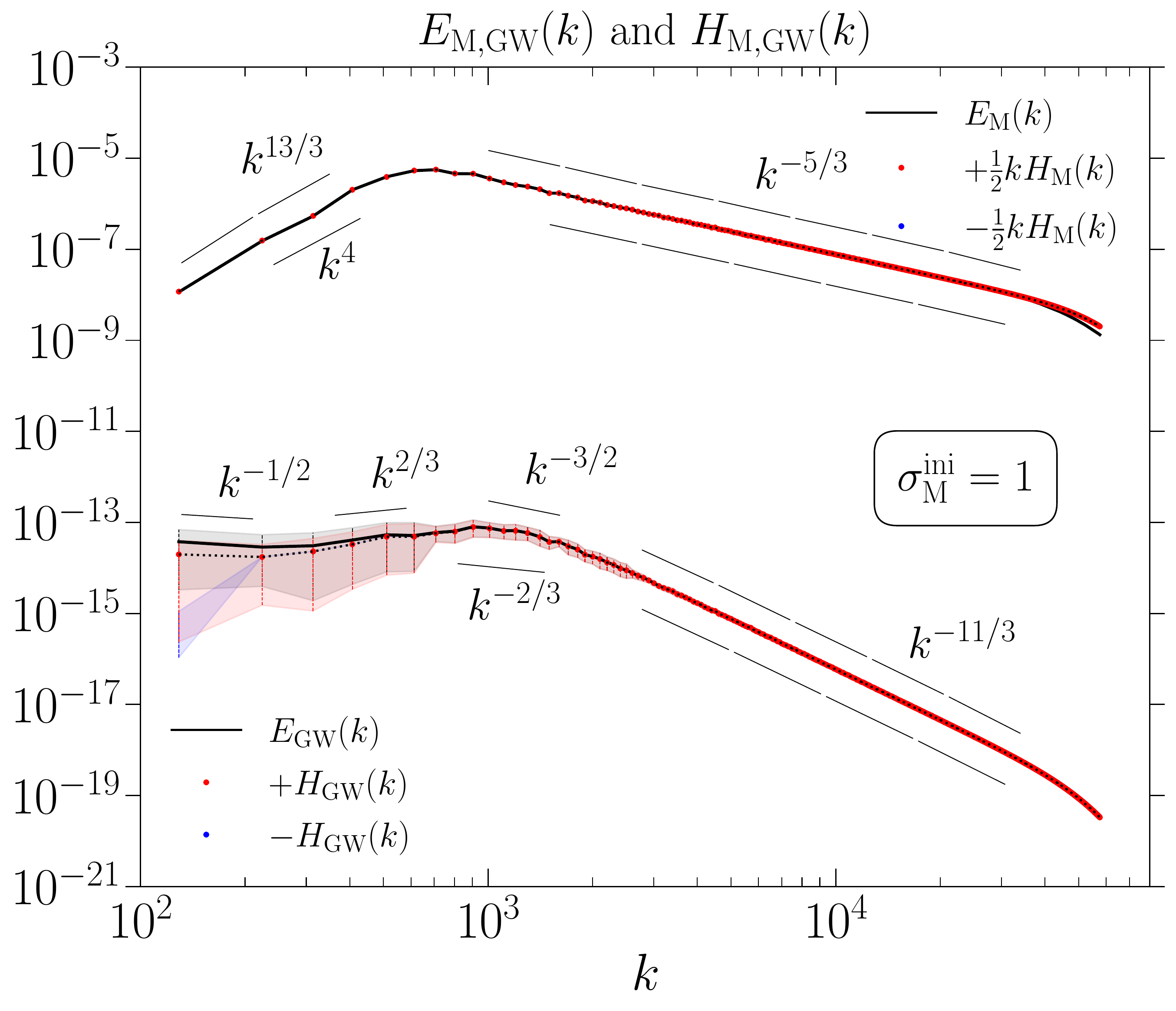}
\end{center}\caption[]{
Magnetic and GW energy (solid lines) and helicity (dashed lines)
spectra for the runs with a magnetic field given at the initial time
of the simulations, normalized to $t_* = 1$, for different values of
the parameter $\sigM$, which is related to the fractional magnetic helicity
as $\PPM = 2\sigM/(1 + \sigM^2)$.
The magnetic spectra are shown at the initial time of the simulation and
the GW spectra are the saturated ones (averaged over times $t > 1.01$),
while the shaded region shows the variation over time between the maxima and minima.
Positive values of the helicity spectra (magnetic and GW) are shown in red, while
negative values in blue.
The discretized intervals show the tangent power laws fitting the spectra
used to compute their local slopes, shown as a fraction of integers, with
a tolerance of $0.1$.
The values of the wave numbers and spectra are comoving and normalized by
the Hubble rate and the radiation energy density at the initial time, respectively.
}\label{pspecm_hel_initial}\end{figure*}

In \Fig{pspecm_hel_initial}, we compare the spectra of the magnetic field
and the resulting GWs, both the symmetric, $\EM(k)$ and $\EGW(k)$, and the antisymmetric
spectra, $\HM(k)$ and $\HGW(k)$, for different values of $\sigM$ at the initial time
of the simulation. 
Since the GW spectra fluctuate around an approximately statistically
steady spectrum,
we show the saturated values of the spectra at each mode computed
by averaging them over times larger than $t = t_* + 1/k_0 = 1.01$, and the shaded region corresponds
to the maxima and minima of the oscillations at every wave number. 
The magnetic spectra are shown at the initial time of the simulation,
when the magnetic energy density has its maximum $\EEMmax$, and it decays for
later times.

The magnetic helicity spectrum shows the same power law scalings as the spectrum
of the magnetic energy density in the inertial range, which correspond to 
$k^{-5/3}$ Kolmogorov-type spectra. 
Thus, this corresponds to HT turbulence; see \Sec{an_spec_pol_section}.
In the subinertial range, we observe that, as we decrease the fractional helicity,
the helical spectrum becomes slightly steeper than the $k^4$ Batchelor spectrum
observed in the magnetic energy density, being both spectra identical in the fully
helical case, as expected.
Note that the magnetic helicity spectra are multiplied by $k/2$ to take into 
account the realizability condition; see \Eq{RealSpec}.

On small length scales or, equivalently, large wave numbers (above the spectral peak $k_*$),
we observe that both the GW energy density and helicity
spectra, $\EGW(k)$ and $\HGW(k)$, follow $k^{-11/3}$ scalings,
which correspond to the Kolmogorov-type magnetic spectra,
in agreement with ref.~\cite{Pol:2019yex}.
On larger scales, we observe an approximately flat spectrum for the GW energy,
as shown from numerical
simulations in ref.~\cite{Pol:2019yex}.
This is a consequence of the $k^4$ Batchelor spectrum for a Gaussian magnetic (hence, 
divergenceless) field, which yields a $k^2$ (i.e., white noise) spectrum
of the magnetic stress \cite{Pol:2019yex,Brandenburg:2019uzj}.
The small deviations from an exact flat spectrum, which increase towards negative
slopes as we decrease the value of the fractional helicity, could be due to
the small
number of points in wave number space when we reach the largest scales of the
simulation, as well as to the oscillations over time.
It is also for large scales (small wave numbers) that we observe a decay of the
GW helical spectrum $\HGW(k)$ with respect to the flat spectrum observed for
$\EGW(k)$.
As we decrease the fractional helicity of the magnetic field, we observe this
decay to be at approximately twice the smallest wave number of the magnetic
field (expected from the source of the GWs that is given through
convolution).
This could be due to the steeper slope of the helical magnetic spectrum at low
wave numbers, impacting the helical spectrum of GWs.
Hence, we can expect the helicity spectrum of GWs to decay with respect to the
flat spectrum in the subinertial range, and omit the smallest wave number of the
simulation $k_0$, since the correct computation of this mode would require the
computation of the magnetic field at $k_0/2$.
We defer further discussion of the GW polarization
$\PPGW(k) = \HGW(k)/\EGW(k)$ to \Secs{spec_pol_section}{an_spec_pol_section}.

\subsection{Runs with forced magnetic field at the initial time}
\label{runs_forced}

\begin{figure*}[t!]\begin{center} 
\includegraphics[width=.49\textwidth]{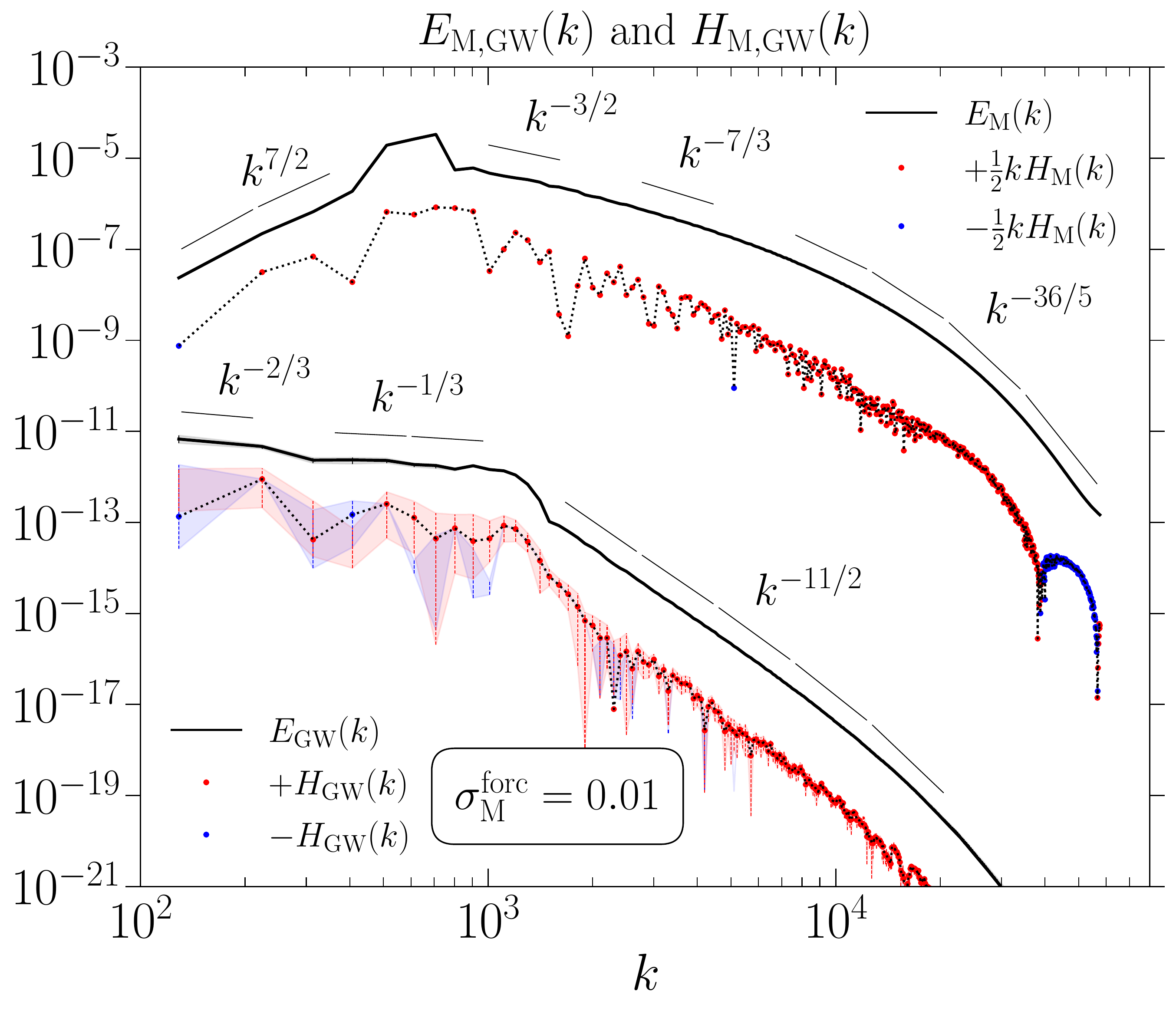}
\includegraphics[width=.49\textwidth]{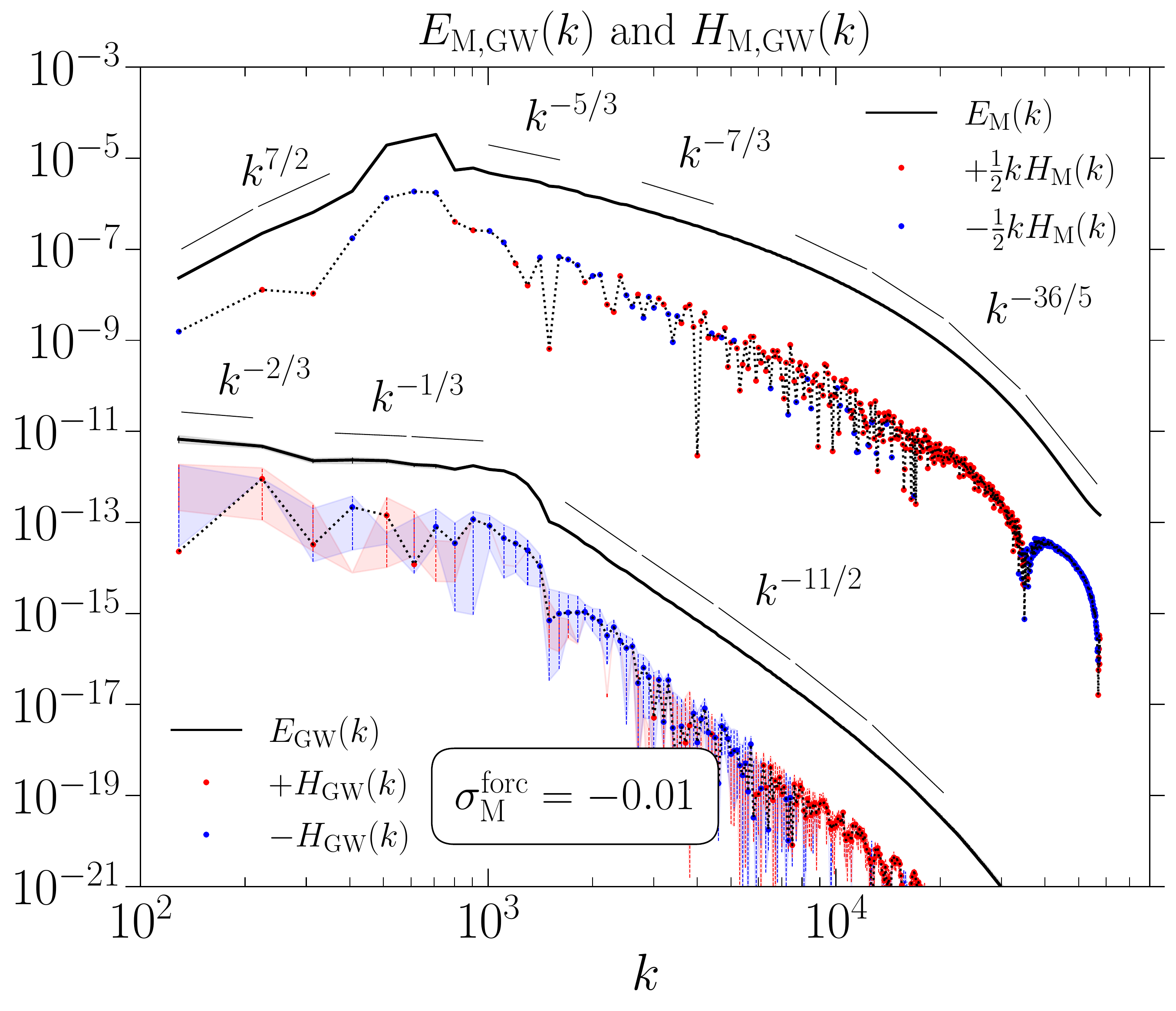}
\end{center}\caption[]{
Magnetic and GW energy (solid lines) and helicity (dashed lines)
spectra, similar to \Fig{pspecm_hel_initial}, for runs with an almost
non-helical forcing ($\sigma_{\rm M} = \pm 0.01$) for short times ($1\leq t \leq 1.1$).
}\label{pspecm_hel_forc001}\end{figure*}

\begin{figure*}[t!]\begin{center} 
\includegraphics[width=.47\textwidth]{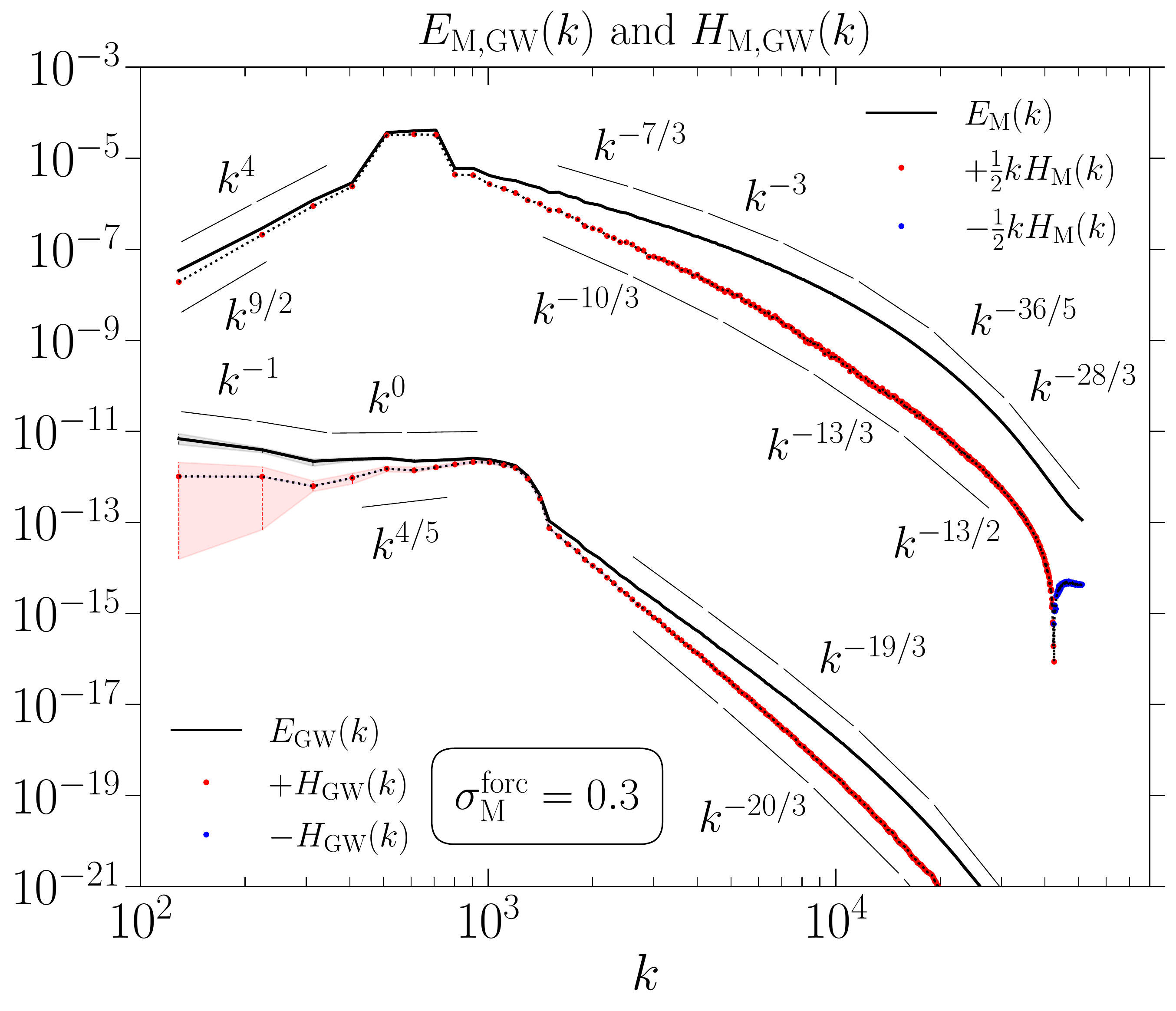}
\includegraphics[width=.47\textwidth]{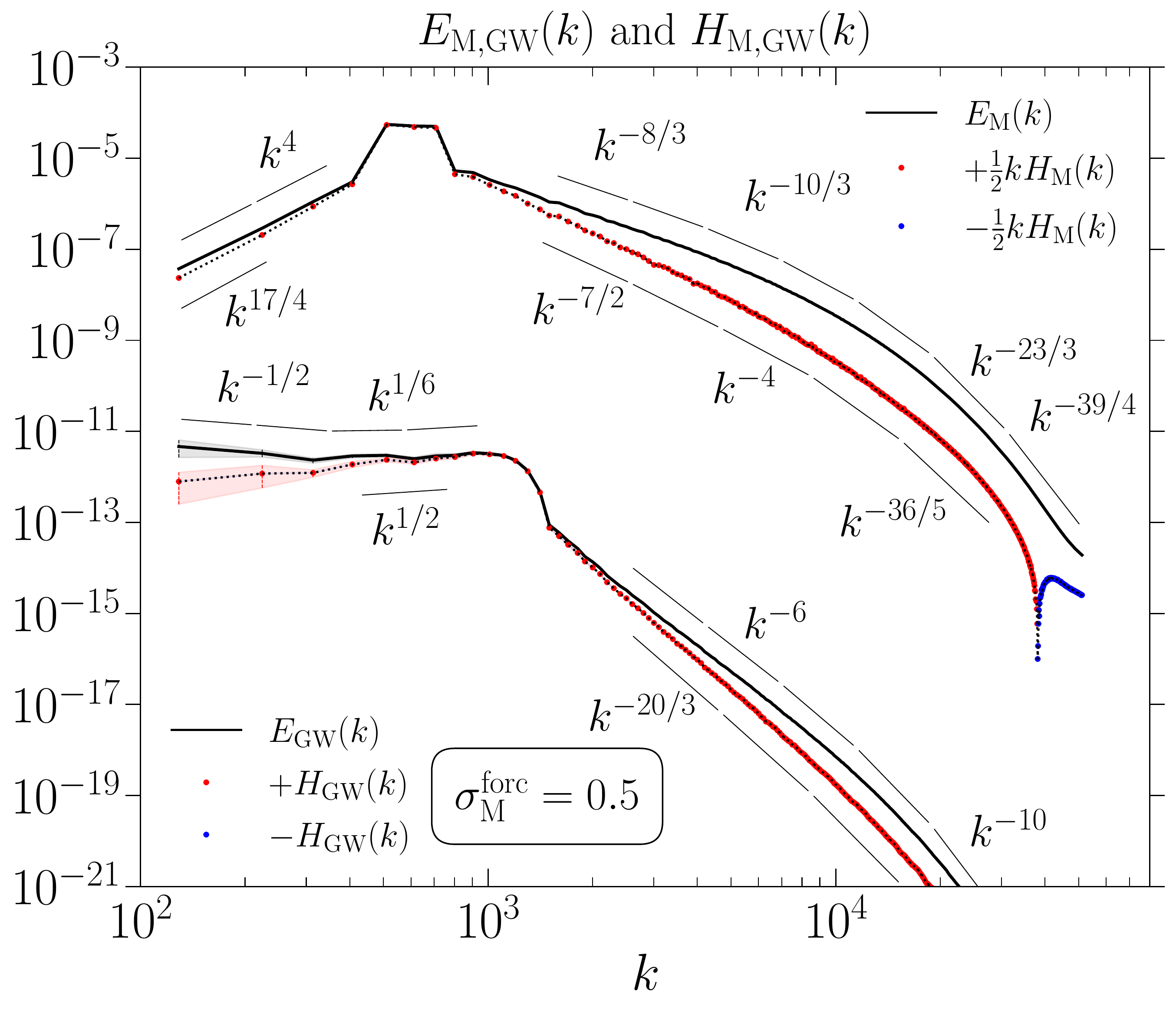}
\includegraphics[width=.47\textwidth]{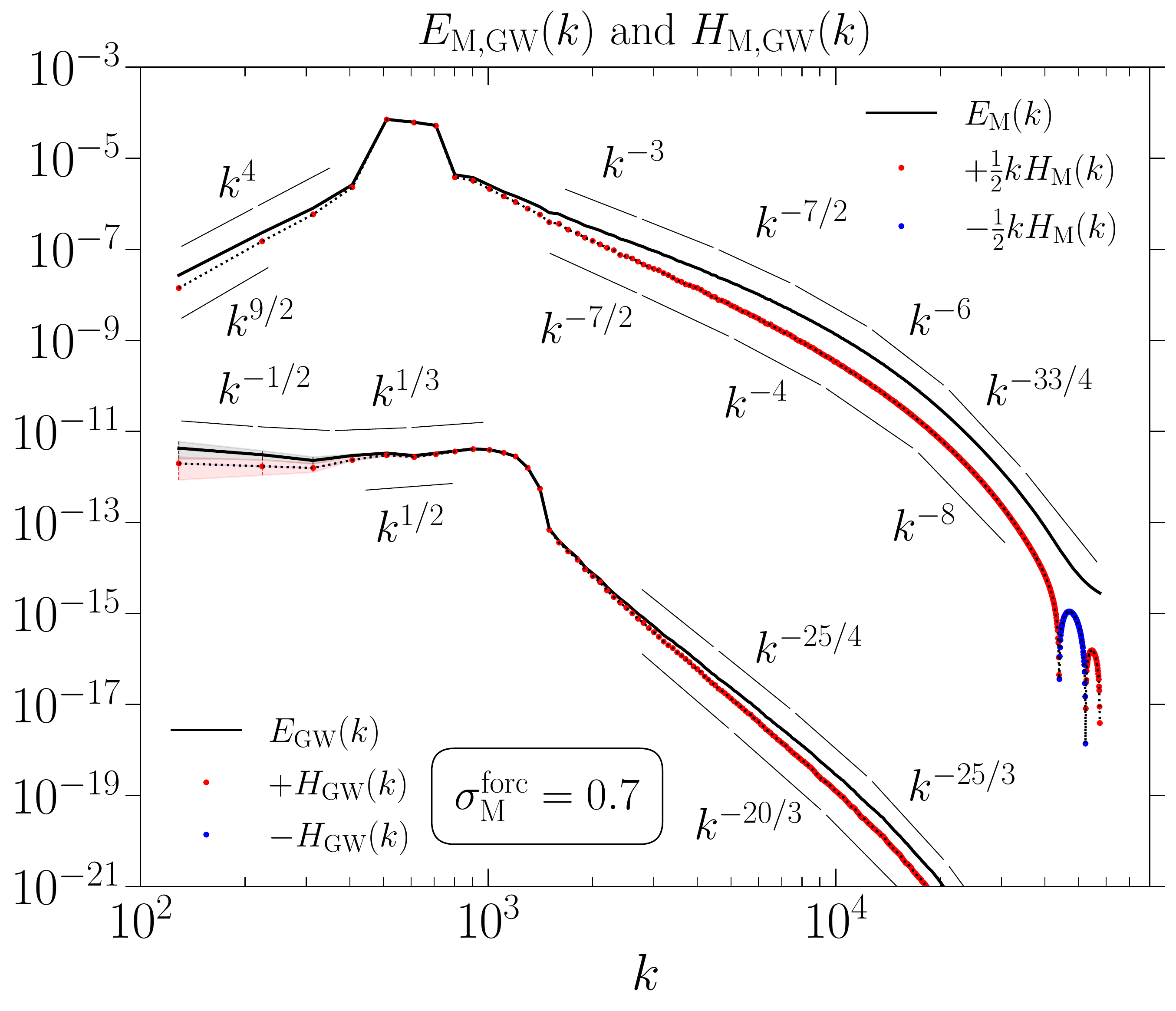}
\includegraphics[width=.47\textwidth]{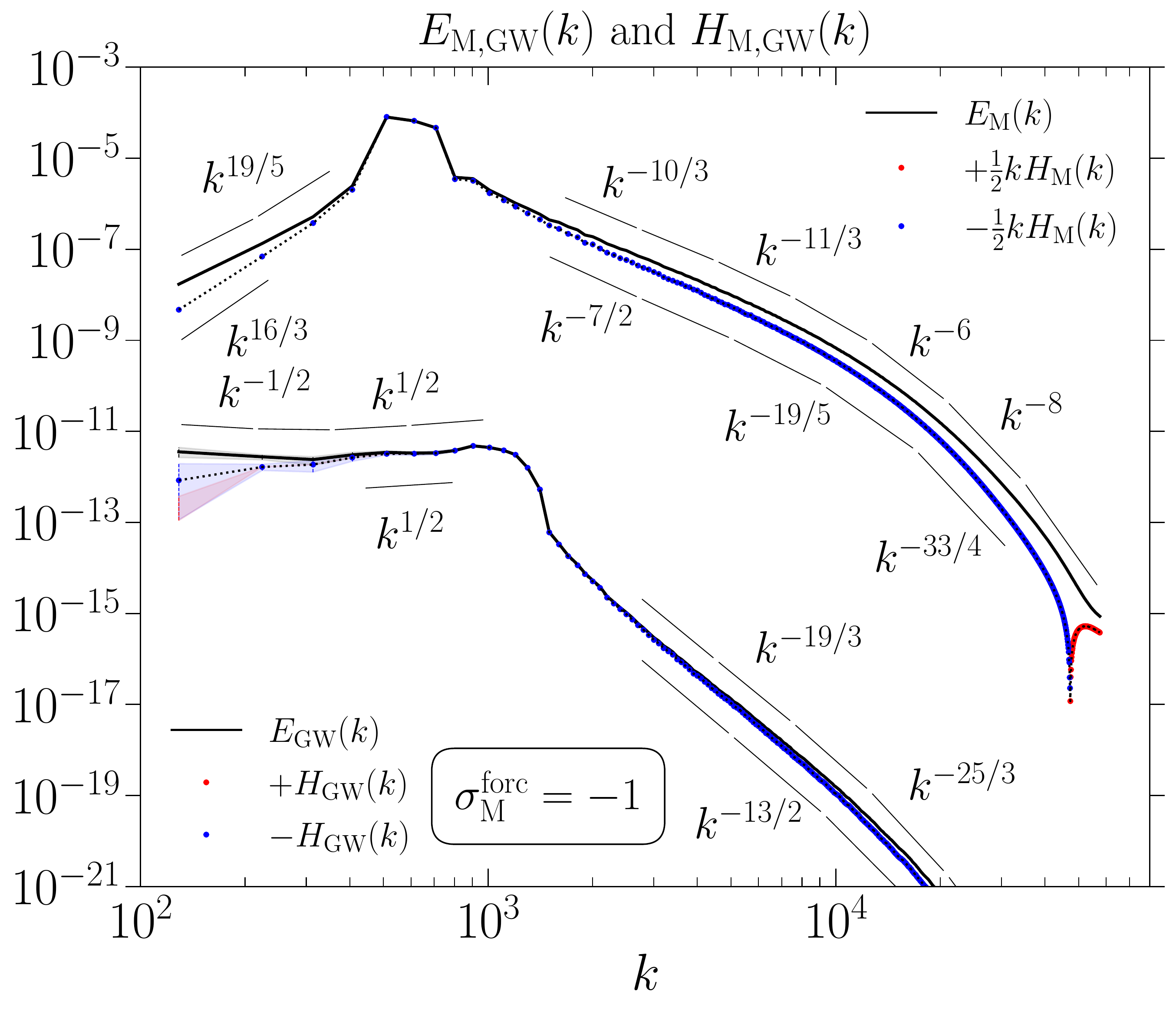}
\end{center}\caption[]{
Magnetic and GW energy (solid lines) and helicity (dashed lines)
spectra, similar to \Fig{pspecm_hel_initial}, for runs with forcing
for short times ($1\leq t \leq 1.1$), for different values of $\sigM$.\\
}\label{pspecm_hel_forc}\end{figure*}

\begin{figure}[t!]\begin{center}
\includegraphics[width=.47\textwidth]{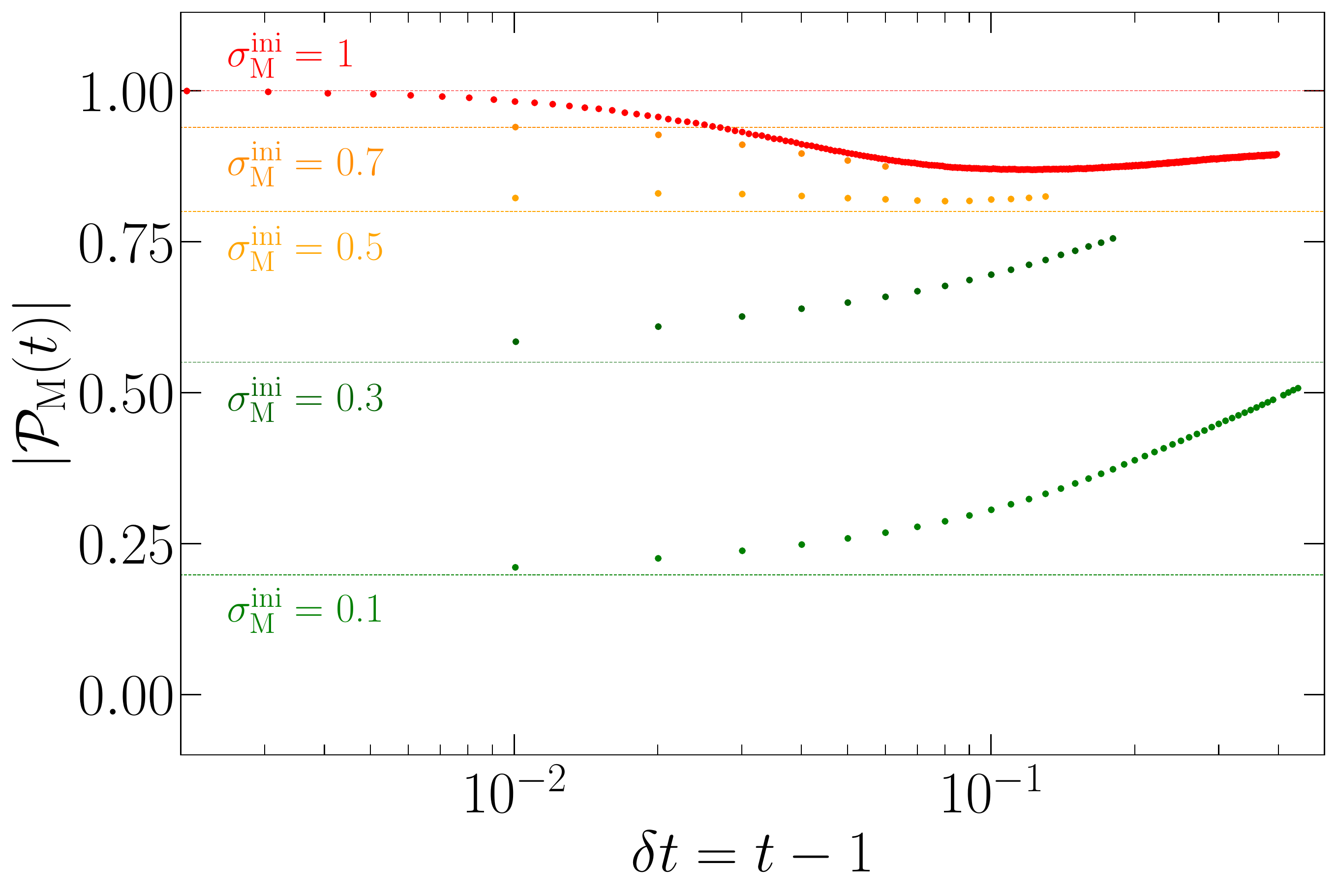}
\includegraphics[width=.47\textwidth]{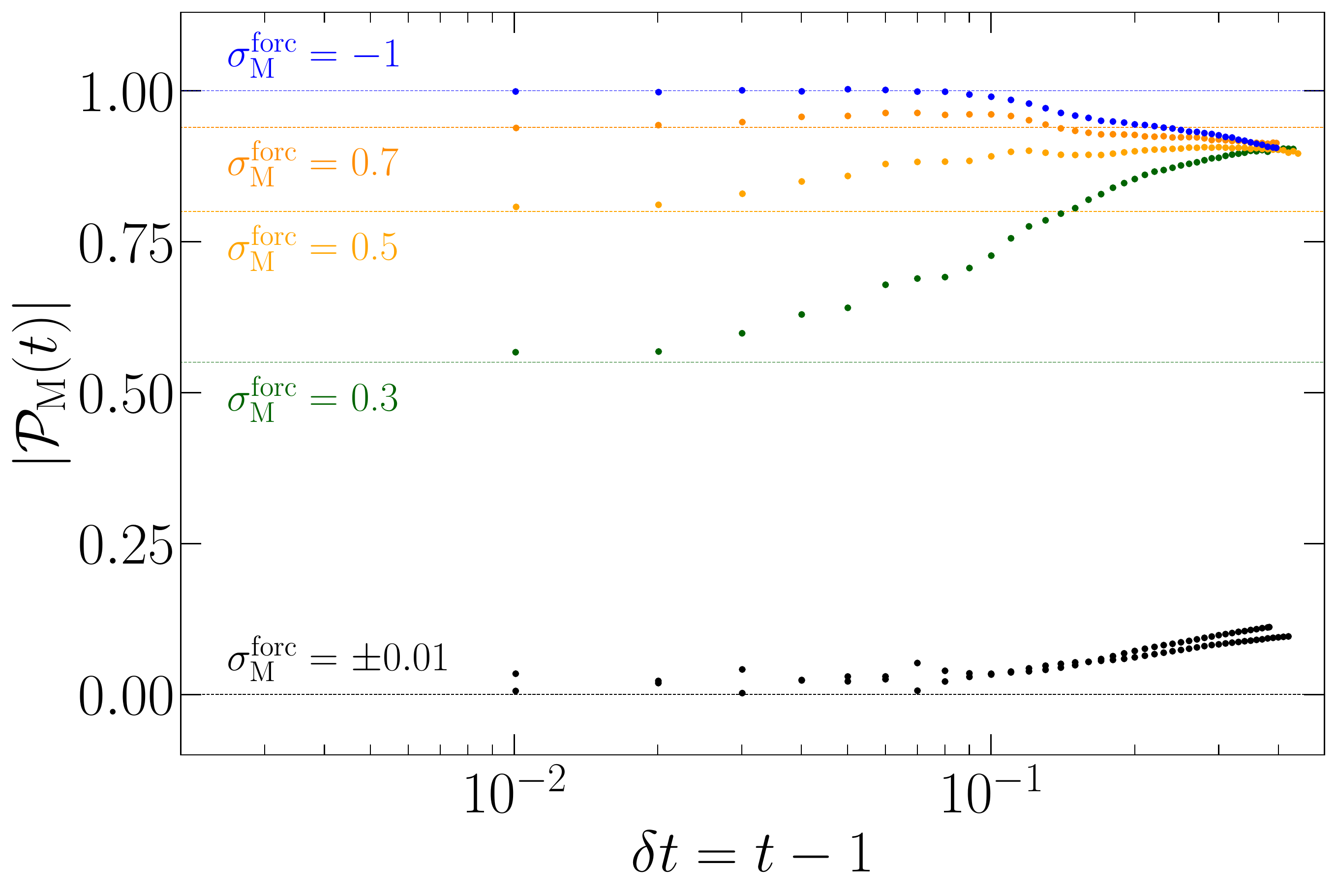}
\end{center}\caption[]{
Time evolution of the fractional helicity of the magnetic field, for the different
values of $\sigM$, for (a) the runs with an initial given magnetic field, and
(b) those with a forced magnetic field.
}\label{PM_vs_t}\end{figure}

In \Figs{pspecm_hel_forc001}{pspecm_hel_forc}, we compare the spectra of the magnetic field
and the resulting GWs for the runs in which we
drive the magnetic field for a short duration $\delta t = \tmax - 1 = 0.1$
(i.e., 10\% of the Hubble time).
In \Fig{pspecm_hel_forc001}, we show runs that are almost non-helical
($\sigM = \pm 0.01$), and in \Fig{pspecm_hel_forc}, we show runs with larger
fractional helicity of the forcing term,\footnote{%
For the case with forced magnetic fields, $\sigma$ corresponds
to the helicity parameter in the forcing term; see \Eq{forcing}, while $\sigM$
is related to the fractional magnetic helicity, defined in \Eq{PPM}.}
up to the fully helical case (which is negative in this case).
We show that magnetic fields with negative helicity drive GW with the same 
GW energy density than those produced by magnetic fields with positive helicity,
but the resulting helicity of the GW spectrum is negative.
The time evolution of the fractional magnetic helicity $\PPM (t)$ is shown
in \Fig{PM_vs_t}.
We observe that at very early times (up to $\delta t \approx 2 \times 10^{-2}$
for the runs with $\sigM = 0.3$ and $0.5$, and up to $\delta t \approx 
10^{-1}$ for the other runs), the value of $\sigM$ corresponds to the
parameter $\sigma$ of the forcing term.

The magnetic spectra are shown at the time $\tmax$, when the magnetic energy
density reaches its maximum value $\EEMmax$, and when we switch off the forcing
term. 
Note that at $\tmax = 1.1$; see \Fig{PM_vs_t},
the value of the fractional magnetic helicity is dynamically evolving
in time, and is already different than its initial value.
At earlier times, the magnetic spectrum shows a spike around the forcing 
wave number $\kf$, which is distributed at later times to a turbulent spectrum due to the MHD
dynamical evolution of the magnetic field.
We see that in all cases, the causal $\EM(k) \propto k^4$ magnetic spectrum
is established.
The spectra are averaged over times $t\geq 1.1$,
which is just after the maximum magnetic energy has been reached and
the GW energy begins to fluctuate around an approximately statistically
steady state.
For $k>k_*$, the slope of the magnetic energy spectrum is
steeper than a $k^{-5/3}$ Kolmogorov-type spectrum.
This is because of the finite time driving during the rather short time
interval, $1\leq t\leq 1.1$.
The consequences of the finite forcing time are the appearance of a smoothed
spike around $\kf$ which has not completely disappeared, and the steeper
slopes, especially in the inertial range.
This effect seems to be enhanced with helicity.
An exponential drop in the magnetic spectra is observed at the
largest wave numbers due to viscosity and magnetic diffusivity.
We can estimate the viscous cutoff wave number $k_\nu$ from the energy dissipation
rate $\epsilon \sim \nu \bra{\omega^2}$, where $\oo = \nab \times 
\uu$ is the fluid vorticity, as $k_\nu \sim (\epsilon/\nu^3)^{1/4}$
(see \Tab{runs}).
Note that for the runs with a given initial magnetic field, the diffusive
scale is around the Nyquist wave number so the inertial range is observed
down to the smallest scales of the simulation.
If the driving was continued over a long time interval
(long enough for the magnetic field to be
processed, i.e., $\tmax \sim 1$) we would recover
the Kolmogorov spectrum \cite{Kahniashvili:2020jgm}.
The magnetic helicity spectrum around the peak $\kf$ is nearly saturated
for all values of $|\sigM|$ considered from 0.3 to 1.
For $\sigM = \pm 0.01$, there is a clear separation between 
$\EM(k)$ and $k\HM(k)/2$ and the sign of the magnetic helicity tends
to fluctuate noticeably, especially at high wave numbers. 
We also see a systematic sign flip at higher $k$ both in $\HM (k)$ and
$\HGW (k)$, which also occurs for the cases with $|\sigM| \geq 0.3$. 
Such sign flips are typical of decaying helical turbulence and are a
consequence of the fact that the fractional magnetic helicity is there
already extremely small. 
The helicity spectrum $k \HM (k)$ is steeper than the magnetic spectrum $\EM(k)$,
while in the runs with an initial magnetic field,
both spectra follow the same power laws in the inertial range.
This is observed for all values of $\sigM$. 
However, for the fully helical case, the difference in slope is smaller, and it
becomes larger for smaller fractional helicity.

We again observe an approximately flat GW spectrum in the subinertial
range, which presents negative slopes that become steeper for smaller
values of $|\sigM|$.
For values of $|\sigM| \geq 0.3$, the spectrum becomes completely
flat near the spectral peak, and then presents an abrupt drop
below the peak, as shown in ref.~\cite{Pol:2019yex}, which is due
to the finite duration of the forcing, and related to the spike
that appears in the magnetic spectra.
The inertial range of the GW spectra also present steeper slopes
than the $k^{-11/3}$ obtained for the Kolmogorov scaling
in the case of initially given magnetic fields.
We observe the slopes $-5.5$, $-6.33$, $-6$, $-6.25$, and $-6.33$
below the spectral peak for $\sigM = \pm 0.01$, $0.3$, $0.5$, $0.7$,
and $-1$, respectively.
The helical GW spectrum is slightly shallower on small scales, and
becomes fully helical around the spectral peak.
Similar to the magnetic field, the helical GW spectrum decays faster
than the GW energy density in the subinertial range after a
characteristic wave number that increases with the fractional magnetic
helicity.
We observe that this decay starts at larger wave numbers for the GW
spectra than for the magnetic spectra.

\subsection{Degree of circular polarization}
\label{spec_pol_section}

\begin{figure*}[t!]\begin{center}
\includegraphics[width=.44\textwidth]{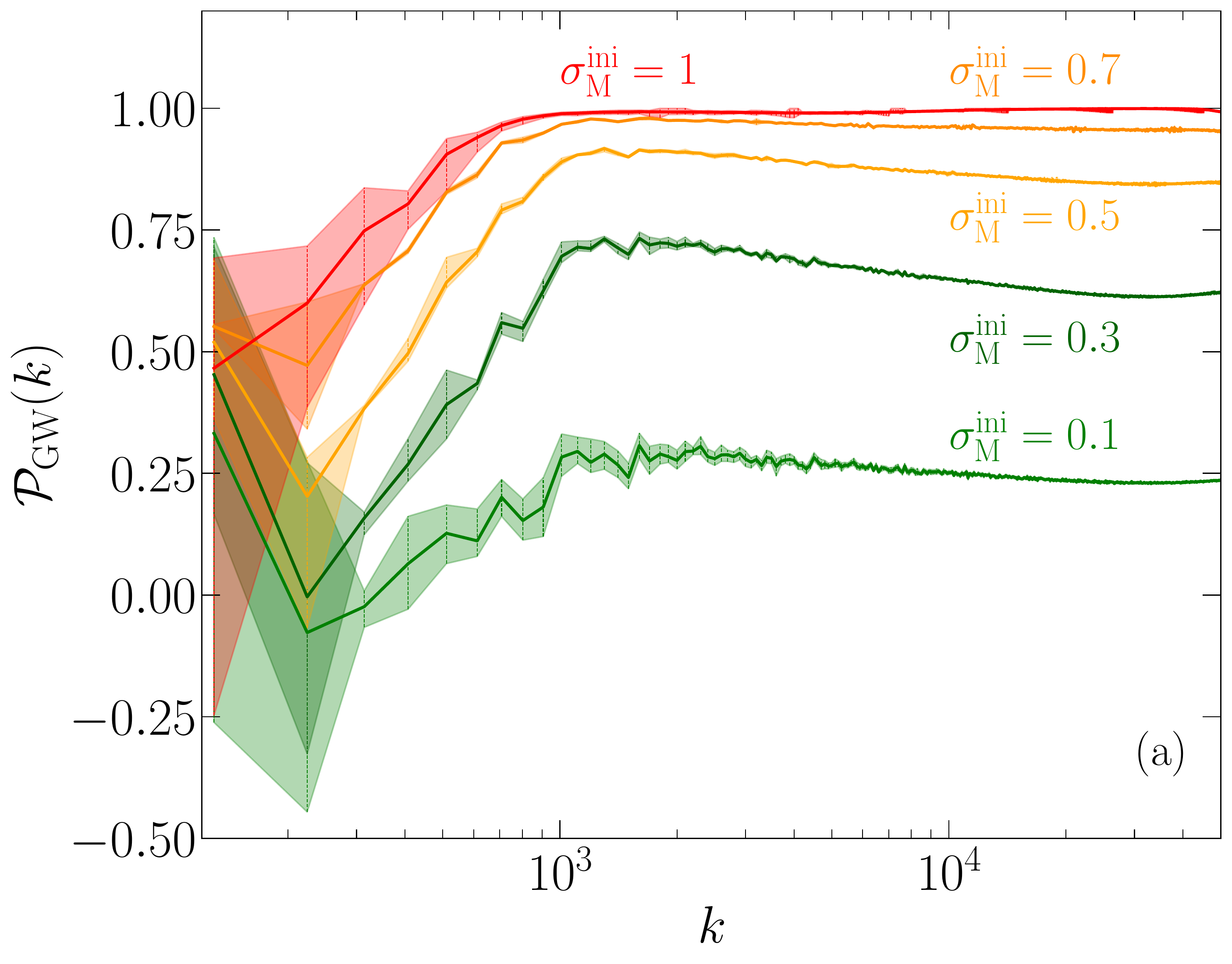}
\includegraphics[width=.44\textwidth]{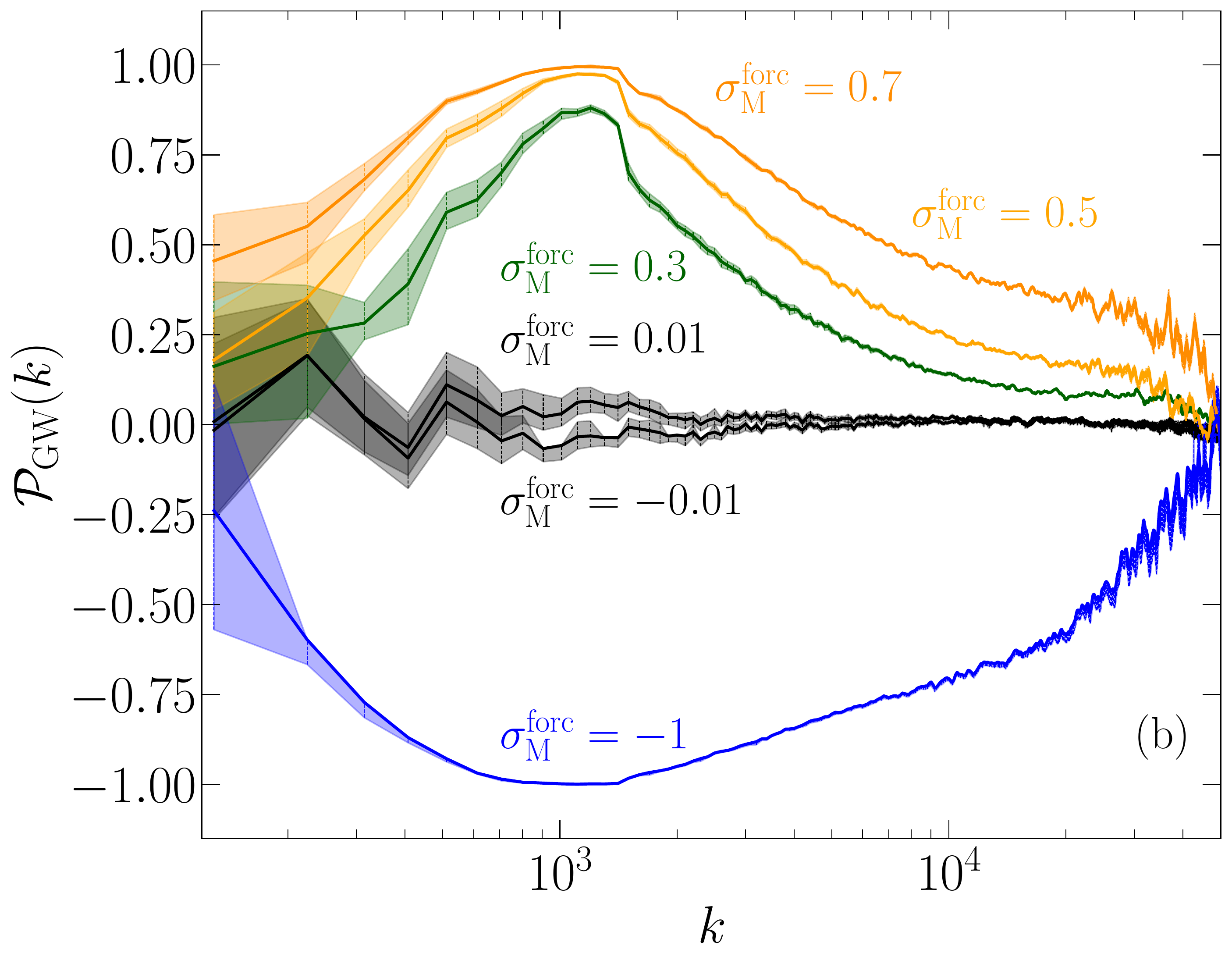}
\includegraphics[width=.44\textwidth]{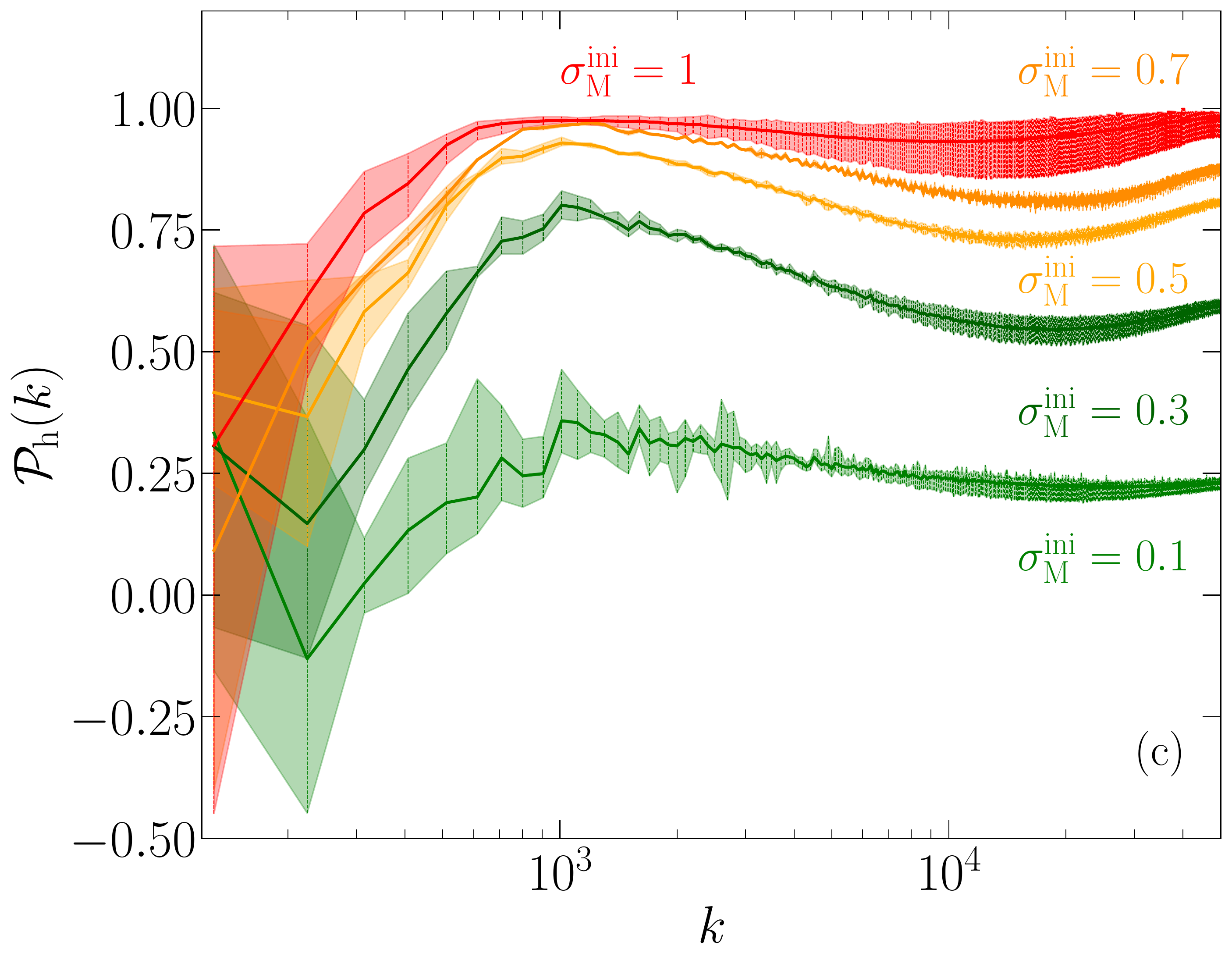}
\includegraphics[width=.44\textwidth]{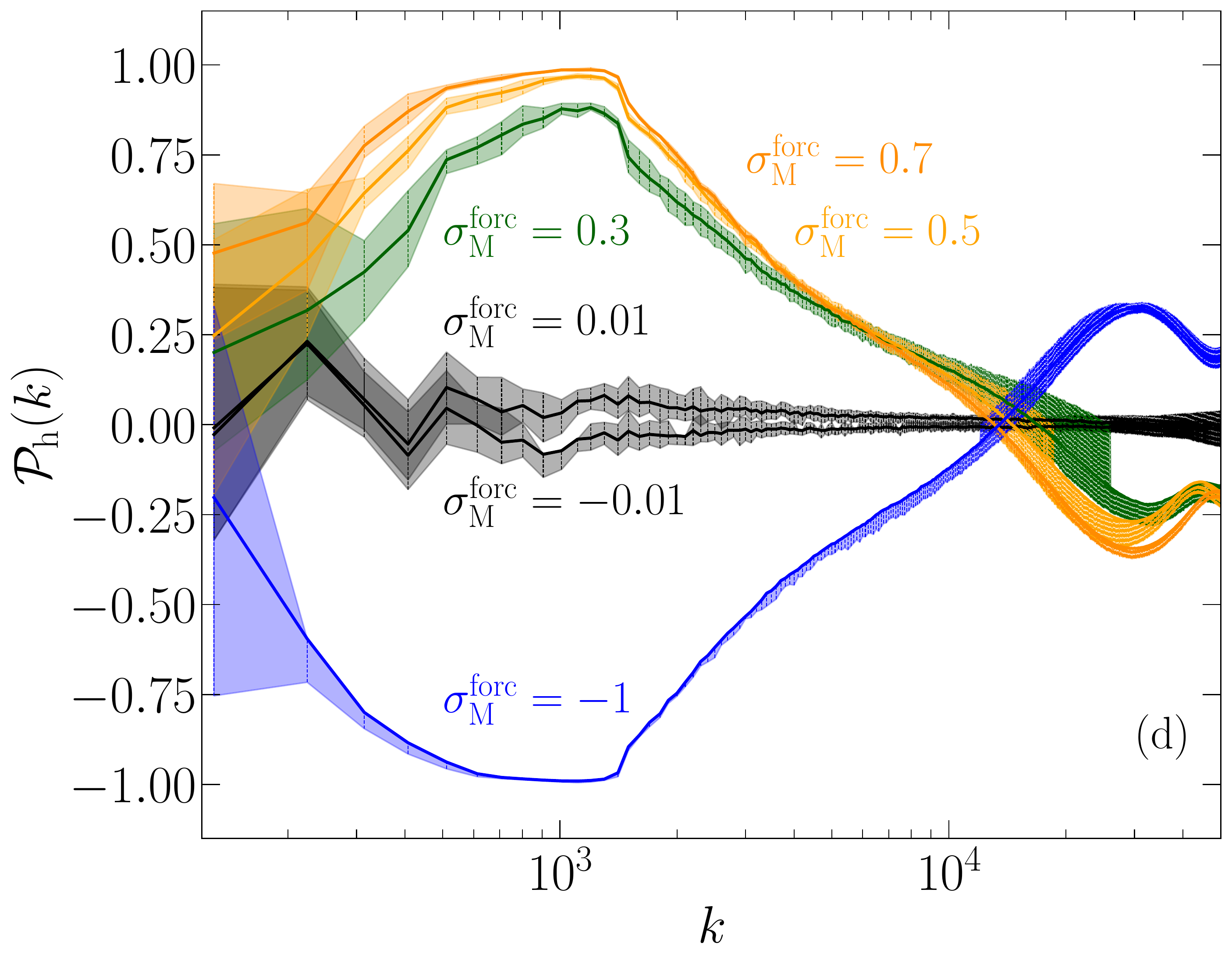}
\end{center}\caption[]{
GW degree of circular polarization, $\PPh (k)$ and $\PPGW (k)$,
for the different values of $\sigM$, 
for the types of simulations: 
runs with a given initial field; see panels (a) and (c), and
runs with a forced magnetic field for a short duration, i.e., for times
$1 \leq t \leq 1.1$; see panels (b) and (d).
The solid line represents the saturated value of the polarization (i.e.,
averaged over times in the oscillatory regime), and the shaded regions
are the maximum and minimum values of the polarization over oscillations.
}\label{ppol_comp}\end{figure*}

In \Fig{ppol_comp}, we plot the GW degree of circular polarization, $\PPh (k)$
and $\PPGW (k)$, for the two types of simulations.
We see that $\PPGW (k)$ reaches $\pm1$ at 
the GW spectral peak $\kGW \approx2k_*\approx1200$ when $\sigM=\pm1$. 
This polarization is larger than what was found in previous analytic predictions; see
ref.~\cite{Kisslinger:2015hua}, and recently used in ref.~\cite{Ellis:2020uid}
in relation to detectability with LISA.
This discrepancy can probably be explained by the use of the
simplified approximations made in the analytic calculations. 
Toward smaller wave numbers, there are systematic fluctuations around
$\PPGW (k)=0$, especially in the case of an initial given magnetic field, 
due to the oscillations of the helical GW spectrum at low wave numbers.

\begin{figure}[t!]\begin{center}
\includegraphics[width=.65\textwidth]{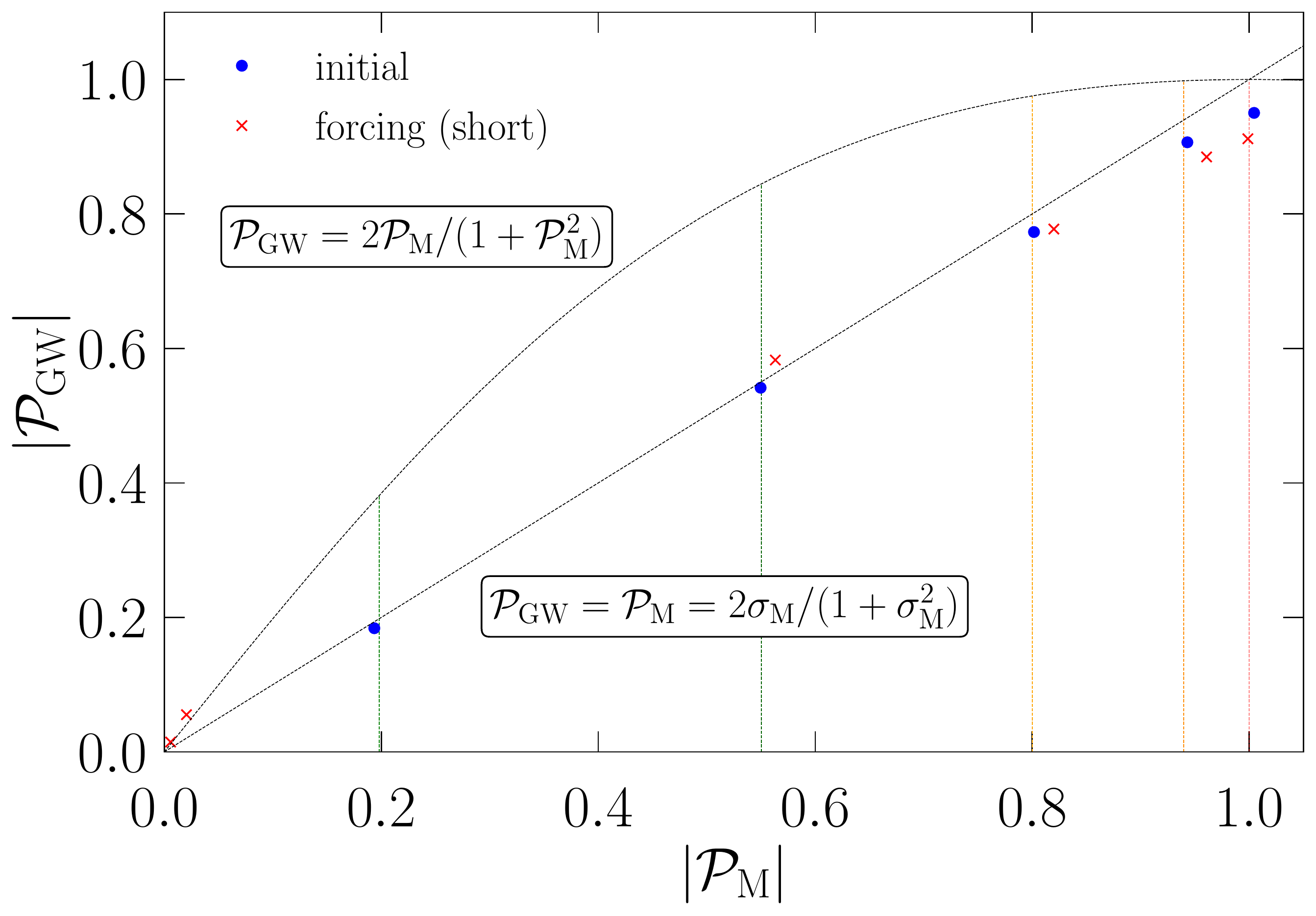}
\end{center}\caption[]{
GW polarization $\PPGW$ versus magnetic polarization $\PPM$;
see \Tab{runs}.
We obtain a numerical fit $\PPGW \approx \PPM = 2\sigM/(1 + \sigM^2)$,
both for the runs with an initial given magnetic field (`initial'), and with an
initially driven field (`forcing (short)').
We compare with the relation obtained for the analytical model
of \App{beltrami_app}, $\PPGW=2\PPM/(1+\PPM^2)$; see \Eq{PPh_bel}.
The vertical lines correspond to $|\sigM|=0.1$ (green), $0.3$ (dark green),
$0.5$ (orange), $0.7$ (dark orange), and $1$ (red).
}\label{pppol1}\end{figure}

The GW polarization integrated over all wave numbers $\PPGW$, as a function of the 
magnetic fractional helicity $\PPM=2\sigM/(1+\sigM^2)$, is shown in \Fig{pppol1}.
The value of the integrated magnetic fractional helicity $\PPM$, as a function
of time is shown in \Fig{PM_vs_t}.
In the runs with an initial given magnetic field, $\PPM$ stays constant 
for a short time interval $\delta t = t - 1 \sim 10^{-2}$,
which is similar to the time that the
GW spectrum takes to enter the stationary regime.
Hence, the value of $\PPM$ does not change while the GW spectrum is established.
However, for the case in which the magnetic field is driven for a duration 
$\delta \tmax = \tmax - 1 = 10^{-1}$, the fractional helicity $\PPM$
has changed more significantly 
when the GW spectrum is established, such that the variation of helicity 
affects the GW polarization; see \Figs{pspecm_hel_forc001}{pspecm_hel_forc}.
At earlier times, as mentioned in \Sec{runs_forced},
the values of $\sigM$ are the same as the values of the parameter
$\sigma$ of the forcing term. 
We observe a dependence $\PPGW \sim \PPM$,
inferred from the numerical results; see \Fig{pppol1}.
This result differs from the analytical model considered
in \App{beltrami_app}, which corresponds to a magnetic
field that depends only on one spatial coordinate with fractional helicity $\PPM=2\sigM/(1+\sigM^2)$.
The predicted dependence of the degree of circular
polarization is $\PPGW = 2\PPM/(1 + \PPM^2)$, given
in \Eq{PPh_bel}, and shown in \Fig{PM_bel}.

\subsection{Comparison to the analytical prediction of the spectrum of polarization}
\label{an_spec_pol_section}

The GW degree of circular polarization of signals produced by primordial
MHD turbulence has been estimated in refs.~\cite{Kahniashvili:2005qi,
Kisslinger:2015hua}.
In this section, we use their model to predict the spectrum of GW
polarization $\PPh$ and compare it to the numerical results, to explore
the validity of their model and the impact of the assumptions made.
In the first place, we define the unequal time correlation (UTC) function of the
magnetic field as
\begin{equation}
\bra{\tilde B_i (\kk, t_1) \tilde B^*_j (\kk', t_2)} = (2 \pi)^6 \delta^3
(\kk - \kk') \left(P_{ij} (\hat \kk) \frac{\FM (k, t_1, t_2)}{4 \pi k^2} +
i \varepsilon_{ijl} \hat k_l \frac{\GM(k, t_1, t_2)}{8 \pi k}\right),
\end{equation}
where we recover the equal time correlation of the magnetic energy 
density and helicity, defined in \Eq{equaltime}, when $t_1=t_2=t$,
$\FM(k, t, t) = \EM(k, t)$, and $\GM(k, t, t) = \HM(k, t)$.
Following ref.~\cite{Kahniashvili:2005qi}, which assumes stationary 
turbulence, we express the spectral functions of the UTC of the freely 
decaying turbulent source as a function of only the time
difference, modelled as $\FM(k, t_1, t_2) =
\FM(k, t_1, t_1 + \delta t) \approx \EM(k, t_1) D_1 (\delta t)$.
Similarly, the helical contribution to the UTC spectrum is
$\GM(k, t_1, t_2) = \GM(k, t_1, t_1 + \delta t)
\approx \HM(k, t_1) D_2 (\delta t).$
The functions $D_1(\delta t)$ and $D_2(\delta t)$ are monotonically decreasing 
functions, with $D_1(0) = D_2(0) = 1$.
To characterize the two types of turbulence considered in the present work,
$t_1$ is considered to be the time when the turbulence starts freely decaying.
This corresponds to the initial time ($t_1=1$) for the cases 
with a given magnetic field, and to the time at which the forcing term
is switched off ($t_1=\tmax$) otherwise.

\begin{figure}[t!]\begin{center}
\includegraphics[width=.44\textwidth]{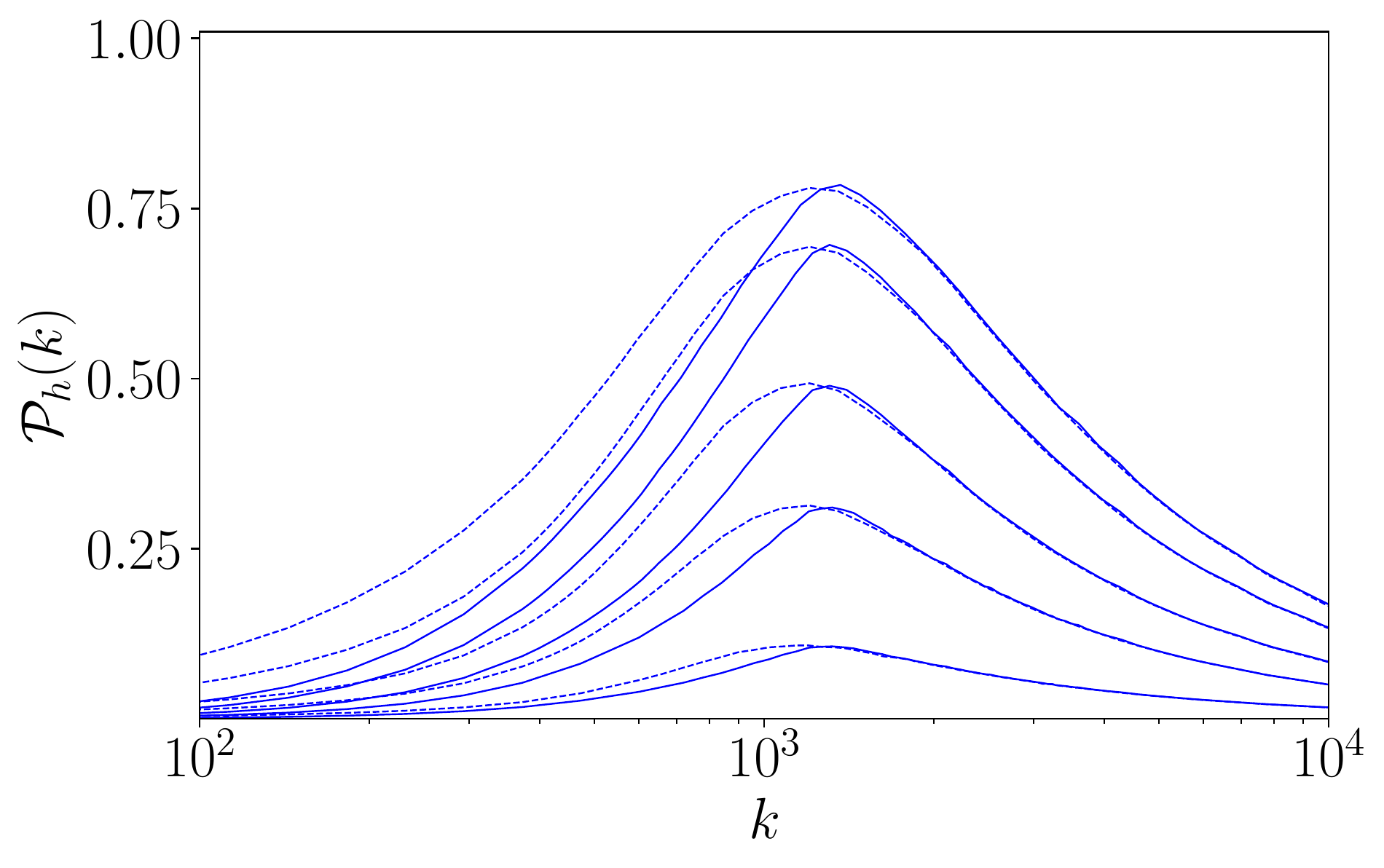}
\includegraphics[width=.54\textwidth]{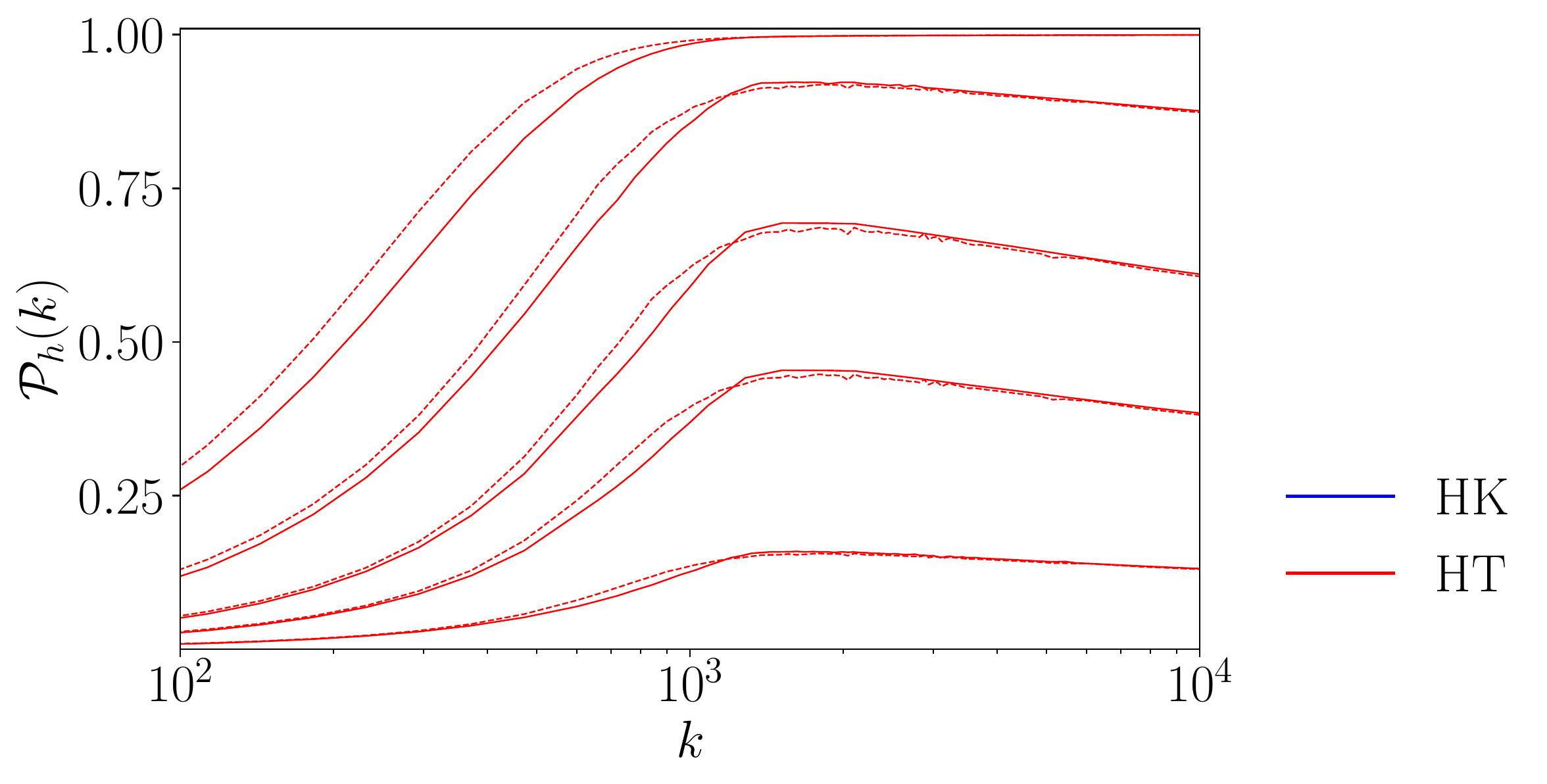}
\end{center}\caption[]{
Polarization spectra $\PPh (k)$ for HK turbulence with
$n_{\rm S} = -5/3$ and $n_{\rm A} = -8/3$ in the subinertial
range (blue), and for HT turbulence with $n_{\rm S} = n_{\rm A} = 
-5/3$ (red), using a single power law in the inertial range (solid lines)
\cite{Kahniashvili:2005qi,Kisslinger:2015hua,Ellis:2020uid},
and using a broken power law with a Batchelor spectrum in the
subinertial range (dotted lines), for fractional helicities
$h=0.1$, $0.3$, $0.5$, $0.8$, and $1$.
We use $\kf=600$ for comparison with the numerical
simulations.
}\label{PGW_analytical}\end{figure}

Assuming that the duration of the turbulence sourcing is short,
i.e., $\tau = t_{\rm fin} - 1 \ll 1$, where $t_{\rm fin}$ corresponds to the
final time of the turbulence,
such that the expansion of the universe
can be neglected, and after averaging over the time oscillations of the 
source, the functions $S_h(k)$ and
$A_h(k)$ can be obtained as
\cite{Kahniashvili:2005qi,Kosowsky:2001xp}
\begin{align}
    S_h(k)= A \frac{\tau}{k^2} \int \dd \ln p_1 \int 
    \dd \ln p_2  \bar \Theta \Big[ & (1 + \gamma^2) (1 + \beta^2)
    \EM(p_1) \EM (p_2) \nonumber \\
    & + 4 \gamma \beta \HM(p_1) \HM (p_2) \Big],
    \label{Sh_an}\\
    A_h(k)=  2 A \frac{\tau}{k^2} \int \dd \ln p_1 \int 
    \dd \ln p_2  \bar \Theta \Big[ & (1 + \gamma^2) \beta
    \EM(p_1) \HM (p_2) \nonumber \\
    & + (1 + \beta^2) \gamma \HM(p_1) \EM (p_2) \Big],
    \label{Ah_an}
\end{align}
where $\bar \Theta = \Theta(p_1 + p_2 -k) \Theta (p_1 + k - p_2)
\Theta (p_2 + k - p_1)$,
$\gamma = (k^2 + p_1^2 - p_2^2)/(2 k p_1)$,
$\beta=(k^2 + p_2^2 - p_1^2)/(2 k p_2)$,
and $A$ is a constant that we omit, since we are interested
in the spectral shapes, and we will use \Eqs{Sh_an}{Ah_an}
to compute the polarization, which does not depend on
$A$.\footnote{\EEqs{Sh_an}{Ah_an} correspond to
Eqs.~(10) and (11) of ref.~\cite{Kahniashvili:2005qi}, in which
$P_{\rm S} (k) = 2\pi^2 \EM(k)/k^2$,
$P_{\rm A} (k) = \pi^2 \HM (k)/k$,
$H(k) = 2 \pi^2 S_h(k)/k^2$,
${\cal H} (k) = 2 \pi^2 A_h(k)/k^2$,
and their value $A$ contains $\tau/k^2$ and differs by constant
coefficients due to the normalization we use in \Eq{GW}.
References~\cite{Kisslinger:2015hua,Ellis:2020uid} use the notation 
${\cal I}_{\rm S}(k)=P_{\rm S} (k)$ and ${\cal I}_{\rm A}=
P_{\rm A} (k)$.}
Previous analytical assumptions \cite{Kahniashvili:2005qi,
Kisslinger:2015hua,Ellis:2020uid} consider two types of turbulence:
\begin{itemize}
    \item  Helical Kolmogorov (HK) turbulence driven by magnetic
    energy dissipation at small scales, resulting in spectral powers
    $n_{\rm S}=-5/3$ and $n_{\rm A}=-8/3$.\footnote{We refer here
    to the spectral slopes of $\EM(k) \propto k^{n_{\rm S}}$ and
    $k \HM(k) \propto k^{n_{\rm A}}$, while ref.~\cite{Kahniashvili:2005qi} uses spectral slopes of
    $P_{\rm S}$ and $P_{\rm A}$, which are divided by $k^2$.}
    \item Turbulence determined by helical transfer (HT) and helicity
    dissipation at small scales, which results in $n_{\rm S} =
    n_{\rm A} = -7/3$, based in ref.~\cite{MC96}.
\end{itemize}
They use power law spectra $\EM(k) \propto k^{n_{\rm S}}$ and
$k \HM (k) \propto h k^{n_{\rm A}}$ in the range
$\kf < k < k_\nu$, with $h$ being the fraction of helicity 
dissipation.
We extend their analytic approach to consider a broken power 
law with a subinertial $k^4$ Batchelor spectrum below $\kf$,
and modify the HT spectrum to $n_{\rm S}=n_{\rm A}=-5/3$,
corresponding to the spectral slopes of the stochastic magnetic
fields that we use in our numerical simulations (see 
\Fig{pspecm_hel_initial}) and that is based in previous MHD
simulations applied to cosmological phase transitions
\cite{Brandenburg:2017neh}.
\FFig{PGW_analytical} shows the resulting polarization degree
$\PPh(k)$ for the HT and HK types of turbulence.
The inclusion of the subinertial range leads to an increase on
polarization at wave numbers right below the peak, which is a more
realistic scenario when compared to the numerical results,
especially for the case of HK turbulence.
The model described above assumes that the partial helicity at $\kf$ is $\PPM(\kf)=h$, 
and describes the magnetic energy and helicity spectra using power laws.
The assumption that the slopes are the same along the inertial range is
accurate in the case with an initial magnetic field; see
\Figs{pspecm_hel_initial}{PGW_analytical_comp_HT}.
\begin{figure}[t!]\begin{center}
\includegraphics[width=.75\textwidth]{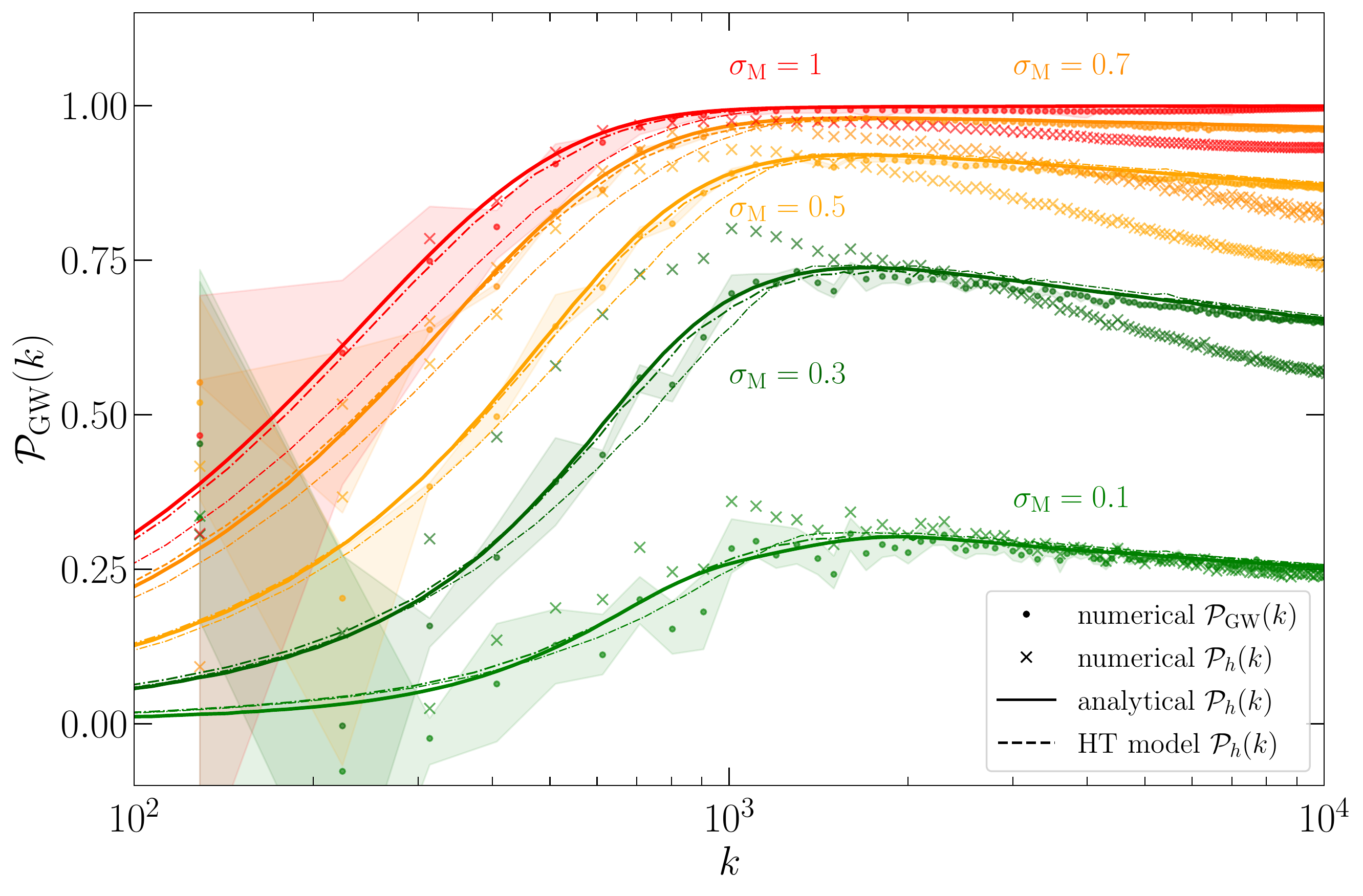}
\end{center}\caption[]{
Polarization spectra $\PPGW (k)$ (dots) and $\PPh (k)$ (crosses)
obtained from the numerical simulations for different $\sigM$,
compared to $\PPh (k)$ computed from
the analytical model (single power law and extended to a broken power law;
see \Fig{PGW_analytical}) using HT type of
turbulence (dashed lines), and from the analytical integrals;
see \Eqs{Sh_an}{Ah_an}, using the numerical spectra of the turbulence
(solid lines).
Similar to \Fig{pppol1}, the shaded regions denote the maximum and minimum
polarizations $\PPGW(k)$ of the fluctuations.
}\label{PGW_analytical_comp_HT}\end{figure}
\begin{figure}[t!]\begin{center}
\includegraphics[width=.49\textwidth]{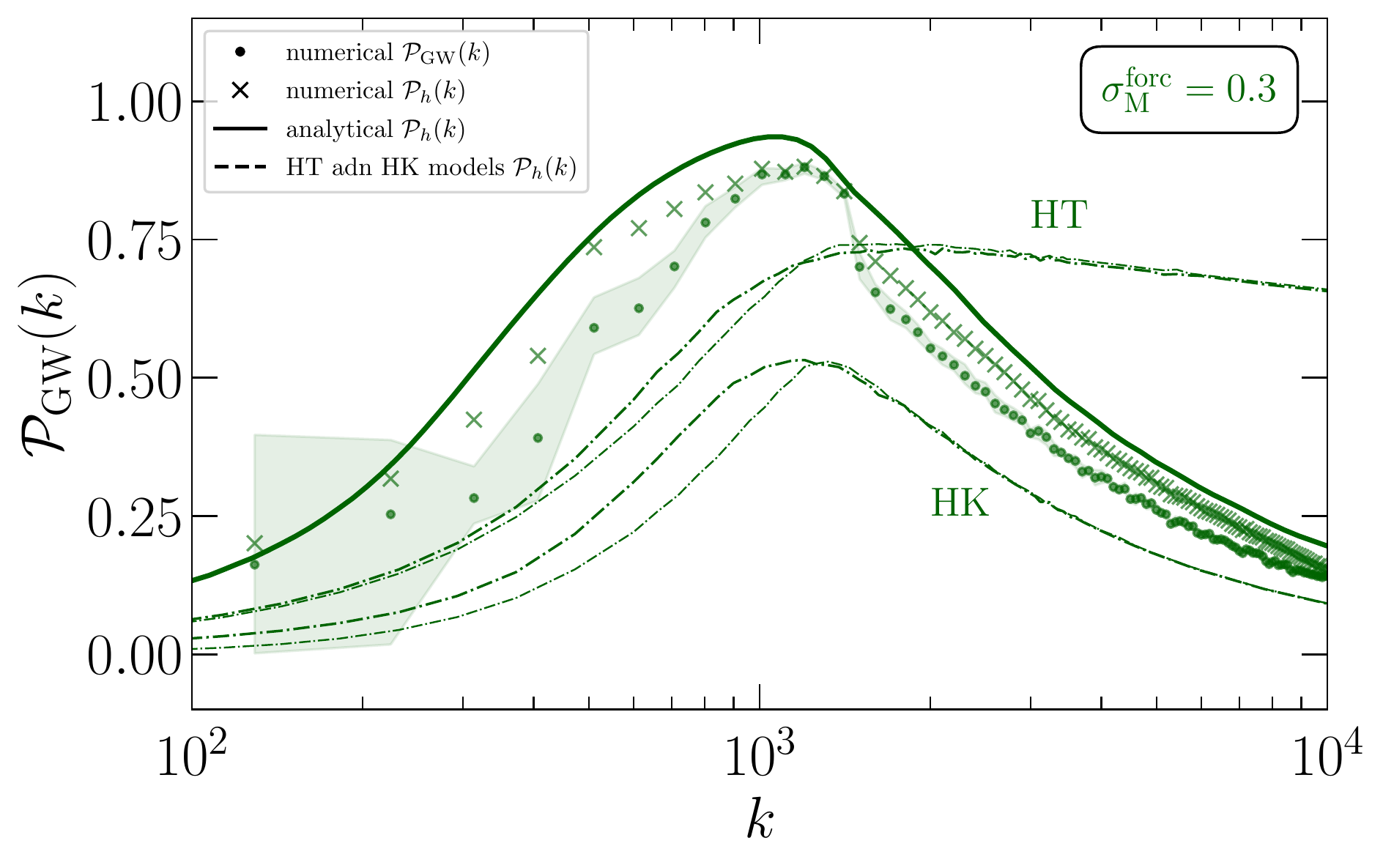}
\includegraphics[width=.49\textwidth]{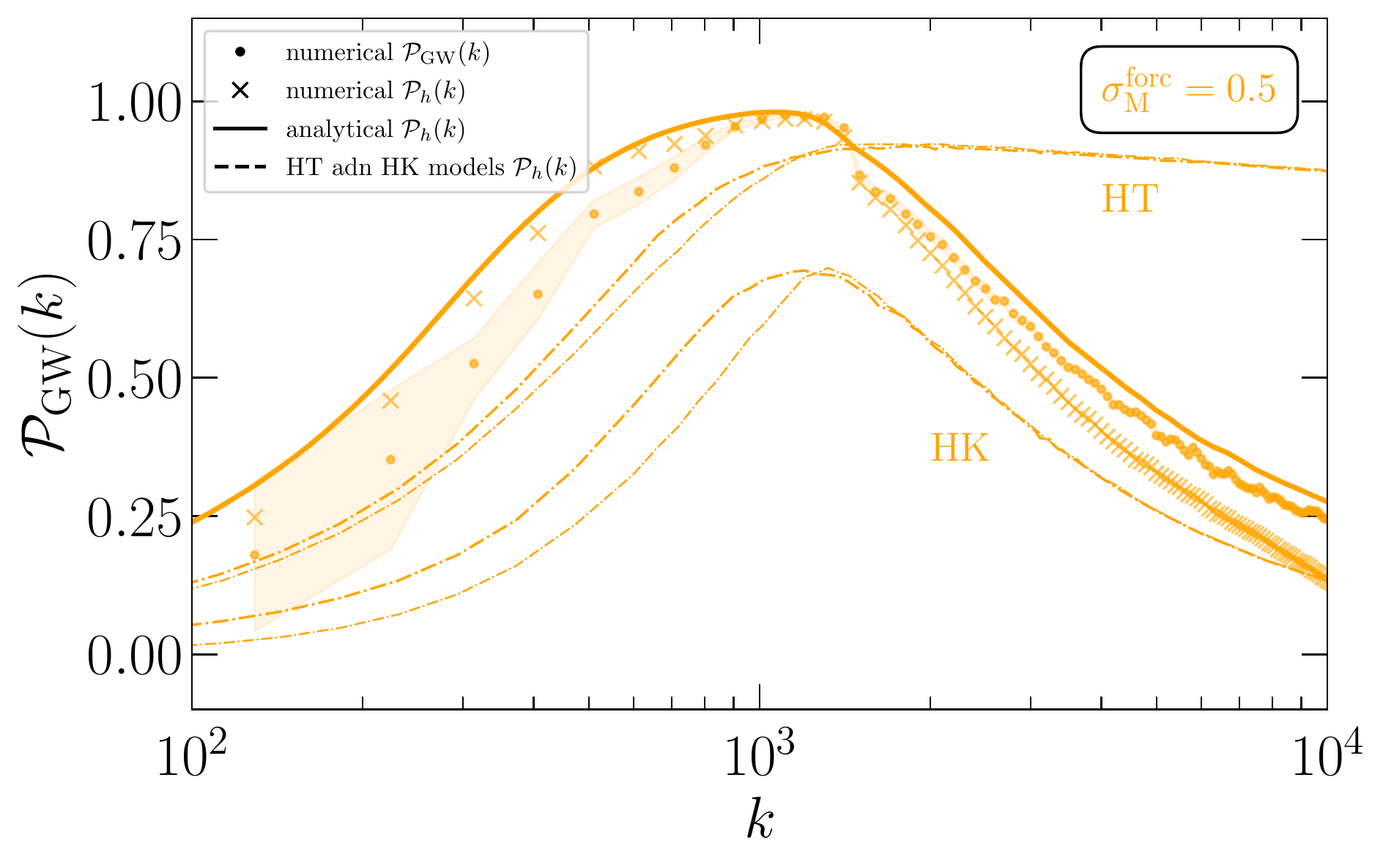}
\includegraphics[width=.49\textwidth]{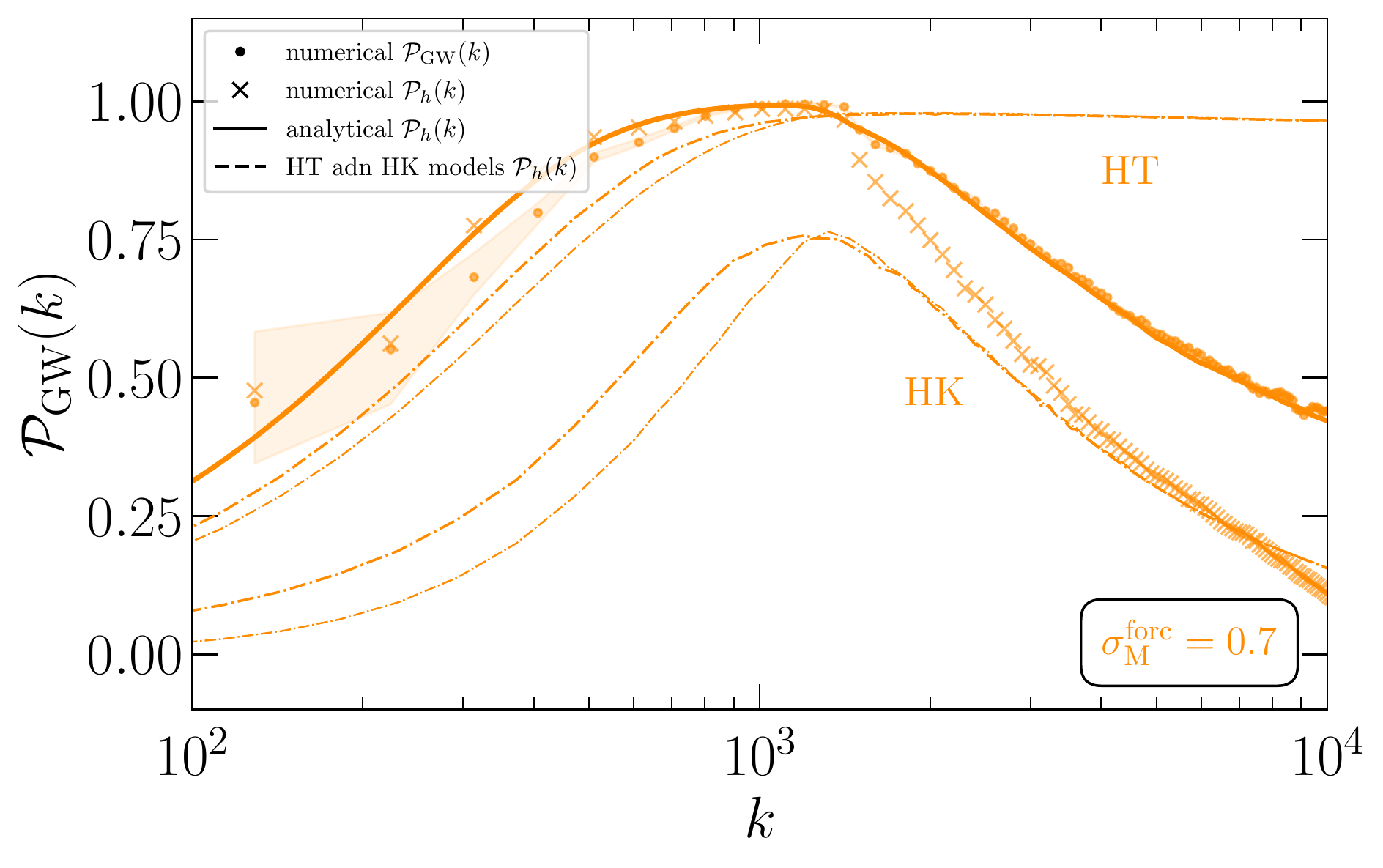}
\includegraphics[width=.49\textwidth]{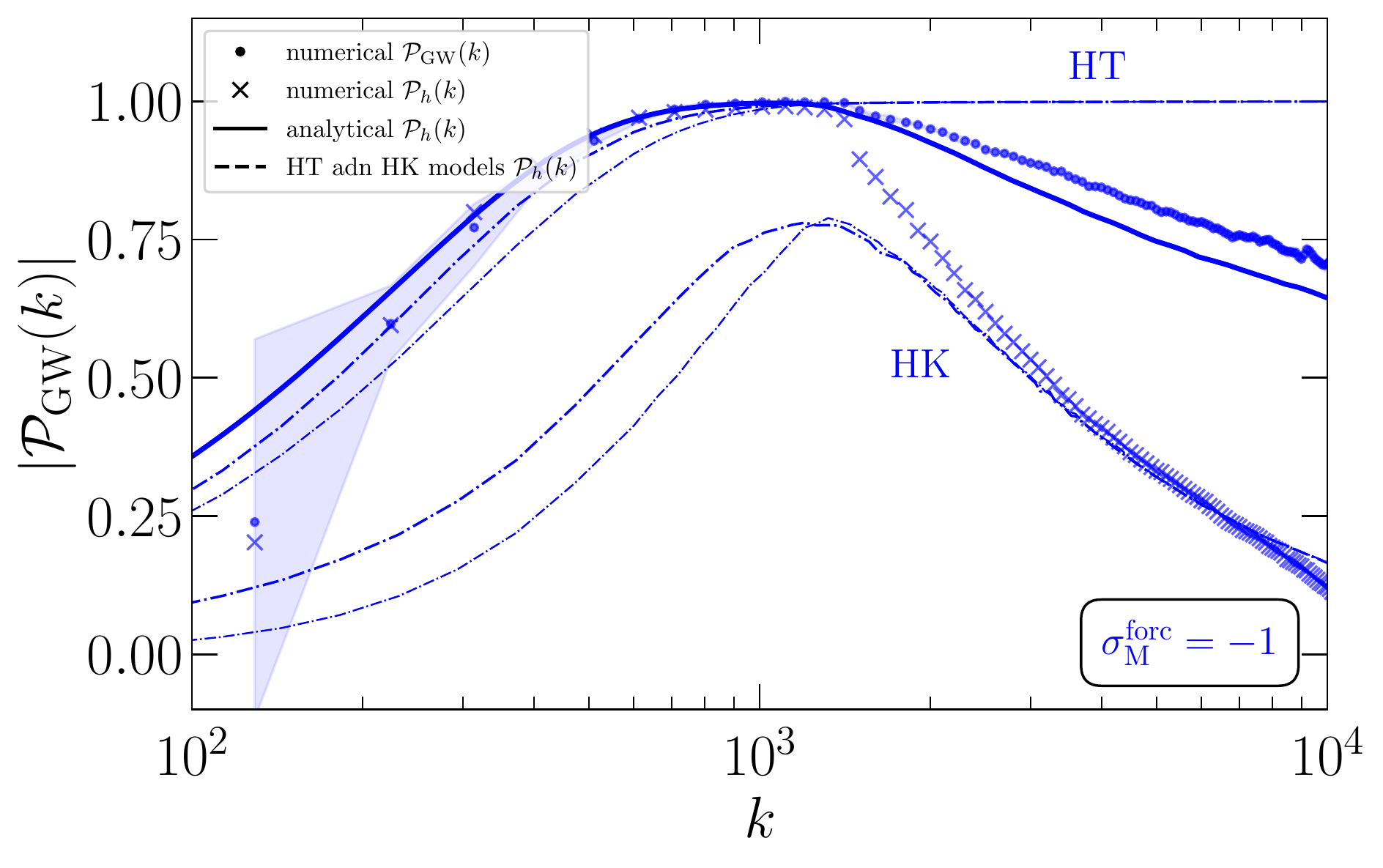}
\end{center}\caption[]{
Similar to \Fig{PGW_analytical_comp_HT}, 
polarization spectra $\PPGW (k)$ and $\PPh (k)$ obtained from
the numerical simulations for different $\sigM$, compared to $\PPh (k)$ obtained from
the analytical model using both HK and HT types of turbulence (dashed lines;
see \Fig{PGW_analytical}),
and obtained from the analytical integral, using the
numerical spectrum of the turbulence (solid lines).
}\label{PGW_analytical_comp_HK}\end{figure}

In \Fig{PGW_analytical_comp_HT}, we compare the analytical results obtained from
HT turbulence (with modified slopes and extended to broken power laws; see
\Fig{PGW_analytical})
with our numerical simulations that consider an initial given magnetic field.
Note that the extension of the turbulent spectra to $k < \kf$ allows one to
get a more accurate position of the polarization spectral peak.
In general, we observe good agreement of the analytical results with the
polarization $\PPGW (k)$ computed from the simulations, while the spectrum
$\PPh(k)$ agrees well at low and intermediate $k$, but decays with respect
to the analytical models for large $k$.
However, in the case in which the magnetic field is forced for a short
duration of time, we observe in \Fig{pspecm_hel_forc} that the energy
and helicity spectra are similar around the spectral peak (where the magnetic
field is maximally helical),
while the helical spectrum starts decaying with a steeper slope 
than the magnetic spectrum at a specific wave number, which increases
with the helicity of the forcing.
This behavior is not captured by previous analytical models.
In \Fig{PGW_analytical_comp_HK}, we compare the results of the polarization
spectra $\PPh(k)$ and $\PPGW(k)$, obtained from the numerical simulations,
with those obtained from HK and HT types of turbulence.
In addition, to take into account the deviations from the assumption 
of constant slopes of the magnetic spectra, we compute the polarization
using \Eqs{Sh_an}{Ah_an} by integrating over the numerically computed
spectra $\EM (k)$ and $\HM (k)$.
We observe in \Figs{PGW_analytical_comp_HT}{PGW_analytical_comp_HK}
that the latter gives an accurate approximation of $\PPGW(k)$
to the numerical results, while $\PPh (k)$ shows a bigger decay at large $k$. 
In general, the polarization spectrum shows features of both HK and HT
types of turbulence.
As we have mentioned, around the spectral peak, the numerical simulations
show that the magnetic field is maximally helical in all the considered 
runs with $|\sigM| \geq 0.3$ or $|\PPM| \geq 0.55$, which is better represented
by the HT spectrum (with $h=1$) in the subinertial range,
while in the inertial range the different
slopes lead to a decrease of the polarization with a scaling similar
to that of the HK spectrum, especially as we decrease the fractional helicity,
although the polarization computed numerically is still larger than that obtained
by the HK model in all cases.

We have confirmed using numerical simulations that in the case of an
initial magnetic field with a Kolmogorov spectrum for both the magnetic
energy density and helicity, the analytical model previously considered in 
refs.~\cite{Kahniashvili:2005qi,Kisslinger:2015hua,Ellis:2020uid}
using HK turbulence gives an accurate prediction of the degree of circular
polarization, which gets better when we consider a broken power law.
However, when we consider the scenario in which the magnetic field is generated
via MHD forcing for a short amount of time, and non-linear interactions appear,
allowing to have different spectral slopes at different scales, previous
analytical estimates underpredict considerably the polarization degree
peak and fail to predict the appropriate shape, which is not given by
either the HK or the HT models, but presents non-linearly combined features of
both. 
In addition, we observe a helical inverse cascade that produces larger degree of circular
polarization at large scales.

\section{Prospects of detecting signals from the electroweak phase
transition}
\label{detectability_sec}

\subsection{Observable GW energy density spectra}

The GW energy density at the present time is obtained from 
the comoving $\OmGW (k)$, defined in \Eq{OmGW_k}, and
expressed as a function of the frequency. 
For a signal that has been produced at the EWPT,
the resulting GW spectrum is
\begin{align}
   \OmGW(k) = &\, \left(\frac{a_0}{a_*}\right)^{-4}
    \left(\frac{H_*}{H_0}\right)^2 k \EGW(k), \nonumber \\
    h_0^2\, \OmGW(f) = & \, 1.652 \times 10^{-5}
    \left(g_*/100\right)^{-1/3}
    \left(2\pi f/\fH\right) \EGW(f),
    \label{OmGW_f}
\end{align}
where we have used the present time values $T_0 = 2.73 \K$, $g_0=3.91$, and
$H_0$, expressed by the Hubble parameter today $h_0$ in units of
$100 \km \s^{-1} \Mpc^{-1}$
\cite{Maggiore:1999vm},
and $g_*$ and $\fH = H_* a_*/a_0$ are the number of degrees of freedom
and the Hubble frequency, respectively, at the electroweak scale.
The ratio of the scale factors is obtained assuming 
adiabatic expansion, i.e., with constant $g\, T^3 a^3$ \cite{Kolb:1990vq},
\begin{equation}
    \frac{a_0}{a_*} = 1.254 \times 10^{15} \left(\frac{\kB T_*}
    {100 \GeV} \right) \left(\frac{g_*(T_*)}{100}\right)^{1/3},
\end{equation} 
with $T_*$ being the electroweak temperature. 
The Hubble rate at the electroweak scale is \cite{Kolb:1990vq}
\begin{equation}
   H_* = 2.066 \times 10^{10} \s^{-1} \left(\frac{\kB T_*}{100
   \GeV}\right)^2 \left(\frac{g_*(T_*)}{100}\right)^{1/2}.
\end{equation}
The resulting spectrum is expressed in terms of the frequency,
which corresponds to the physical comoving wave number (from the GW dispersion 
relation $2 \pi f = k$) shifted 
to the present time; see ref.~\cite{Kolb:1990vq},
\begin{equation}
    f = \frac{H_*}{2 \pi} \left(\frac{a_*}{a_0}\right) k
    = \frac{\fH}{2 \pi}k,
    \quad \text{with }
    \fH=1.646 \times 10^{-5} \Hz \left(\frac{\kB T_*}{100
    \GeV}\right) \left(\frac{g_*(T_*)}{100}\right)^{1/6},
\end{equation}
being the Hubble frequency at the electroweak scale,
which corresponds to $k_{\rm H}=2\pi$ according to our normalization. 
The helical spectrum of GWs $h_0^2\, \XiGW (f)$, defined in \Eq{XiGW_k},
is computed in the same way as $h_0^2 \OmGW(f)$ in \Eq{OmGW_f},
substituting $\EGW(k)$ by $\HGW(k)$.

\begin{figure}[t!]\begin{center}
\includegraphics[width=.75\textwidth]{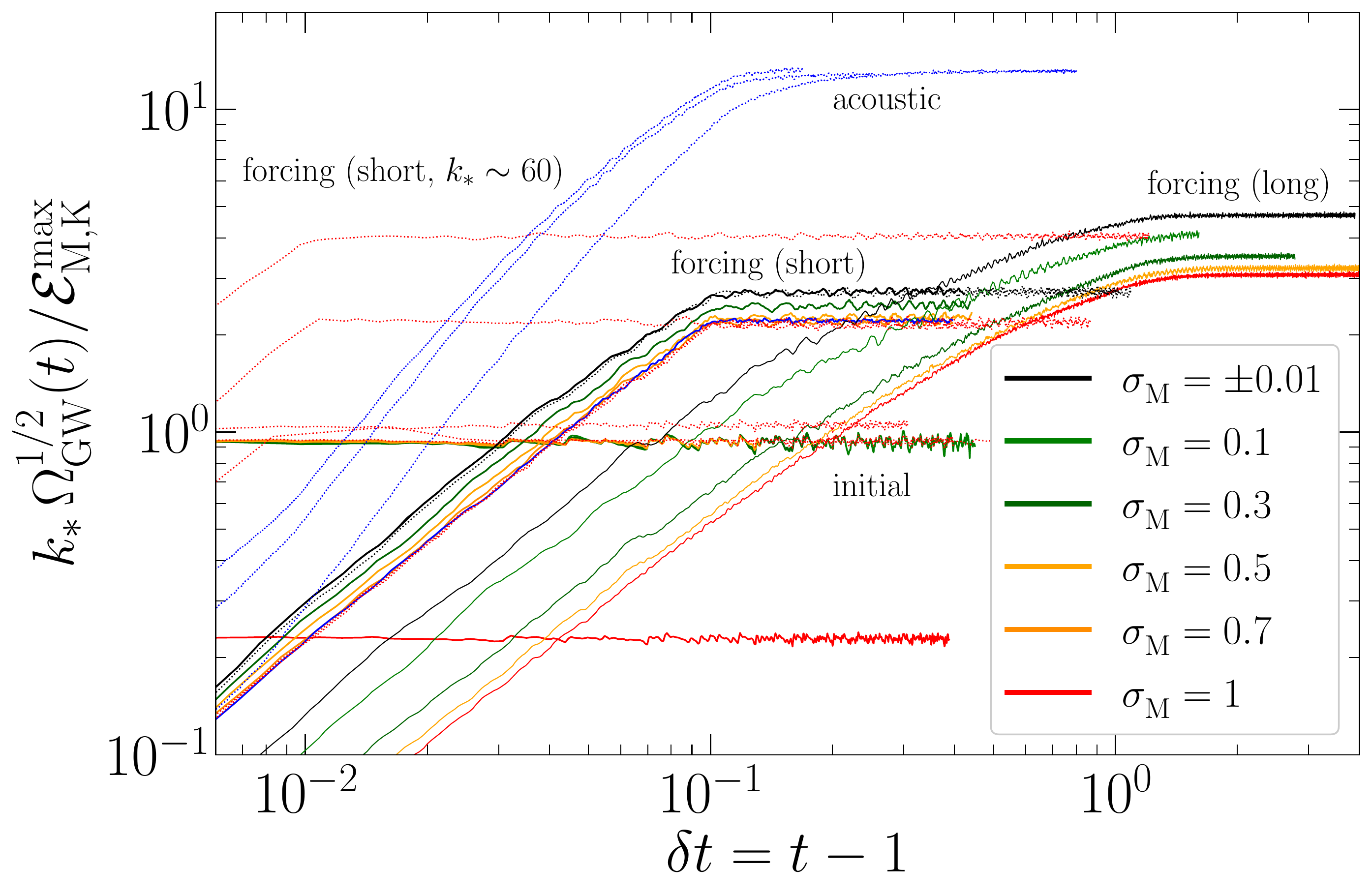}
\end{center}\caption[]{
Efficiency of GW energy density $q (t) =\kf \OmGW^{1/2} (t)/{\cal E}_i^{\rm max}$
for $i=$ M (magnetic) and K (kinetic) in units of $a^{-4} (H_*/H_0)^2 = 1.644
\times 10^{-5}\, (g_*/100)^{-1/3}$;
see \Eqs{OmGW}{OmGW_f}, for the runs with an initial
given magnetic field (`initial'), and for the runs with a forced magnetic field 
(`forcing (short)'), with $\tmax = 1.1$.
Added for comparison are the runs in ref.~\cite{Kahniashvili:2020jgm}
(`forcing (long)'), in which the magnetic field is forced for longer times
($\tmax = 3$), and the runs of ref.~\cite{Pol:2019yex}, which contain cases
with initial given magnetic field, with forced magnetic field
at $\kf = 60$, $600$, and $6000$, and runs of acoustic turbulence (`acoustic').
}\label{OmGW_efficiency}\end{figure}

Recent numerical simulations of GWs produced by MHD turbulence 
have found a dependence of the GW energy density $\OmGW$ on 
the square of the magnetic energy density $\EEM^2$ and the inverse of the 
square of the magnetic spectral peak $\kf^{-2}$
\cite{Pol:2019yex,Kahniashvili:2020jgm,Brandenburg:2021bvg,Brandenburg:2021tmp}; see \Eq{OmGW}.
In the present work, our numerical simulations follow this scaling, and we define the GW
efficiency $q^2(t)=\kf^2 \OmGW (t)/(\EEMmax)^2$, shown in \Fig{OmGW_efficiency}.
Such scaling was obtained in early analyses of MHD turbulent production of GWs;
see, e.g., refs.~\cite{Kamionkowski:1993fg,Kosowsky:2001xp,Apreda:2001us,Dolgov:2002ra},
which assumed stationary turbulence,
and reported in ref.~\cite{Niksa:2018ofa} in the case when the decay
of the magnetic field does not impact the evolution of GWs.
We compute this scaling
in the analytical model presented in \App{beltrami_app}; see also
the Beltrami field studied in ref.~\cite{Pol:2018pao}. 
However, ref.~\cite{Niksa:2018ofa} proposes a general $\EEM^{3/2}$ scaling for
MHD turbulence; see their equation~(82),
which was previously obtained in refs.~\cite{Caprini:2009yp,Caprini:2010xv},
and assumes that the magnetic field decay impacts the GW dynamics.
This is often used when considering GW signals from MHD
turbulence in the LISA band \cite{Caprini:2019egz}.
In general, the exact scaling depends on the dynamical evolution of the
magnetic field and, 
in particular, on the UTC of the stress, which is modelled in
previous estimates (see reviews \cite{Binetruy:2012ze,Caprini:2018mtu}), while 
the direct numerical simulations of MHD turbulence allow to obtain the final
GW spectrum with no assumptions on the magnetic stress UTC.
We observe that the GW production is similar
for all values of $\sigM$ if the magnetic field is present at
the initial time of generation with a GW efficiency of $q = 0.95$.
For the case in which the magnetic field is forced at initial times,
we observe an enhancement of the GW production by a factor of
$\sim 5$, and larger GW energy densities for smaller helicities.
The dependency of the GW production on the helicity is
consistent with that for the runs of ref.~\cite{Kahniashvili:2020jgm},
in which the forcing term is
present for longer times, and the GW production is larger.
In the case of acoustic turbulence (e.g., sound waves), the GW 
production is larger by a factor of $\sim 200$, as reported
in ref.~\cite{Pol:2019yex}.
The GW production obtained in the numerical simulations is smaller than the
estimated amplitudes computed in previous analytical estimates
\cite{Gogoberidze:2007an,Ellis:2020uid}.
This is probably due to simplifying assumptions made.
We defer the study of the scaling of the GW amplitudes with the characteristic scale
and the amplitude of the turbulence sourcing to future work.

\subsection{Interferometry of GW detectors LISA and Taiji}
\label{interferometry_text}

The GW signals produced at the EWPT
are expected to be detectable with future planned space-based GW detectors,
e.g., LISA \cite{Audley:2017drz}, Taiji \cite{Guo:2018npi}, TianQin \cite{Luo:2015ght},
DECIGO \cite{Seto:2001qf}, and BBO \cite{Crowder:2005nr}.
We revisit the interferometry of this type of detectors in \App{interferometry},
and apply the analysis to LISA and Taiji to consider the potential
detectability of the circular polarization of GW signals produced by 
primordial magnetic fields.
In general, it is necessary that the GW background presents anisotropies to
measure its circular polarization \cite{Seto:2006hf,Seto:2006dz}. 
Hence, a priori,
parity-violating effects cannot be detected if the system
of GW detectors is coplanar, which is the case for space-based
GW detectors, and the GW background is isotropic \cite{Seto:2007tn}. 
However, different approaches have recently been proposed
to detect the circular polarization of a statistically isotropic
GW background \cite{Domcke:2019zls,Seto:2020zxw,Orlando:2020oko}.
On the one hand, a statistically isotropic GW background, such as that expected
from cosmological sources, can present anisotropies that have been
kinematically induced due to the proper motion of the solar system,
and the induced anisotropies allow one to detect
the circular polarization of the background
\cite{Seto:2006hf,Seto:2006dz}.
On the other hand, the combination of a network of GW detectors
breaks the coplanarity of the detectors allowing one to detect
circular polarization.
This has been considered in the case of ground-based GW detectors;
see, e.g., refs.~\cite{Martinovic:2021hzy,Seto:2007tn,Seto:2008sr,
Crowder:2012ik,Domcke:2019zls}, and for the LISA--Taiji network
\cite{Seto:2020zxw,Orlando:2020oko}.

The total and the polarization signal-to-noise ratios (SNR) of a stochastic GW 
background with energy density $\OmGW(f)$ and
helical spectra $\XiGW(f)$, for a duration $T$ of the observations, are
\begin{align}
    \SNR = &\,
    2 \sqrt{T} \left[\int_{0}^\infty
    \dd f \left(\frac{\OmGW(f)}{\Omega_{\rm s} (f)}\right)^2\right]^{1/2},
    \label{SNR_tot}\\
    \SNR_{\rm pol}^{\rm dip}=&\, 2 \sqrt{T} \left[
    \int_0^\infty \dd f \left(\frac{\XiGW(f) - \quarter
    \dd \XiGW(f)/\dd \ln f} {\Xi_{\rm s}^{\rm dip} (f)}\right)^2\right]^{1/2}, \label{SNR_pol_1}\\\
    \SNR_{\rm pol}^{\text{comb}}=&\, 2 \sqrt{T} \left[
    \int_0^\infty \dd f \left(\frac{\XiGW(f)} {\Xi_{\rm s}^{\rm comb} (f)}\right)^2\right]^{1/2},
    \label{SNR_pol_2}
\end{align}
where the GW sensitivity $\Omega_{\rm s} (f)$ in \Eq{SNR_tot} refers to
LISA, $\Omega_{\rm s}^A (f)$, Taiji, $\Omega_{\rm s}^C (f)$, or
the combined LISA--Taiji network, $\Omega_{\rm s}^{\rm comb} (f)$; see \Eqs{OmsA}{Omega_s_LISATAiji}.
The polarization $\SNR_{\rm pol}$ given in \Eq{SNR_pol_1} is
obtained by using the
anisotropies induced by the polarization of the GW background, due to our proper
motion, yielding a dipolar response in the LISA $A$ and $E$ channels
or the Taiji $C$ and $D$ channels, with the sensitivity
$\Xi_{\rm s}^{\rm dip} (f) = \Xi_{\rm s}^{AE} (f)$ or
$\Xi_{\rm s}^{CD} (f)$ given in \Eq{XisAE}.
On the other hand, the polarization $\SNR_{\rm pol}$
given in \Eq{SNR_pol_2} is obtained by
cross-correlating the different channels between LISA and Taiji
with the sensitivity
$\Xi_s^{\rm comb} (f)$ given in \Eq{Xi_s_comb}.
Further details on the dipole response function and
the LISA--Taiji cross-correlations are given in
\Apps{interferometry_SignalToNoise}{App_LISA_Taiji}, respectively,
and the GW sensitivity functions are shown in \Fig{GW_sensitivity}. 
Assuming flat GW energy density $\OmGW(f)$ and helicity spectra $\XiGW(f)$ 
of the background, we get
\begin{align}
    h_0^2\, \OmGW^{\rm LISA} (f) = h_0^2\, \Omega_{\rm flat}^A=&\,1.65 \times 10^{-13}\left(\frac{\SNR}{10}
    \right)\sqrt{\frac{4 \yr}{T}}, \\ 
    h_0^2 \, \OmGW^{\rm Taiji} (f) = h_0^2 \, \Omega_{\rm flat}^C= &\, 6.81 \times 10^{-14}\left(\frac{\SNR}{10}
    \right)\sqrt{\frac{4 \yr}{T}}, \\
    h_0^2\,\XiGW^{\rm LISA} (f)= h_0^2 \, \Xi_{\rm flat}^{AE} =
    &\,10^{-10}\left(\frac{\SNR_{\rm pol}}{10}
    \right) \sqrt{\frac{4 \yr}{T}} \left(\frac{1.23 \times 10^{-3}}{v/c}
    \right), \label{XiGW_LISA_dip_flat} \\
    h_0^2\,\XiGW^{\rm Taiji} (f)= h_0^2 \, \Xi_{\rm flat}^{CD} =
    &\, 4.16 \times 10^{-11}\left(\frac{\SNR_{\rm pol}}{10}
    \right) \sqrt{\frac{4 \yr}{T}} \left(\frac{1.23 \times 10^{-3}}{v/c}
    \right),
    \label{XiGW_Taiji_dip_flat}
\end{align} 
where $v$ is the solar system's proper motion. 
Using the values of $\SNR_{\rm pol}=1$, $v/c=10^{-3}$, and $T=3\yr$, we recover the amplitude
$h_0^2\,\Xi_{\rm flat}^{AE} = 1.4 \times 10^{-11}$, reported in ref.~\cite{Domcke:2019zls}.
The exact value of the SNR necessary to claim that the signal is detectable with
a large likelihood is not trivial, and requires a detailed analysis
that depends on the spectral shape of the GW signal.
For a simplified treatment, we follow refs.~\cite{Caprini:2019pxz,Caprini:2015zlo},
in which a value of $\SNR=10$ is proposed.
To study the potential detectability of the GW signals produced 
by primordial turbulence, we compute the power law sensitivities (PLS)
\cite{Caprini:2019pxz}, assuming a spectral shape defined by a
power law of generic slope; see \App{interferometry}. 
\FFig{GW_sensitivity} shows the PLS
of LISA and Taji, corresponding to the GW
energy density, $\Omega_{\rm PLS}^{A}(f)$ and $\Omega_{\rm PLS}^{C} (f)$,
and to the helicity, $\Xi_{\rm PLS}^{AE} (f)$ and $\Xi_{\rm PLS}^{CD} (f)$,
using the dipole response function, and for the combined LISA--Taiji
network, $\Xi_{\rm PLS}^{\rm comb} (f)$.
The reconstruction of the signal for more complex spectral shapes
is an active topic of research; see, e.g., the review \cite{Romano:2016dpx}
or the work by the LISA cosmology working group \cite{Caprini:2019pxz},
and we defer it to future work.
In general, the values of the polarized GW signal $h_0^2\,\XiGW (f)$ that 
can be detectable by using the dipole response function of a single
space-based GW mission, either LISA or Taiji,
require very large
amplitudes of the magnetic fields
generating GWs.\footnote{%
We find that magnetic energy densities of $\sim\!75\%$ the radiation energy density
are required for a polarized SNR of 10 with LISA if we assume that the scaling
of $\OmGW(f)$ with $\EEM^2$ is still valid in the highly relativistic limit and
we use the results for magnetic fields that are initially driven; see \Fig{XiGW_detectors}.
Our result is consistent with the results reported in ref.~\cite{Ellis:2020uid}, which
require strong first-order phase transitions, i.e., $\alpha \sim 1$, for a detectable
polarized signal; see the first right panel of their figure~(8).
The strength of the transition $\alpha$ is the ratio of vacuum to radiation
energy density and it is related to the kinetic energy induced in the
plasma by the efficiency $\kappa$, which becomes $\sim\!55\%$ of
the radiation energy density for $\alpha=1$ 
\cite{Kamionkowski:1993fg,Espinosa:2010hh}.}
Hence, we consider the combination of LISA and Taiji to detect such polarized
GW signals, following ref.~\cite{Orlando:2020oko}.
Assuming a flat GW polarized spectrum of the background, we get
\begin{equation}
    h_0^2\,\XiGW^{\text{LISA--Taiji}} (f) = h_0^2\,\Xi_{\rm flat}^{\rm comb} = 
    5.1 \times 10^{-13} \left(\frac{\SNR_{\rm pol}}{10}
    \right)\sqrt{\frac{4 \yr}{T}},
    \label{XiGW_LISA_Taiji_flat}
\end{equation}
which shows an improvement in the potential detectability of the helicity
by a factor of $\sim 80$ with respect to the dipole response function
of Taiji.

\begin{figure}[t!]\begin{center}
\includegraphics[width=.49\textwidth]{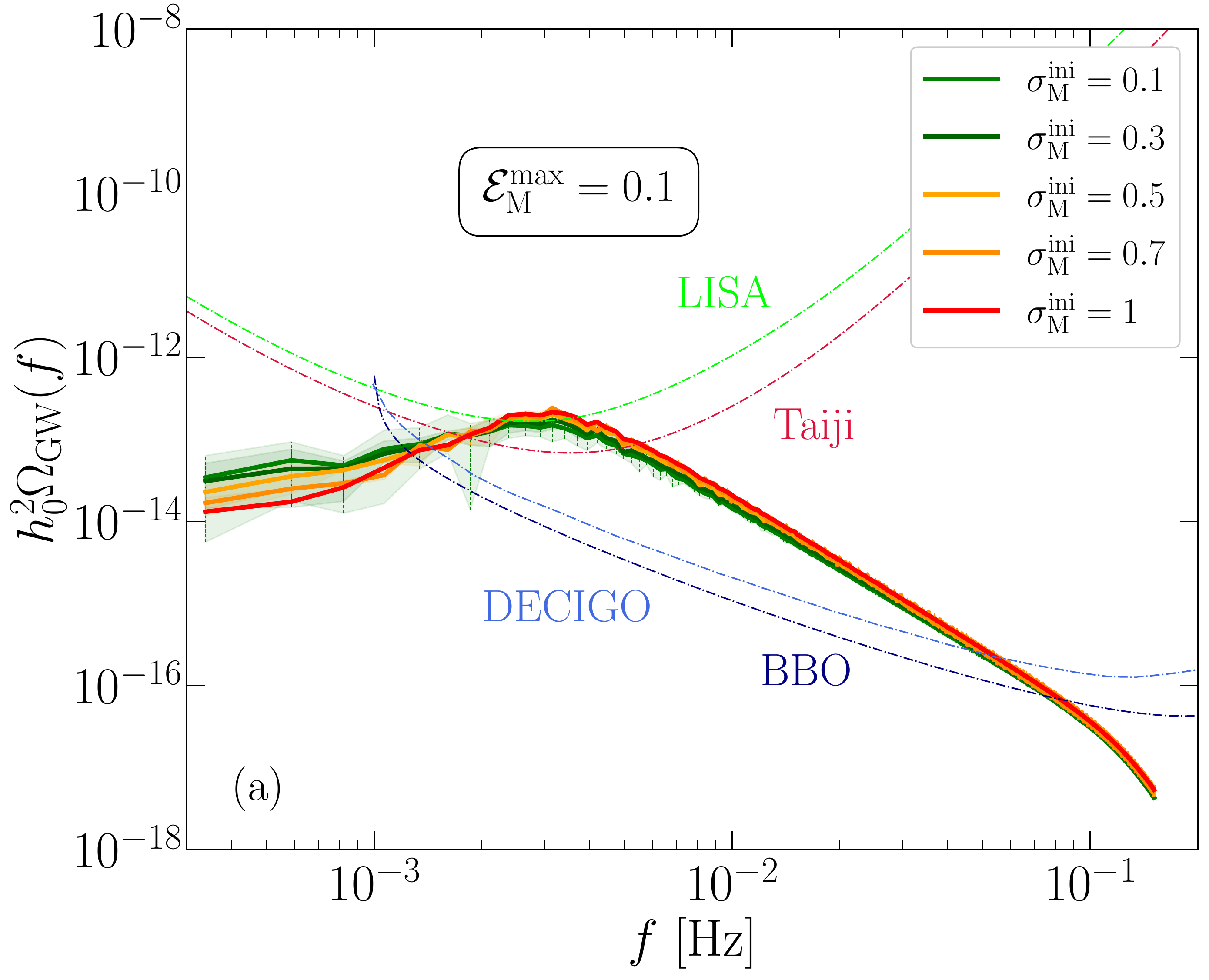}
\includegraphics[width=.49\textwidth]{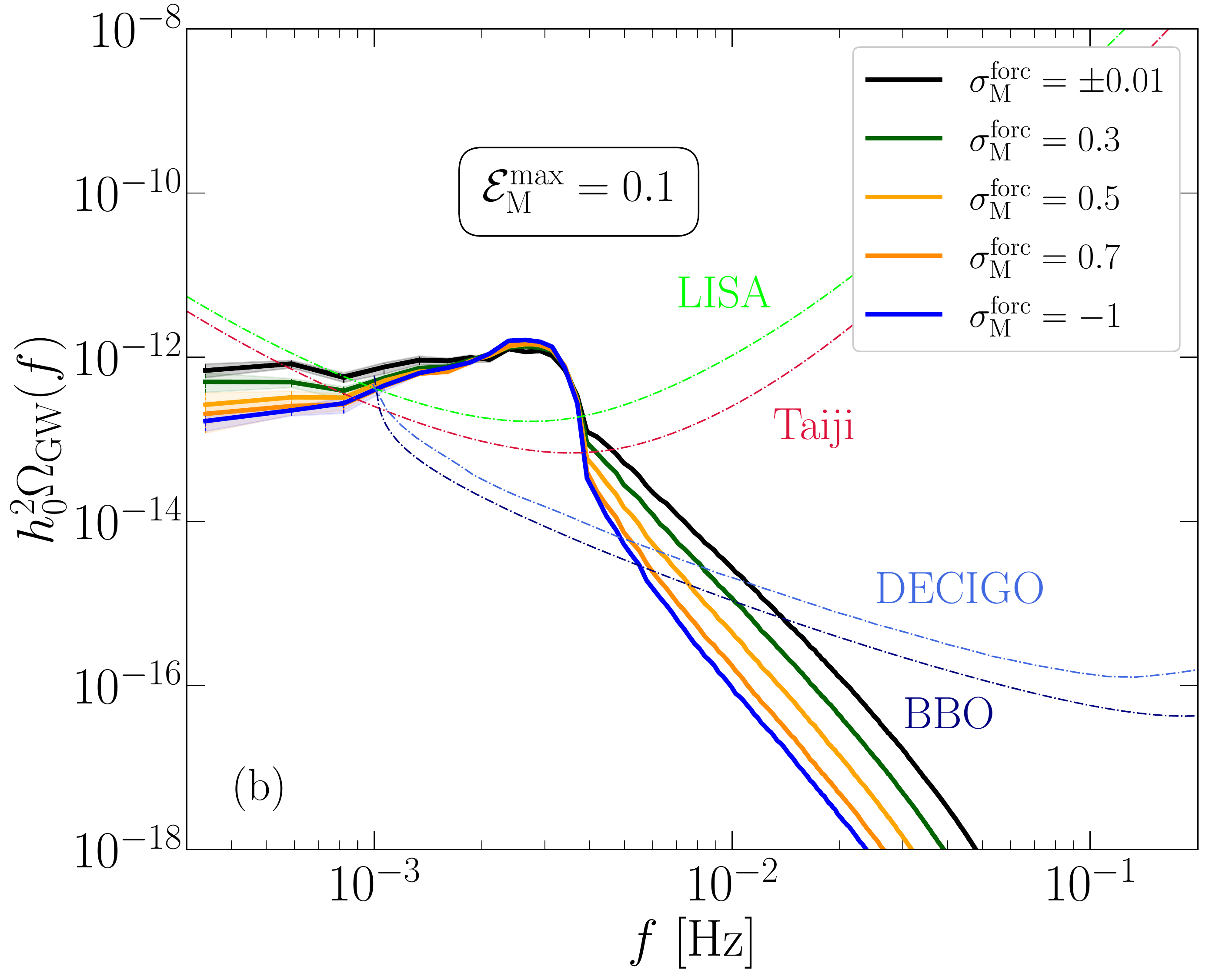}
\includegraphics[width=.49\textwidth]{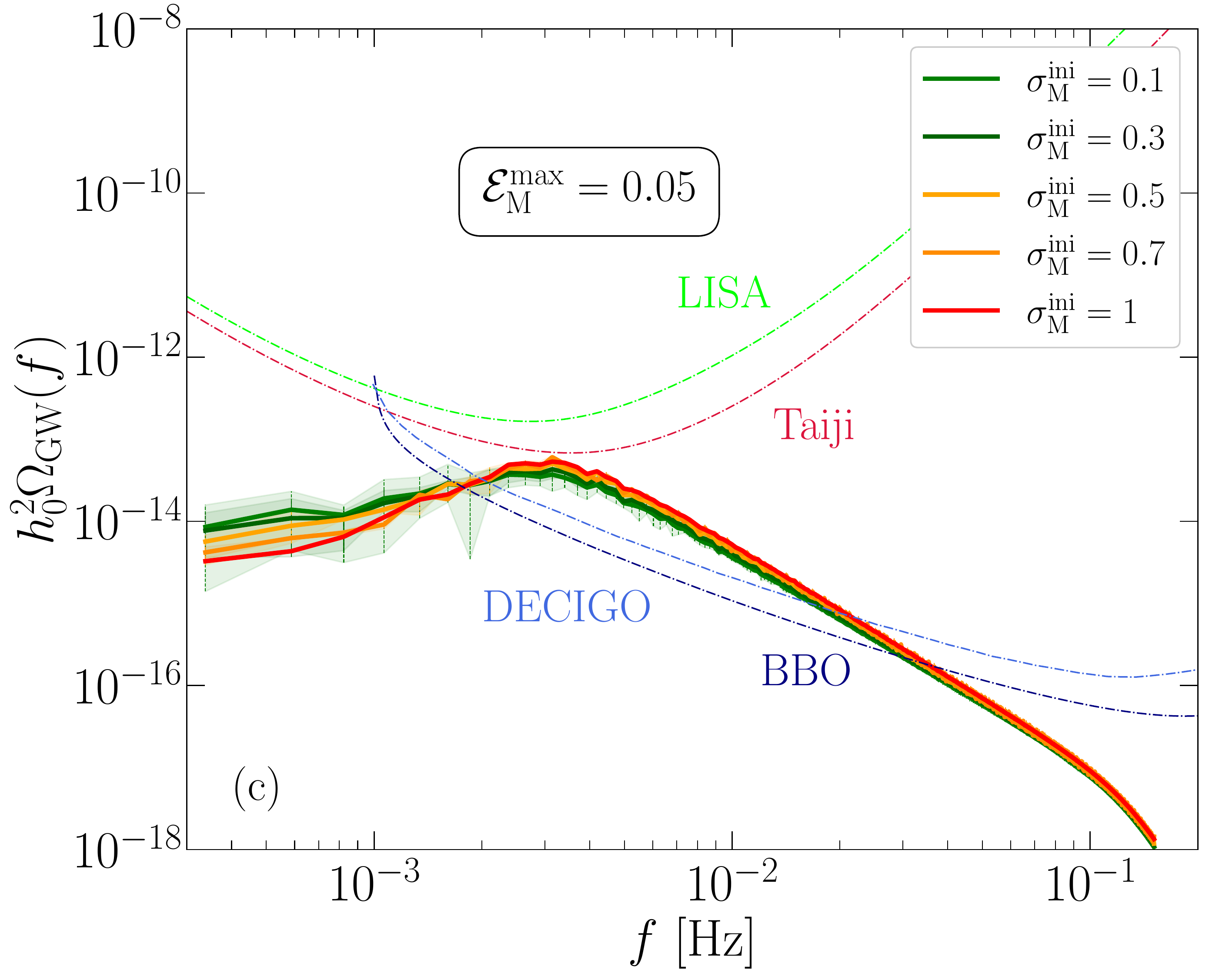}
\includegraphics[width=.49\textwidth]{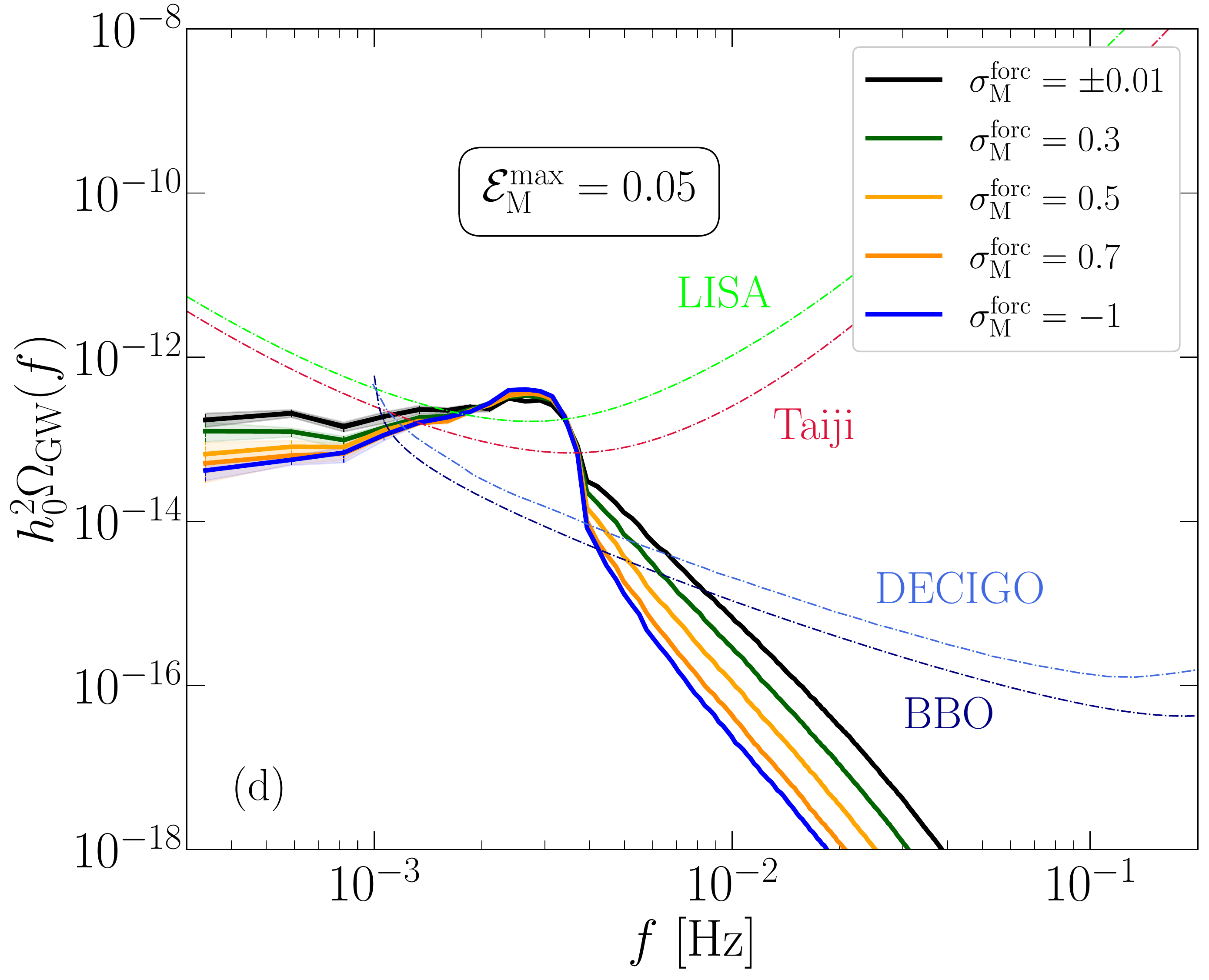}
\end{center}\caption[]{GW spectrum $h_0^2\, \OmGW (f)$ for
signals produced at the EWPT from magnetic fields 
with $\EEMmax = 0.1$ for the case with an initial given field (panel a), and
the case with a driven field (panel b),
and with $\EEMmax = 0.05$ for both cases (panels c and d). 
The PLS of the GW detectors assume a SNR
of 10 for LISA, Taiji,
BBO, and DECIGO, for an observation duration
of 4 years; see refs.~\cite{Caprini:2019pxz,Schmitz:2020syl}
and \Fig{GW_sensitivity}.
}\label{OmGW_detectors}\end{figure}

\subsection{Detectability of GW energy density and polarization}
\label{detection_sec}

To assess the observational prospects of detecting GWs, 
we plot in \Fig{OmGW_detectors} the resulting GW signal computed numerically;
see \Sec{numerical_sec},
for different values of $\sigM$, for runs with an initial
given magnetic field (left panels) and an initially
driven magnetic field (right panels),
and compare with the expected PLS of
LISA \cite{Caprini:2019pxz},
DECIGO, and BBO \cite{Schmitz:2020syl}.
We use the results from the numerical simulations; see 
\Tab{runs}, shifted to $\EEMmax = 0.1$ and $0.05$, using
the computed $\EEM^2$ scaling; see \Fig{OmGW_efficiency}.
The value of $0.1$ has been reported as an upper bound on the combined
magnetic, velocity, and GW energy density (as a fraction of the total
energy density) from BBN \cite{Shvartsman:1969mm,Kahniashvili:2009qi}.
The spectra are similar, but we now also see that for smaller values
of $|\sigM|$, the jump in $\OmGW (f)$ near the peak is
less pronounced, so for larger frequencies, i.e., to the right of the peak,
$\OmGW (f)$ increases (decreases) for smaller (larger)
values of $|\sigM|$. 
For smaller frequencies, we have the aforementioned shallow spectrum
$\OmGW(f)\propto f$, which is approximately independent of
the value of $\sigM$. 
We see that, for an initial magnetic energy density of $\EEMmax=0.1$,
the GW signal produced is detectable by LISA with a SNR
larger than 10 for both types of turbulence.
For $\EEMmax=0.05$, only
the case in which the magnetic field is driven at initial times 
has a SNR above 10, while the case with an initially given magnetic field
has a SNR between 1 and 10.

\begin{figure}[t!]\begin{center}
\includegraphics[width=.49\textwidth]{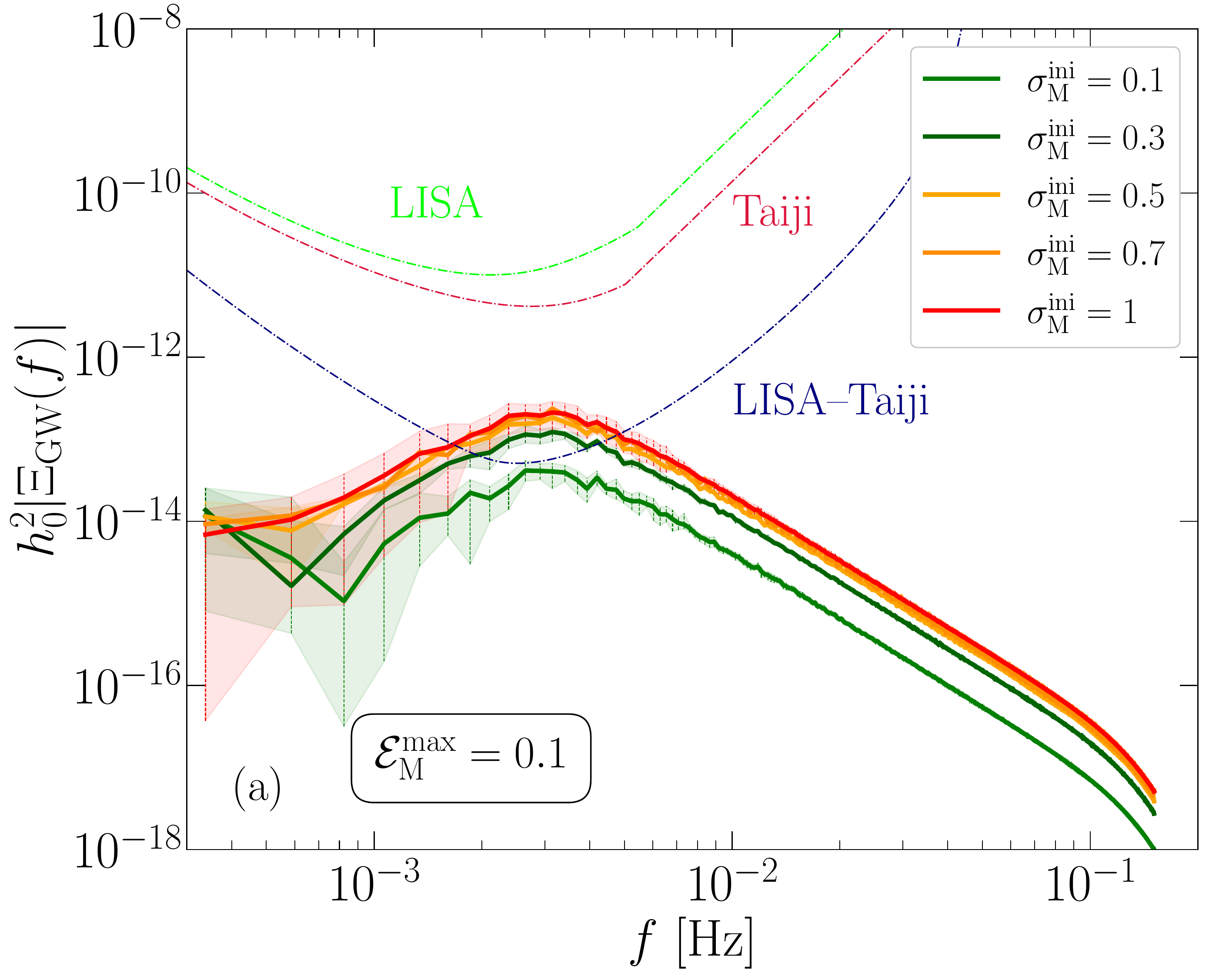}
\includegraphics[width=.49\textwidth]{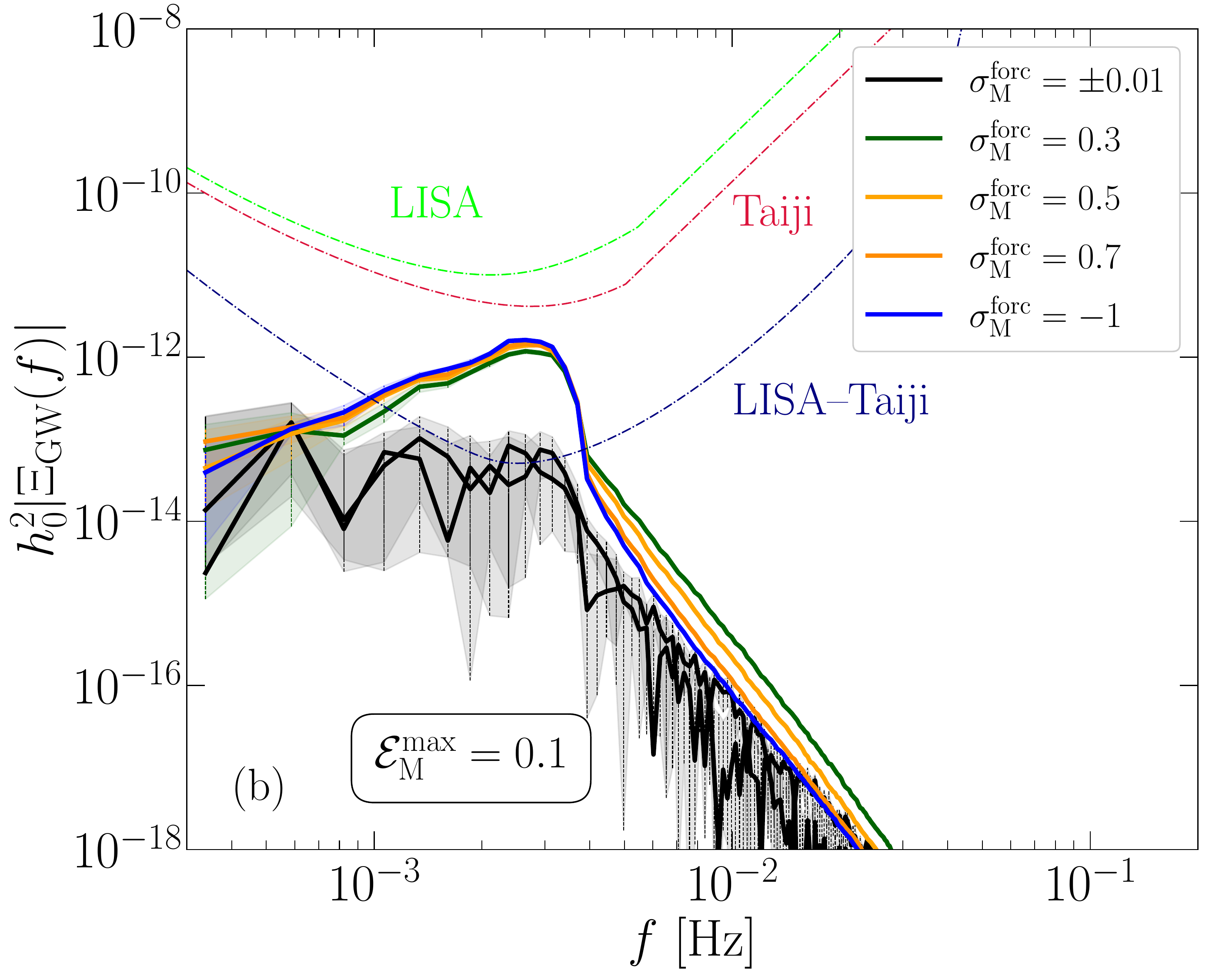}
\end{center}\caption[]{Helical GW 
spectrum $h_0^2 |\XiGW(f)|$ for
signals produced at the EWPT from magnetic fields
with $\EEMmax=0.1$ 
for the case with an initial given field (panel a) and
the case with a driven field (panel b).
The PLS to a polarized GW signal for LISA and Taiji
correspond to SNR of 1 for an observation duration
of 4 years; see \Fig{GW_sensitivity} and ref.~\cite{Ellis:2020uid}.
The LISA--Taiji curves correspond to the PLS with a SNR of 1
and 4 years of mission obtained by cross-correlating the LISA
and Taiji channels; see \Fig{GW_sensitivity}.
}\label{XiGW_detectors}\end{figure}

We show in \Fig{XiGW_detectors}
the helicity spectra $|\Xi_{\rm GW}(f)|$ obtained from
our numerical simulations together with
the PLS obtained using the dipole response
function of LISA, and obtained by cross-correlating LISA and Taiji channels.
As in \Fig{OmGW_detectors}, we shift the numerical GW signal
to $\EEMmax=0.1$ and $0.05$, according to the scaling $\OmGW(f) \sim 
\EEM^2$.
The helical GW spectrum, as the GW energy density, is smaller for the
cases in which the magnetic field is given at the initial time than
those in which it is initially driven.
In this case, since the degree of circular polarization of the GW
spectrum is proportional to the magnetic helicity, the fully helical
runs result in the largest polarized GW signals. 
This is clearly seen in the case with initially given magnetic fields.
In the driven case, the larger efficiency for smaller fractional 
magnetic helicity compensates for moderate values of the helicity,
and the polarized signal is comparable for all runs with 
$|\sigM| \geq 0.3$.
Moreover, in the inertial range, since the drop of GW energy
is larger for runs with larger helicity, we can observe that the
helical GW spectrum becomes smaller for larger values of $|\sigM|$.
We observe that in the limit of a non-helical magnetic field, i.e.,
$\sigM=\pm 0.01$, this is no longer the case since the helical spectrum
is proportional to the magnetic helicity.
In all the cases, we observe that the degree of circular polarization
of the GW signals is not large enough to be detectable using the
dipolar response function of LISA due to our proper motion.
However, the detection of circular polarization of an isotropic GW background can be
improved by cross-correlating two space-based detectors, e.g., LISA and Taiji
\cite{Seto:2020zxw,Orlando:2020oko}, 
which are both planned to be launched around 2034 \cite{Audley:2017drz,Guo:2018npi}.
We have discussed this briefly in \Sec{interferometry_text}, and in
more detail in \App{App_LISA_Taiji}.
The resulting SNR of the cross-correlated channels of LISA and Taiji is given 
in \Eq{SNR_pol_2}.
\FFig{XiGW_detectors} shows that the combined LISA--Taiji
network could lead to the detection of parity-violating
signals produced from primordial magnetic fields around the
electroweak scale.
In the case of an initial given magnetic field, the SNR is
between 1 and 10 for the upper bound estimate of $\EEMmax = 0.1$, while for a driven
magnetic field, the SNR is above 10.
For smaller $\EEMmax$, e.g., $0.05$, the former can only reach a SNR of 1,
while the latter case yields values of the SNR close to 10
in the case with moderate values of the fractional magnetic helicity.

\begin{figure}[t!]\begin{center}
\includegraphics[width=.49\textwidth]{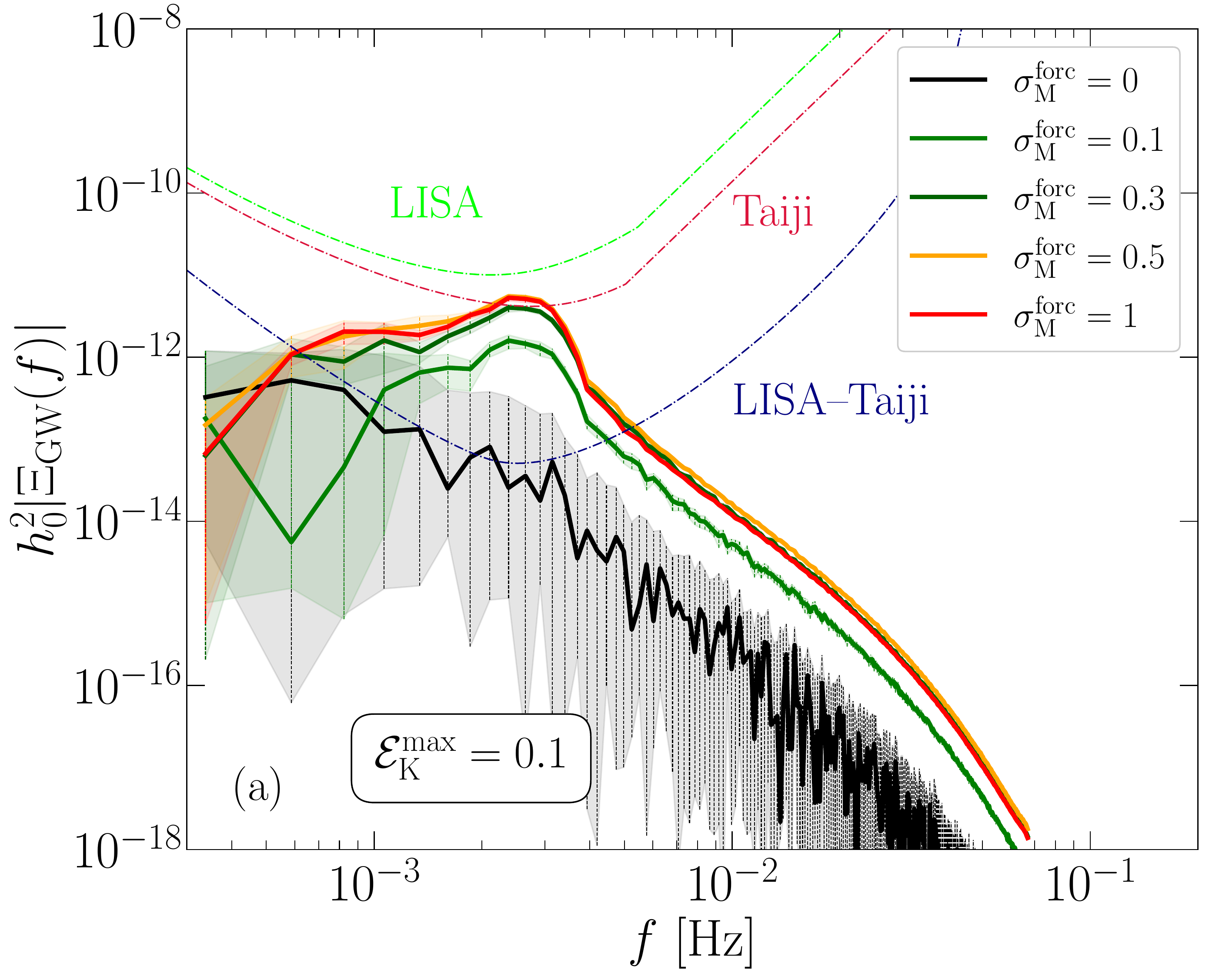}
\includegraphics[width=.49\textwidth]{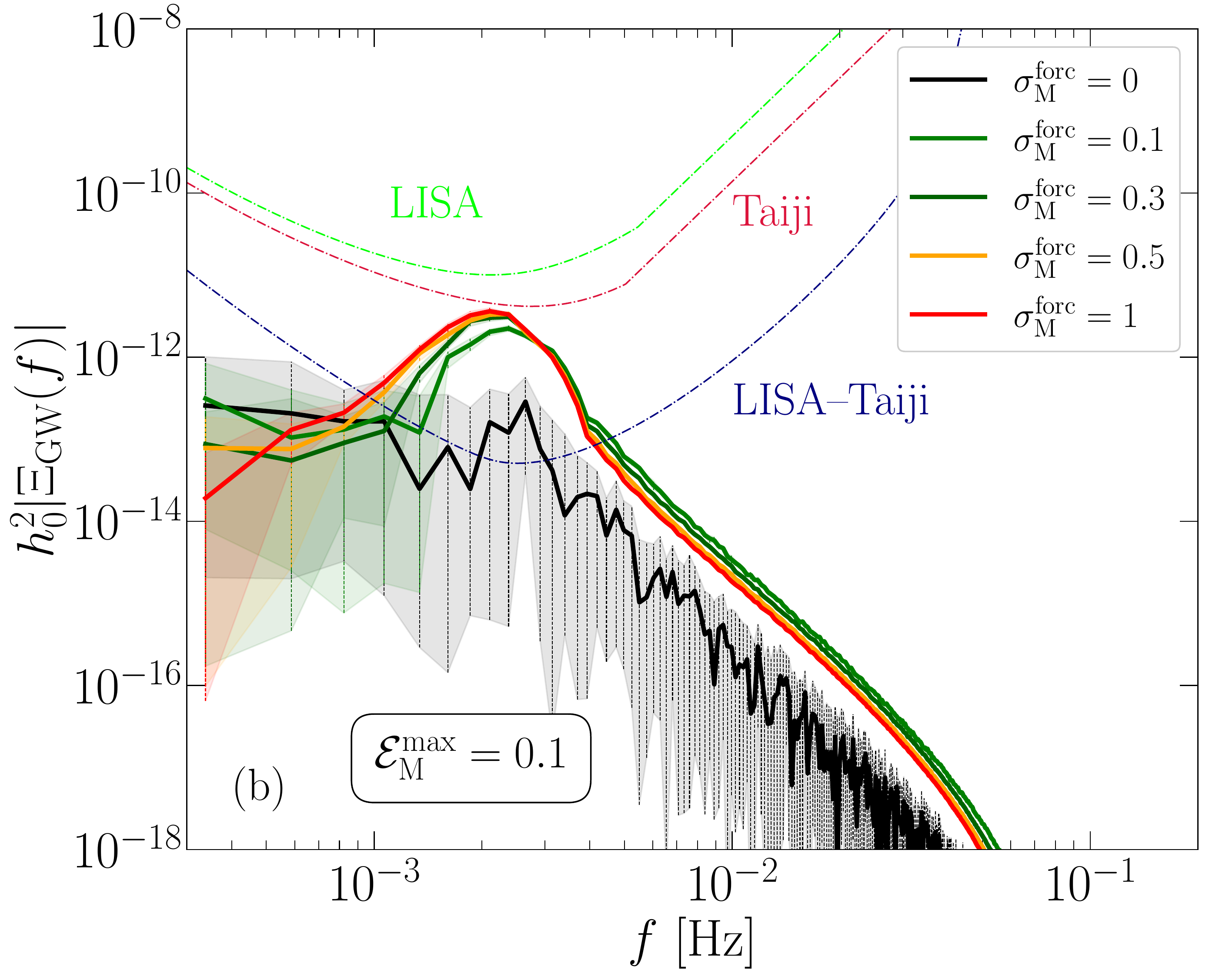}
\end{center}\caption[]{Similar to \Fig{XiGW_detectors}, helical GW
spectrum $h_0^2 |\XiGW (f)|$ for
signals produced at the EWPT from kinetically dominated (panel a),
and magnetically dominated turbulence
(panel b), using the results from the numerical 
simulations presented in ref.~\cite{Kahniashvili:2020jgm},
shifted to ${\cal E}_i^{\rm max}=0.1$, with $i=$ M or K. 
}\label{XiGW_detectors_GWCirc}\end{figure}

\FFig{XiGW_detectors_GWCirc} shows the helical GW
spectrum $h_0^2 \, \XiGW(f)$ computed from the simulations
presented in ref.~\cite{Kahniashvili:2020jgm}.
In this case, for $\EEMmax=0.1$, 
we find a SNR close to but still below unity when only considering
LISA self-correlations.
The GW signals are larger in this case because the
forcing term is acting for longer times, leading to
a larger GW production; see \Fig{OmGW_efficiency}.
We observe that the kinetically dominated turbulence leads to
larger values of the helical GW spectrum than the magnetically
dominated one around the spectral peak.
This is due to the larger production of GW amplitude in the kinetic
case, since the degree of circular polarization is larger
at low wave numbers for the magnetic case; see figure~(3)
of ref.~\cite{Kahniashvili:2020jgm}.
Hence, the helical inverse cascade is more efficient in the
magnetically dominated case, leading to larger $\PPGW (f)$
at low frequencies for magnetically dominated turbulence.
However, the resulting GW signal $\OmGW(f)$ is stronger for
kinetically dominated turbulence in this range of frequencies,
compensating the stronger inverse cascade.
This is not seen in figure~(4) of ref.~\cite{Kahniashvili:2020jgm}, 
since the energy density of the turbulent sourcing ${\cal E}_i^{\rm max}$,
with $i=$ M or K,
is not the same for all runs (see their table~II), and hence, 
the resulting GW spectra cannot be directly compared.
Note that this relies on the result that the GW spectrum
scales with ${\cal E}_i^2$, which the numerical 
simulations seem to indicate \cite{Pol:2018pao,Pol:2019yex,
Kahniashvili:2020jgm,Brandenburg:2021aln,Brandenburg:2021bvg,
Brandenburg:2021tmp}.

\section{Conclusions}
\label{conclusions}

In the present work, we have studied the generation of polarized 
stochastic GW backgrounds produced by partially and fully
helical turbulent sources, in particular, primordial magnetic fields.
Our numerical simulations have confirmed that there is a direct correspondence 
between the magnetic helicity and the degree of circular polarization
of the GWs produced from the resulting magnetic stress. 
We have calculated the GW spectra,
both the energy density $\OmGW(f)$ and the helicity $\XiGW(f)$, 
that are produced by primordial magnetic fields for different values of the 
fractional magnetic helicity, and assuming two types of turbulence
production.
On the one hand, we have studied the production of GWs due to the presence of a fully
developed turbulent magnetic field at the
initial moment of GW production with a characteristic scale well defined
by the turbulent process.
On the other hand, we have studied the production of GWs due to
a primordial magnetic field that is driven by an electromotive force
that models magnetogenesis by injecting energy at a characteristic
scale for a short duration (around a 10\% of one Hubble time). 
In both cases, the resulting magnetic energy and helicity spectra
are typical of fully developed turbulence, but their spectral shapes
depend on the type of turbulence and the fractional helicity.

\subsection{Numerical GW spectra}

Our work confirms a shallow GW spectrum in the subinertial
range that was obtained in previous numerical simulations \cite{Pol:2019yex}.
We observe that the spectrum can instead possess even a small negative slope,
instead of the flat spectrum reported in ref.~\cite{Pol:2019yex},
especially for low values of the helicity and for cases when the magnetic
field is produced via  turbulence forcing for a short duration; see figures 
\ref{pspecm_hel_initial}--\ref{pspecm_hel_forc}.
Such spectral slopes have also been reported in other recent numerical simulations
which model the generation of magnetic fields via a forcing term
\cite{Kahniashvili:2020jgm,Brandenburg:2021tmp},
and in ref.~\cite{Brandenburg:2021aln}, in which the magnetic field 
is produced by the chiral magnetic effect.
As suggested in ref.~\cite{Brandenburg:2021tmp}, this could be due to a finite
size of the simulation domain or it could indeed be physical due
to inverse transfer, analogous to that in helical \cite{Frisch:75}
and non-helical hydromagnetic turbulence \cite{Brandenburg:2016odr}.
Additionally, it has been pointed out in
refs.~\cite{Brandenburg:2021bvg,Brandenburg:2019uzj}
that the spectrum of the stress becomes shallower than white noise when the magnetic (or velocity)
field is non-Gaussian, which would lead to a shallower GW spectrum.
In the case when the magnetic field is driven, the MHD evolution might indeed generate a
stochastic magnetic field with deviations from a Gaussian field.
However, the exact shape of the GW spectrum at large scales requires further study.

\subsection{GW polarization spectrum of magnetic fields
initially present}

In the runs with an initially given magnetic field,
we show that the magnetic spectral shape 
does not depend on helicity, and the helicity spectrum has the same
slopes as the magnetic spectrum, with a strength proportional to its
fractional helicity (see \Fig{pspecm_hel_initial}).
Both $\EM (k)$ and $k \HM(k)$ are characterized by a $k^4$
Batchelor spectrum below the 
peak at $\kf$ (taken to be $600$ in our simulations) and by a $k^{-5/3}$
Kolmogorov spectrum in the inertial range. 
The resulting GW spectrum $\EGW(k)$ (defined per linear wave number interval)
has an inertial range slope of $-11/3$, as was already
shown in the numerical simulations of ref.~\cite{Pol:2019yex},
with a spectral peak at around $2\kf$, as expected, since GWs are sourced by the magnetic
stress, which is obtained by convolution of the magnetic spectrum with itself. 
The helical GW spectrum $\HGW(k)$ also has a similar spectral shape as the GW
energy density spectrum.
The exception is precisely at large scales (or low wave numbers), where we observe
a decay of the helical GW spectrum in the smallest modes of the simulation 
(affecting the second and/or third largest wave numbers of the simulation).
This becomes 
more noticeable for small values of the fractional magnetic helicity $|\sigM|$.
The reason for this is not clear, since this is not observed in the magnetic helicity
spectrum, but it might be due to the finite size of the domain.
In \Fig{ppol_comp}, we show that there are strong fluctuations at low wave numbers,
which induce uncertainty on the actual value of the circular degree of polarization 
$\PPGW(k)$
that is much reduced for larger wave numbers.
In the case of an initial given magnetic field, the polarization degree follows the
description given in refs.~\cite{Kahniashvili:2005qi,Kisslinger:2015hua,Ellis:2020uid},
after a few minor modifications, 
for turbulence dominated by helical transfer (HT), which
assumes the same spectral slopes for the
energy and helicity spectra, and relies on the assumptions of stationary
turbulence and short sourcing; see
figures~\ref{PGW_analytical} and \ref{PGW_analytical_comp_HT}. 
However, we showed that the polarization degree is not the same if computed
directly from the strains, $\PPh (k) = \Ah(k)/\Sh(k)$, or
from the time derivatives of the strains, $\PPGW (k) = \Ahd(k)/\Shd(k) =
\HGW(k)/\EGW(k)$,
the latter being the one that is in 
better agreement with the analytical description.
We find a linear relation between the magnetic polarization degree
$\PPM$ and the resulting
GW polarization $\PPGW$, both obtained as the ratio of the total helicity
of the magnetic or GW field 
to the total energy density; see \Fig{pppol1}. 
The linear relation $\PPGW \sim \PPM$, obtained from the numerical simulations,
deviates from the analytical model considered in \App{beltrami_app}; see \Fig{PM_bel}.
However, the model in \App{beltrami_app} corresponds to a single-mode 
magnetic field of fractional magnetic helicity.
Its generalization to a three-dimensional fully developed turbulent field is used
for the numerical simulations; see \Eq{Bikk}.

\subsection{GW polarization spectrum of magnetic fields
initially driven}

When the turbulence is forced for a finite duration, the situation changes drastically. 
Due to the quasi-monochromatic sourcing, a spike appears in the magnetic spectrum
$\EM(k)$ at initial times.
When the magnetic energy density has reached its maximum value $\EEMmax$, and it starts
to decay (which is taken to be at $\tmax=1.1$, given in Hubble times), the spike has
smoothed around the spectral peak $\kf$ but it is still not completely gone;
see \Fig{pspecm_hel_forc}.
In the subinertial range, we observe a Batchelor spectrum with slopes very close to $4$,
while in the inertial range, the spectrum has negative slopes, steeper than
$-5/3$ (which corresponds to Kolmogorov turbulence).
The spectrum of magnetic helicity $\HM (k)$ presents a similar shape below and around the peak
for all values of the fractional helicity above $|\sigM|=0.3$ (shown in \Fig{pspecm_hel_forc})
with a fractional polarization of almost 1 in this range of wave numbers.
The exception to this is for almost non-helical runs; see \Fig{pspecm_hel_forc001},
for which the spectrum of helicity is negligible compared to the magnetic spectrum at
all scales.
At wave numbers above the peak, the helicity spectrum shows a sharper decrease with $k$ than the
magnetic energy density, which becomes less pronounced as we increase the helicity,
such that in the fully helical case the slopes of the magnetic and helicity spectra
become very similar.
This is plausibly explained by a forward cascade of current helicity \cite{Brandenburg:2005xc}. 
The GW spectrum $\EGW(k)$ shows a drop on amplitude at scales just below the peak that has also been observed
in other recent numerical simulations \cite{Pol:2019yex,Brandenburg:2021tmp} due
to the finite sourcing of the magnetic field. 
This drop appears also in the antisymmetric or helical spectrum of GWs $\HGW(k)$, as we show in
\Fig{pspecm_hel_forc}, and it does not depend on the fractional helicity. 
The GW degree of circular polarization $\PPGW(k)$ in this case is shown in \Fig{ppol_comp}, 
where we show that it reaches the fully polarized case at the peak for large values
of the fractional magnetic helicity $|\sigM|$ and then decays at large wave numbers.
The case with different slopes of the magnetic and helicity spectra was studied in
previous analytical works \cite{Kahniashvili:2005qi,Kisslinger:2015hua,Ellis:2020uid}, 
leading to a maximum degree of circular polarization of 80\%, underpredicting it when
compared to our numerical simulations. 
We compared in \Fig{PGW_analytical_comp_HK} the prediction of the polarization degree from the
analytical model with our numerical results, showing that the numerical
simulations do not follow the degree of circular polarization obtained by
the analytical models in any of the two models considered previously, i.e., 
helical Kolmogorov (HK) or HT turbulence.
Previous analytical assumptions were using a single power law for the magnetic
energy and helicity spectra that did not depend on the wave number,
besides the assumptions used to model the unequal time correlator and
on the turbulence duration, as discussed in \Sec{an_spec_pol_section}.
However, we showed in \Figs{PGW_analytical_comp_HT}{PGW_analytical_comp_HK}
that the predictions by the analytical models are more accurate,
when compared to the numerical results,
if we use the proper magnetic spectra obtained from
the numerical simulations
and evaluate the analytical model to compute the GW polarization degree using \Eqs{Sh_an}
{Ah_an}.
This shows that the assumption of a single power law for the magnetic spectra
affects more strongly the resulting polarization degree than the assumptions of
short duration and stationary turbulence.
We observe that in this case, the analytical model yields values of unity for the 
polarization degree at the peak, larger than those obtained in previous analytical 
estimations, and in agreement with the numerical simulations. 
Reference \cite{Kahniashvili:2020jgm} studied the case of stationary turbulence, and 
computed the resulting GW degree of circular polarization.
They also showed
different values than previous analytical estimations for sources forced during larger 
times (around two Hubble times).
It is unclear whether the discrepancies are due to the long duration of the forcing or due to the
consideration of a single power law for the magnetic spectra.
This aspect is left for future work.
The precise form of the polarization degree can be important if one wants to infer the
magnetic helicity from circular polarization measurements of GWs.
Cosmological causal magnetic fields may well be close to fully helical
because the magnetic helicity is a conserved quantity while the
magnetic energy decays 
(and the correlation length increases), so the ratio always increases until it
reaches nearly 100\% if the magnetic field dynamically evolves for long enough,
with a fractional helicity growth rate depending quadratically on its initial 
value: longer (shorter) period is needed for a field with smaller (higher) initial 
fractional magnetic helicity to become fully helical \cite{Tevzadze:2012kk}.

\subsection{Detectability of the GW polarization with space-based GW detectors}

Finally, we have explored the potential detectability of the GW signals by
future space-based GW detectors if primordial partially or fully
helical magnetic fields were present or produced at the EWPT.
The resulting GW energy densities (shown in \Fig{OmGW_detectors})
are detectable by LISA with a
SNR of 10 for magnetic energy densities of 10\% of the radiation energy
density if the magnetic field is initially given, 
and magnetic
energy density ratios of at least 3\% and 2\%
if the magnetic field is initially driven for a short
(around 10\% of the Hubble time) and a long duration (about two Hubble times), respectively.
Even smaller magnetic amplitudes yield GW signals that can be detectable
with second-generation space-based detectors, e.g., BBO and DECIGO.
Using the dipole response induced by the proper motion of the solar system in the 
LISA interferometer channels, as recently studied in 
refs.~\cite{Domcke:2019zls,Ellis:2020uid},
a detectable polarized GW signal with a SNR of at least 10
requires magnetic energy densities above 10\% of the radiation energy density
(as shown in \Figs{XiGW_detectors}{XiGW_detectors_GWCirc}),
which marginally coincides with the upper bound imposed by the BBN on the energy densities
of additional relativistic components based on the abundance of light elements
\cite{Grasso:1996kk,Shvartsman:1969mm,Kahniashvili:2009qi}.
This is consistent with other turbulent sources; see, e.g., 
ref.~\cite{Ellis:2020uid}, in which they
require strong first-order phase transitions ($\alpha \sim 1$) to obtain a SNR 
of 10; see their figure~8.
We can reach a maximum SNR of unity in the limit of 10\% of energy density 
transformed in magnetic
fields only in the case where the magnetic field is fully helical and
forced for a long duration,
following the numerical results of ref.~\cite{Kahniashvili:2020jgm};
see \Fig{XiGW_detectors_GWCirc}.
Therefore, using the dipole response function of a planar
space-based GW detector as LISA 
is not enough to detect the signals computed in the present work,
although we highlight that our investigation is limited by the consideration 
of sub-relativistic velocities,
which possibly leads to an underestimation
of the signal strength \cite{Kosowsky:2001xp}.
In addition, following ref.~\cite{Orlando:2020oko}, we have computed the power law sensitivity
corresponding to a polarized GW signal by cross-correlating two space-based GW detectors,
e.g., LISA and Taiji, which breaks the coplanarity of the detectors and allows one to detect polarization
in the monopole response functions. 
We have shown that the GW degree of circular polarization produced by primordial 
magnetic fields generated at the EWPT can yield polarization SNR up to 20 if
they are initially driven for a short time (about 10\% of one Hubble time)
with a maximum magnetic energy density of 10\%,
as long as the fractional magnetic helicity is $|\sigM| \geq 0.3$ or,
equivalently, $|\PPM| \geq 0.5$; see \Fig{XiGW_detectors}.
This is due to the fact that the resulting GW amplitudes are larger
for smaller values of the fractional helicity; see
\Fig{OmGW_efficiency},
which compensates for the decrease in magnetic helicity.
When the magnetic field is initially given, the polarization SNR of the
GW signal stays approximately between 1 and 5, such that its potential
detectability is more challenging.

\section*{Data availability}
The source code used for the simulations of this study,
the {\sc Pencil Code}, is freely available \cite{PC}.
The simulation data are also available at Ref.~\cite{DATA}.
The calculations, the simulation data, and the routines generating the
plots are also available on \href{https://github.com/AlbertoRoper/GW_turbulence/tree/master/JCAP_2107_05356}
{GitHub} \cite{GH}.
 
\section*{Acknowledgements}

We thank Arthur Kosowsky for useful discussion.
ARP is supported by the French National Research Agency (ANR) project MMUniverse
(ANR-19-CE31-0020).
Support through the Swedish Research Council (Vetenskapsr{\aa}det),
grant 2019-04234, and the Shota Rustaveli 
National Science Foundation (SRNSF) of Georgia (grant FR/18-1462)
are gratefully acknowledged.
We acknowledge the allocation of computing resources provided by the
Swedish National Allocations Committee at the Center for Parallel
Computers at the Royal Institute of Technology in Stockholm,
and the A9 allocation of GENCI at the Occigen supercomputer
to the project ``Opening new windows on
Early Universe with multi-messenger astronomy.''

\appendix

\section{Analytical model for polarization}
\label{beltrami_app}

\begin{figure}
    \centering
    \includegraphics[width=.6\textwidth]{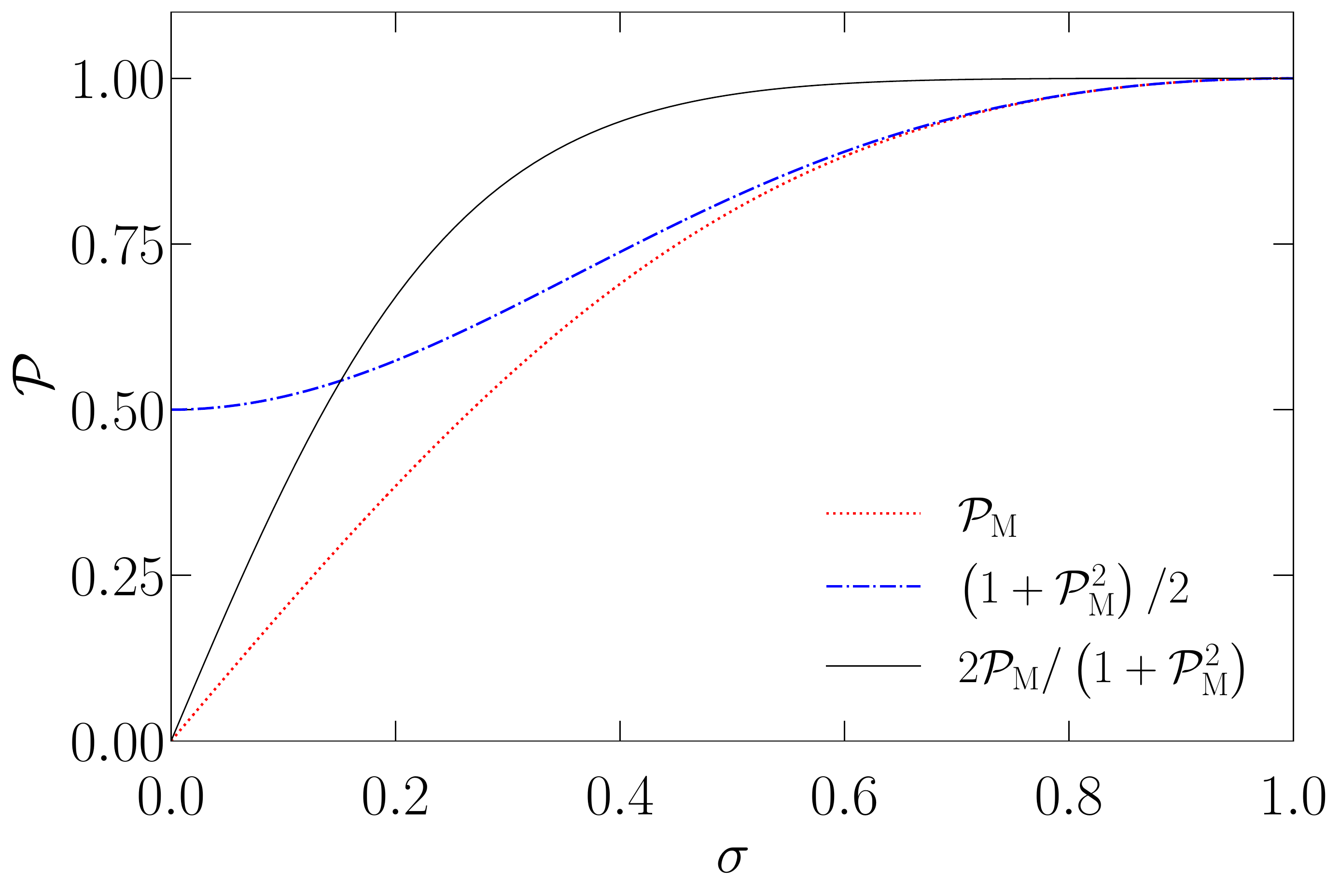}
    \caption{Functions $\half (1 + \PPM^2)$ and $\PPM$, that appear in the symmetric  and antisymmetric
    functions of the GW spectra, respectively, and degree of circular polarization
    $\PPGW = 2 \PPM/(1 + \PPM^2)$ obtained for the 1D analytical model.
    We obtained the numerical fit $\PPGW \sim \PPM$ in the
    numerical simulations of MHD turbulence; see \Fig{pppol1}.}
    \label{PM_bel}
\end{figure}

We present in the current section the calculations of the GW
spectra, both the symmetric $\EGW(k)$ and the antisymmetric $\HGW(k)$
functions, and the degree of circular polarization $\PPGW (k)$,
for a magnetic field with arbitrary fractional helicity that varies in one spatial direction,
determined by the parameter $\sigma=\sigM$.
This model allows one to show analytically that fully helical
magnetic fields induce circularly polarized GW signals, and to predict
the dependence of the GW amplitude and polarization on the
magnetic amplitude and fractional helicity.
We start with a transverse
magnetic field given as
\begin{equation}
\BB(x, t) = \sqrt{\frac{2}{1 + \sigma^2}} B_0\, \Theta(t-1)
\left( \begin{array}{c}
0 \\ \sigma \sin k_0 x \\ \cos k_0 x
\end{array} \right),
\end{equation}
where $\sigma \in [0, 1]$ is a parameter that modifies the 
helicity of the magnetic field, $k_0$ is the characteristic
wave number, and $\EEM = \half \bra{\BB^2} = \half B_0^2$ is the magnetic
energy density, with $B_0$ the magnetic field amplitude.
The Heaviside function $\Theta(t-1)$ is included to indicate that
this field is zero for $t \leq 1$.
This field is a monochromatic 1D simplification of the general
function used in the turbulence simulations; see \Eq{forcing}.
Note that when $\sigma=\pm 1$, we have a Beltrami (fully helical) field, that was
studied in ref.~\cite{Pol:2018pao} in the context of GW generation, and used
to study numerical accuracy of the {\sc Pencil Code} simulations.
The vector potential $\AAA$ is defined such that $\nab \times \AAA = \BB$,
\begin{equation}
    \AAA(x,t)=\sqrt{\frac{2}{1 + \sigma^2}} \frac{B_0}{k_0}
    \, \Theta(t - 1) \left( \begin{array}{c}
    0 \\ \sin k_0 x \\ \sigma \cos k_0 x \end{array} \right)
    + \AAA_g,
\end{equation}
where $\AAA_g$ is gauge-dependent,
and the helicity of the magnetic field is
\begin{equation}
    \HHM=\bra{\BB \cdot \AAA}=\frac{2\sigma}{1+\sigma^2}\frac{B_0^2}{k_0},
\end{equation}
which is not gauge-dependent.
The magnetic energy density $\EM(k)$ and the helicity $\HM(k)$ spectra are
\begin{align}
\EM (k, t) =& \, \half \int_{\Omega_1} \tilBB (\kk, t)
\cdot \tilBB^*(\kk ,t)\,
\dd \Omega_1 =\EEM \, \Theta(t - 1)\, \delta(k - |k_0|), \\
\HM (k, t)=&\, \int_{\Omega_1} \tilBB (\kk, t) \cdot \tilAA^* 
(\kk, t) \dd
\Omega_1 = \frac{2\sigma}{1 + \sigma^2} \frac{2 \EEM}{k_0}\,
\Theta(t - 1) |k_0| \, \delta(k - |k_0|),
\end{align}
where $\Omega_1 = 2$ is the one-dimensional solid angle
and $\delta(k)$ is the one-dimensional Dirac's delta function.
The fractional magnetic helicity is
\begin{equation}
\PPM = \frac{k_0\bra{\BB(x, t) \cdot \AAA (x, t)}}{\bra{\BB^2 (x, t)}} = 
\frac{k_0\HHM}{2 \EEM} = 
\frac{2 \sigma}{1 + \sigma^2},
\label{PPM_bel}
\end{equation}
which reduces to $+1$, $0$, and $-1$,
for $\sigma = +1$, $0$, and $-1$, respectively.
The stress tensor of the magnetic fields is
\begin{equation}
T_{ij} (x, t) = - B_i B_j + \frac{1}{2} \delta_{ij} \BB^2,
\end{equation}
with
\begin{equation}
\BB^2 (x, t) = 2 \EEM \, \Theta(t - 1)
\left(1 + \frac{1 - \sigma^2}{1+\sigma^2} \cos 2k_0 x \right),
\label{B2_bel}
\end{equation}
and
\begin{equation}
-B_i B_j (x, t) = - \frac{2\EEM}{1+\sigma^2} \, \Theta(t - 1)
\left( \begin{array}{c c c}
0 & 0 & 0 \\
0 & \sigma^2 (1 - \cos 2 k_0 x) & \sigma \sin 2 k_0 x \\
0 & \sigma \sin 2k_0 x & 1 + \cos 2 k_0 x
\end{array} \right).
\label{BiBj_bel}
\end{equation}
Combining \Eqs{B2_bel}{BiBj_bel}, we obtain the stress tensor
\begin{equation}
T_{ij} (x, t) = \EEM \, \Theta(t - 1)\left(
\begin{array}{c c c} T_{11} (x) & 0 & 0 \\
0 & T_{22} (x) & T_{12} (x) \\ 0 & T_{12} (x)
& -T_{22} (x) \end{array} \right),
\label{Tij}
\end{equation}
with
\begin{align}
T_{11}(x)=&\,1 + \frac{1 - \sigma^2}{1+\sigma^2} \cos 2k_0 x,
\nonumber \\
T_{22}(x)=&\,1 + \frac{1 - \sigma^2}{1+\sigma^2} \cos 2k_0 x
- \frac{2\sigma^2}{1+\sigma^2}  (1 - \cos 2k_0 x) =
\frac{1-\sigma^2}{1+\sigma^2} + \cos 2k_0 x, \nonumber \\
T_{12}(x)=&\, - \frac{2\sigma}{1+\sigma^2} \sin 2 k_0 x,
\end{align}
which becomes $T_{11}(x) = 1$, $T_{22} (x) = \cos 2k_0 x$,
and $T_{12} (x) = \mp \sin 2k_0 x$ in the fully helical case
(i.e., $\sigma = \pm 1$) studied in ref.~\cite{Pol:2018pao}.
Since GWs are produced by linear perturbations over the metric tensor,
and the stress tensor components are also perturbations over background
fields, constant modes ($\kk = \nullvector$) do not produce GWs.
For this reason, we can neglect the terms that are homogeneous in space,
and we write
\begin{equation}
T_{ij}^{\rm TT}(x, t) = \frac{\EEM}{1+\sigma^2} \, \Theta(t - 1) \left( \begin{array}{c c c}
0 & 0 & 0 \\
0 & (1 + \sigma^2) \cos 2k_0 x & - 2\sigma \sin 2k_0 x \\
0 & - 2\sigma \sin 2k_0x & - (1 + \sigma^2) \cos 2k_0 x
\end{array}\right),
\end{equation} 
where we noted that taking $T_{11}$ to zero, the stress tensor becomes 
traceless and transverse (TT), with the $+$ and $\times$ modes,
\begin{equation}
T_+ (x, t) = \EEM \, \Theta(t - 1) \cos 2 k_0 x, \quad
T_\times(x, t) = \EEM \PPM \, \Theta(t - 1)
\sin 2k_0x,
\end{equation}
where we have used $\PPM$, given in \Eq{PPM_bel}.
In Fourier space, the resulting stress tensor components are
\begin{equation}
\tilde T_+(k_x=\pm 2k_0, t) = \half \EEM \, \Theta(t - 1), \quad
\tilde T_\times (k_x=\pm 2k_0, t) =  \mp \half i \EEM  \PPM \, \Theta(t - 1).
\end{equation}
The GW strains $\tilde h_+ (k_x, t)$ and $\tilde h_\times (k_x, t)$ are computed 
from the GW \Eq{GW} with initial condition $h_{+,\times} = \partial_t h_{+,\times} = 0$
at $t=1$, and assuming flat
non-expanding space-time
for the radiation-dominated epoch
(see ref.~\cite{Pol:2018pao} for more details),
\begin{align}
\tilde h_+ (\pm 2k_0, t) = &\, \frac{3}{k_0} \int_1^t \tilde T_+ (\pm 2k_0, \tau)
\sin [2 k_0 (t - \tau)] \, \dd \tau = \frac{3}{4 k_0^2} \EEM \, \Theta(t - 1)
\left( 1 - \cos [2 k_0 (t - 1)] \right) \nonumber \\
= &\, \frac{3}{2 k_0^2} \EEM \sin^2 [k_0(t-1)] \, \Theta(t-1), \\ 
\tilde h_\times (\pm 2k_0, t) = &\, \frac{3}{k_0} \int_1^t \tilde T_\times (\pm 2k_0, \tau)
\sin [2 k_0 (t - \tau)] \, \dd \tau
\nonumber \\
= & \, \mp \frac{3i}{2 k_0^2} \PPM \EEM
\sin^2 [k_0(t-1)]\, \Theta(t-1).
\end{align}
The spectral functions $\Sh(k,t)$ and $\Ah(k,t)$ are
\begin{align}
\Sh(2 |k_0|, t)  = &\, \int_{\Omega_1} \left[ \tilde h_+ (\kk, t)
\tilde h_+^* (\kk, t)  + \tilde h_\times (\kk, t)
\tilde h_\times^* (\kk, t)\right] \dd \Omega_{\kk} \nonumber \\
= &\, \frac{9}{2 k_0^4} \left(1 + \PPM^2
\right)
\EEM^2 \sin^4 [k_0 (t-1)] \, \Theta(t - 1), \\
i \Ah(2 |k_0|, t) = &\, \int_{\Omega_1} \left[ \tilde h_+ (\kk, t)
\tilde h_\times^* (\kk, t)  -
\tilde h_+^* (\kk, t) \tilde h_\times (\kk, t)\right] \dd \Omega_{\kk}
\nonumber \\
= &\, i \frac{9}{k_0^4} \PPM
\EEM^2 \sin^4 [k_0 (t - 1)] \, \Theta(t - 1),
\end{align}
and zero for all the other values of the wave number $k$. 
The spectrum of the characteristic amplitude $\hc(k,t)$,
and its total value integrated over $k$, are
\begin{align}
    \hc(2|k_0|, t) = &\, \sqrt{k\Sh(k,t)} \nonumber \\
    = &\, \frac{3}{k_0^2}
    \sqrt{|k_0| \left(1+\PPM^2\right)} \EEM \sin^2 [k_0(t-1)]
    \, \Theta(t-1),\\
    \hc(t) = &\, \left(\int \Sh(k, t) \, \dd k\right)^{1/2}\nonumber \\
    = &\, \frac{3}{k_0^2}
    \sqrt{\half\left(1+\PPM^2\right)}
    \EEM \sin^2 [k_0 (t-1)] \, \Theta(t-1).
\end{align} 
The characteristic amplitude averaged over oscillations in time
is $\hc = \threehalf \sqrt{\half (1 + \PPM^2)} \EEM/k_0^2$, 
and the polarization $\PPh$ is
\begin{equation}
\PPh = \frac{\Ah(2 |k_0|, t)}
{\Sh(2 |k_0|, t)} = \frac{2 \PPM}{1 + \PPM^2}.
\label{PPh_bel}
\end{equation}
The spectral functions $\Shd (k, t)$ and $\Ahd (k, t)$ are
\begin{align}
    \Shd(2 |k_0|, t) = &\, \int_{\Omega_1} \left[\tilde{\dot{h}}_+ (\kk, t)
    \tilde{\dot{h}}_+^* (\kk, t) + \tilde{\dot{h}}_\times (\kk, t)
    \tilde{\dot{h}}_\times^* (\kk, t) \right] \dd \Omega_{\kk}\nonumber \\
    = &\,\frac{9}{2 k_0^2}\left(
    1 + \PPM^2 \right) \EEM^2 \sin^2 [2 k_0 (t-1)] \, \Theta(t-1), \\
    i \Ahd(2 |k_0|, t) = &\, \int_{\Omega_1} \left[\tilde{\dot{h}}_+ (\kk, t) \tilde{\dot{h}}_\times^* (\kk, t)
    - \tilde{\dot{h}}_+^*(\kk, t) \tilde{\dot{h}}_\times (\kk, t) \right]
    \dd \Omega_{\kk}\nonumber \\
    = &\, i \frac{9}{k_0^2}\PPM \EEM^2 \sin^2 [2 k_0 (t-1)] \, \Theta(t-1),
\end{align}
such that $\PPGW = \Ahd(k, t)/\Shd (k, t) =
\Ah(k, t)/\Sh (k, t) = \PPh$; see \Eq{PPh_bel}. 
The GW spectrum $\OmGW(k, t)$
and the total GW energy density
are
\begin{align}
    \OmGW(2|k_0|, t) = \frac{k \Shd(k, t)}{6}
    =  &\, \frac{3}{2 |k_0|} \left(
    1 + \PPM^2 \right)
    \EEM^2 \sin^2 [2k_0(t-1)]\, \Theta(t-1),
    \\
    \OmGW(t) = \frac{1}{6}\int_0^\infty
    \Shd (k, t) \, \dd k =  &\,
    \frac{3}{4k_0^2} \left(
    1 + \PPM^2 \right) \EEM^2 \sin^2
    [2k_0(t-1)]\, \Theta(t-1).
\end{align}
The GW energy density averaged over oscillations in time
is $\OmGW = \threeeigth (1 + \PPM^2) \EEM^2/k_0^2$.
The energy ratio is $\OmGW/\EEM = \half \hc \, \sqrt{\half (1 + \PPM^2)}$,
which was reported in ref.~\cite{Pol:2018pao} for the fully helical field,
i.e., $\PPM = \pm 1$.
For larger (smaller) values of the fractional magnetic helicity $|\PPM|$, this
result predicts larger (smaller) amplitudes of the GW energy density by a factor
$1 + \PPM^2 \in [1,2]$; see \Fig{PM_bel}.
In turbulent simulations; see \Fig{OmGW_efficiency}, we observe that
$\OmGW$ is not noticeably dependent on $\PPM$ if the magnetic
field is initially given, and that it decreases for larger values of $|\PPM|$
if the magnetic field is initially driven.

We find that the symmetric functions $\Sh(k, t)$ and $\Shd (k, t)$, and hence
the characteristic amplitude $\hc (k, t)$ and the GW energy density $\OmGW (k, t)$,
are proportional to the function $\half(1 + \PPM^2)$, while the antisymmetric
functions $\Ah(k, t)$ and $\Ahd (k, t)$, are proportional to $\PPM$.
These functions and the degree of circular polarization are shown in \Fig{PM_bel}
compared to the empirical relation $\PPGW \approx \PPM = 2\sigM/(1+\sigM^2)$, obtained from the
numerical simulations; see \Fig{pppol1}.

\section{LISA and Taiji interferometry}
\label{interferometry}

\subsection{Time-delay interferometry}

We derive here some of the expressions that are useful to compute the
response functions and sensitivity curves of LISA and Taiji.
Both space-based GW detectors have triangular configurations with three arms of the same
length, $L=2.5 \times 10^6 \km$ for LISA \cite{Audley:2017drz},
and $L=3 \times 10^6 \km$ for Taiji \cite{Ruan:2020smc}.
The combination of two arms with a common mass at their
vertices is a Michelson interferometer.
Hence, the 3 arms of LISA or Taiji lead to three interferometers 
that correspond to the physical channels $X$, $Y$, and $Z$. 
These channels are linearly combined to obtain the time-delay interferometry
(TDI) channels of LISA,
commonly known as $A$, $E$, and $T$ \cite{Adams:2010vc}, and
we call $C$, $D$, and $S$, the analogous Taiji channels, as done in 
ref.~\cite{Orlando:2020oko}.
Following ref.~\cite{Domcke:2019zls}, 
the time delay $\delta t$ induced by a gravitational wave in each of
the detector arms is
\begin{equation} 
    \sigma_i = \frac{c\delta t}{2L} = \sum_{\lambda=+,-} \int
    \tilde h_\lambda (\kk, t - L/c) \,
    e_{ab}^\lambda (\hat \kk) \Qi (\kk),
    \label{td_sigma}
\end{equation} 
where $+$ and $-$ are the helical polarization modes,\footnote{%
In the rest of the text we have used the linear
$+$ and $\times$ polarization modes, instead of the helical $+$ and $-$ modes.
The latter are useful in this section since the
GW energy density and the helical spectra, previously defined 
in the linear basis; see \Eqss{OmGW_k}{APiij}, can be expressed in the helical
basis as $\OmGW(f)=\OmGW^+(f)+\OmGW^-(f)$
and $\XiGW(f)=\OmGW^+(f)-\OmGW^-(f)$.} 
$e_{ab}^\pm (\hat \kk)$ are the helical polarization basis tensors,\footnote{%
The helical polarization basis tensors $e_{ab}^\pm (\hat \kk)$ are related to
the linear polarization basis tensors $e_{ab}^{+,\times} (\hat \kk)$,
defined in \Eq{basis_pol}, as \cite{Caprini:2003vc,Domcke:2019zls}
\begin{equation}
    e_{ab}^\pm (\hat \kk) = \sqrt{\half} \left(e_{ab}^+ (\hat \kk)
    \pm i e_{ab}^\times (\hat \kk)\right). \nonumber
\end{equation}
}
$i$ indicates each of the interferometers $i=1$, $2$, $3$, and $\Qi (\kk)$ are
the interferometer response functions,\footnote{%
Reference~\cite{Domcke:2019zls} uses the convention $ck=f$, while in
the present work we use $ck=2\pi f$.
Hence, \Eqs{response_Q}{transfer_T}, when compared to their
equivalents in ref.~\cite{Domcke:2019zls},
present a factor of $2\pi$ dividing $k$.}
\begin{align}
    \Qone(\kk)=e^{-i kL \hat \kk \cdot \hat \xx_1}
    \left[{\cal T}(kL, \hat \kk \cdot \hat \UU_1)\, \hat \UU_1^a \hat \UU_1^b -
    {\cal T}(kL,-\hat \kk \cdot \hat \UU_3)\, \hat \UU_3^a \hat \UU_3^b\right],
    \nonumber \\
    \Qtwo(\kk) =e^{-i kL \hat \kk \cdot \hat \xx_2}
    \left[{\cal T}(kL, \hat \kk \cdot \hat \UU_2)\, \hat \UU_2^a \hat \UU_2^b -
    {\cal T}(kL,-\hat \kk \cdot \hat \UU_1)\, \hat \UU_1^a \hat \UU_1^b\right],
    \nonumber \\
    \Qthree(\kk)=e^{-i kL \hat \kk \cdot \hat \xx_3}
    \left[{\cal T}(kL, \hat \kk \cdot \hat \UU_3)\, \hat \UU_3^a \hat \UU_3^b -
    {\cal T}(kL,-\hat \kk \cdot \hat \UU_2)\, \hat \UU_2^a \hat \UU_2^b\right],
    \label{response_Q}
\end{align}
with $\cal T$ being the $i$-th detector transfer function,
\begin{align}
    {\cal T}(kL, \hat \kk \cdot \hat \UU_i) = &\,
    e^{-\half i kL(1+\hat \kk \cdot \hat \UU_i)} \,
    {\rm sinc} \left[\half k L (1 - \hat \kk \cdot \hat \UU_i)\right] 
    \nonumber \\
    + & \, e^{+\half i kL(1-\hat \kk \cdot \hat \UU_i)} \,
    {\rm sinc} \left[\half k L (1 + \hat \kk \cdot \hat \UU_i)\right],
    \label{transfer_T}
\end{align}
where ${\rm sinc}(x)=\sin(x)/x$. 
The vectors $\hat \UU_i$ are the unit vectors following the direction of the 
arms, i.e., the vector pointing from spacecraft $i$ to $i+1$ (modulo 3),
with $i=1$, $2$, $3$. 
We can define a reference frame in which the interferometer is located in the
$xz$-plane,
with the three spacecraft located at $\xx_1=(0,0,0)$, $\xx_2=L(0,0,1)$, and
$\xx_3=\half L(\sqrt{3},0,1)$, chosen for simplicity.
It can be shown that rotations of the plane do not affect the 
response function after integrating over
all directions in the sky \cite{Bartolo:2018qqn}.
Hence, the unit vectors along the arms are
\begin{align}
\hat \UU_1 & \, = \hat \xx_2 - \hat \xx_1 =  (0,0,1), \nonumber \\
\hat \UU_2 & \, =
\hat \xx_3 - \hat \xx_2 =  \half (\sqrt{3},0,-1), \nonumber \\
\hat \UU_3 & \, = \hat \xx_1 - \hat \xx_3 = -\half (\sqrt{3},0,1).
\end{align}
We describe the wave vectors in spherical coordinates,
\begin{equation}
    \kk = k(\cos \phi \sin \theta, \sin \phi \sin \theta, \cos \theta),
    \label{kvec}
\end{equation} 
with $\phi \in [0, 2\pi]$ and $\theta \in [0, \pi]$,
such that the terms $\hat \kk \cdot \hat \xx_i$ that appear in the response functions $\Qi (\kk)$;
see \Eq{response_Q}, are
\begin{equation}
    \hat \kk \cdot \hat \xx_1 = 0, \quad
    \hat \kk \cdot \hat \xx_2 = \cos \theta, \quad
    \hat \kk \cdot \hat \xx_3= \half\left(\sqrt{3} \cos \phi \sin \theta + \cos \theta\right),
\end{equation}
and the products $\hat \kk \cdot \hat \UU_i$ that
appear in the interferometer transfer functions $\cal T$; see \Eq{transfer_T}, are
\begin{align}
    \hat \kk \cdot \hat \UU_1 = & \, \cos \theta, \nonumber \\
    \hat \kk \cdot \hat \UU_2 = & \, \half\left(\sqrt{3}\cos \phi
    \sin \theta - \cos \theta \right), \nonumber \\
    \hat \kk \cdot \hat \UU_3 = & \, - \half \left(\sqrt{3}\cos \phi
    \sin \theta + \cos \theta \right).
\end{align}
The signals of the channels $A$, $E$, and $T$\,\footnote{%
The following results and discussion are applicable to Taiji using
$C$, $D$, and $S$, instead of $A$, $E$, and $T$.}
are obtained by linearly combining
the $X$, $Y$, and $Z$ interferometer channels (or $i=1$, $2$, $3$, in \Eq{td_sigma}),
\begin{equation}
    \Sigma_A=\frac{1}{3}\left(2\sigma_X-\sigma_Y-\sigma_Z\right),
    \quad
    \Sigma_E=\frac{1}{\sqrt{3}}\left(\sigma_Z-\sigma_Y\right),
    \quad
    \Sigma_T=\frac{1}{3}\left(\sigma_X+\sigma_Y+\sigma_Z\right).
    \label{sigmaA}
\end{equation}
Combining \eqref{td_sigma} and \eqref{sigmaA},
we can define $\Qab (\kk)$ as a function of $\Qi (\kk)$,
\begin{equation}
    \Qab (\kk) =\frac{1}{3}\left(\begin{array}{ccc}
         2 & -1 & -1  \\
         0 & - \sqrt{3} & \sqrt{3} \\
         1 & 1 & 1 
    \end{array}\right) \Qi (\kk),
\end{equation}
where $O=A$, $E$, and $T$.

\subsection{Signal and response functions}

\begin{figure}
    \centering
    \includegraphics[width=.49\textwidth]{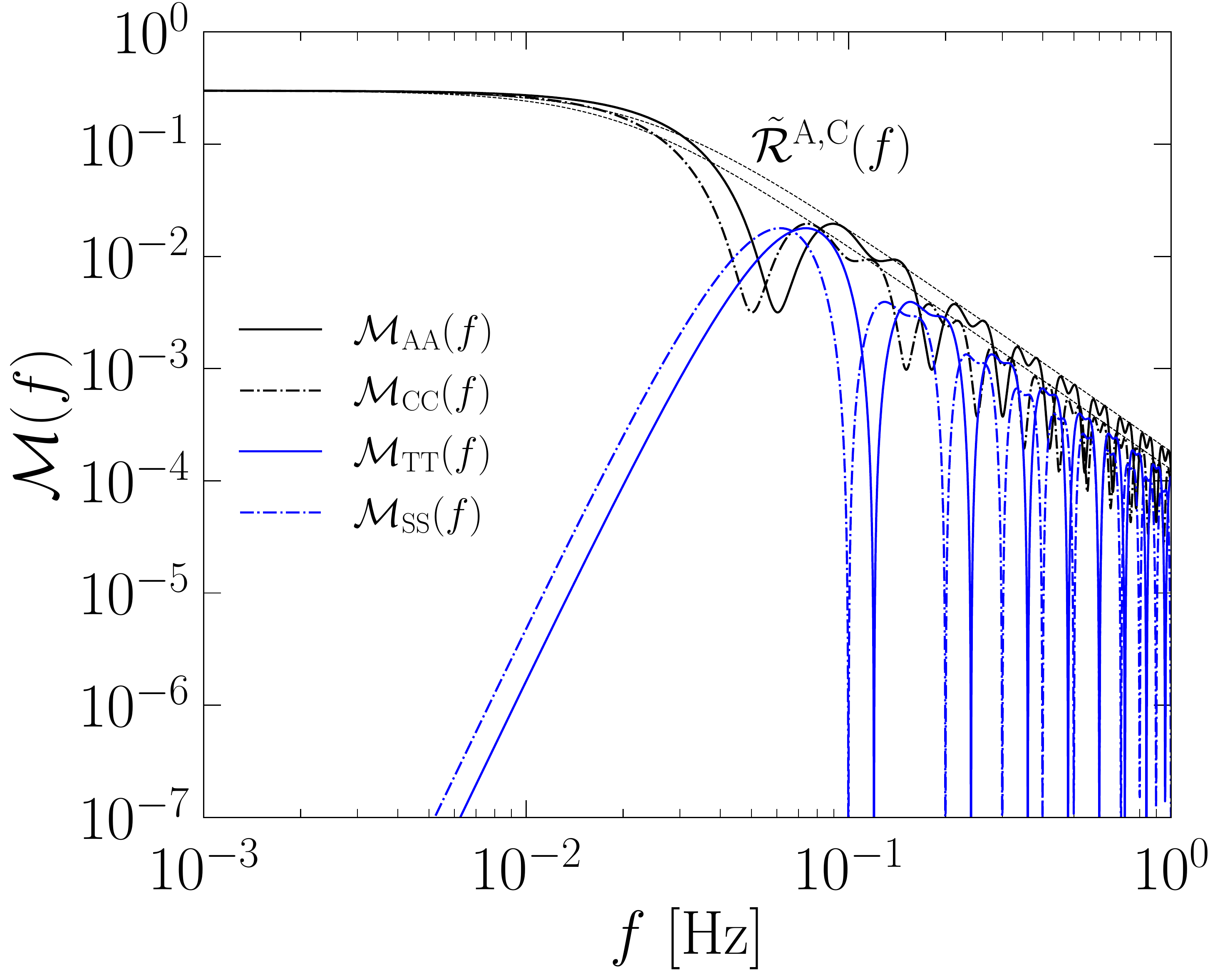}
    \includegraphics[width=.49\textwidth]{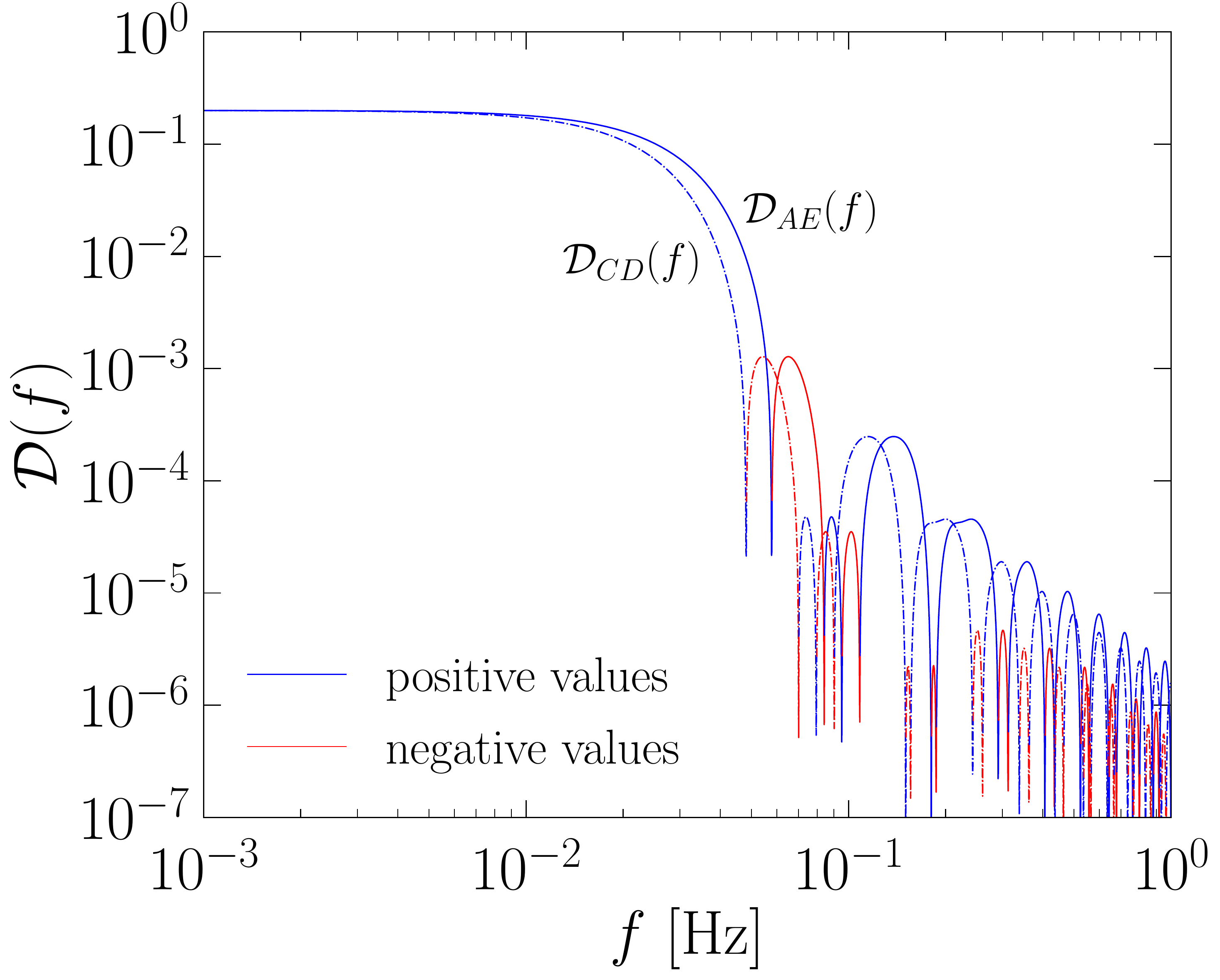}
    \caption{Monopole response functions (left panel) of the LISA $A$ and
    $E$ channels, ${\cal M}_{\rm AA} (f)$,
    the Taiji $C$ and $D$ channels, ${\cal M}_{CC}(f)$,
    and the LISA $T$ and Taiji $S$ null channels,
    ${\cal M}_{TT} (f)$ and ${\cal M}_{SS} (f)$ respectively,
    compared with
    the analytical fit of the LISA and Taiji geometric function $\tilde{\cal R}^{A, C} (f)$.
    Dipole response functions
    (right panel) induced by the peculiar velocity of the solar 
    system in the LISA correlated $A$ and $E$ channels,
    ${\cal D}_{AE}(f)$,
    and the Taiji $C$ and $D$ channels,
     ${\cal D}_{CD}(f)$.}
    \label{LISA_response}
\end{figure}

The two-point correlation function of the signals that a stochastic
GW background produces in the time domain in two of the channels, $O$ and $O'$,
can be expanded as a function of the peculiar
velocity of the solar system, $v/c=1.23 \times 10^{-3}$ \cite{Domcke:2019zls},
\begin{align}
    \bra{\Sigma_O(t) \Sigma_{O'}(t')}= & \, \frac{1}{8}
    \sum_{\lambda=+,-} \int \dd k \Biggl[\Mon(k) S_h^\lambda(k)
    \cos \left[ck(t-t')\right] \nonumber \\
    & + \frac{v}{c} \Don(k, \hat \vv) \left(S_h^\lambda(k)
    - k \frac{\dd S_h^\lambda (k)}{\dd k}\right) \sin \left[ck(t-t') \right] + {\cal O}\left(\frac{v^2}{c^2}\right) \Biggr],
    \label{sig_f}
\end{align}
where the strain spectral functions $S_h^\pm (k)$\,\footnote{%
The autocorrelation function of the signal in \Eq{sig_f} corresponds to
equation~(21) of ref.~\cite{Domcke:2019zls} in terms of
the spectral functions $P_\pm (k)$, defined as; see their equation~(5),
\begin{equation}
\bra{\tilde h_\pm (\kk, t) \tilde h_\pm (\kk', t)} = (2\pi)^6
\delta^3 (\kk - \kk') \frac{P_\pm (k)}{4\pi k^3}. \nonumber
\end{equation}
In the present work we use the spectral functions $S_h^\pm (k)$,
defined as in \Eqs{SPiij}{APiij},
such that we can relate them to each other with $S_h^\pm (k) = 2 P_\pm (k)/k$.
}
are defined using the $+$ and $-$ polarization basis,
such that $S_h(k)=S_h^+(k)+S_h^-(k)$ and $A_h(k)=S_h^+(k)-S_h^-(k)$.
$\Monpm (k)$ and $\Donpm (k, \hat \vv)$ are the monopole and dipole quadratic 
interferometer response functions of the channels  $O$ and $O'$,
\begin{align} 
    \Monpm (k)=&\,4 \int \frac{\dd \Omega_k}{4\pi} e_{ab}^\pm (\hat \kk)
    e_{cd}^\pm (-\hat \kk) \Qab (\kk) \Qcd (-\kk), \\ 
    \Donpm (k, \hat \vv)=&\,4i \int \frac{\dd \Omega_k}{4\pi}
    e_{ab}^\pm (\hat \kk) e_{cd}^\pm
    (-\hat \kk) \Qab (\kk) \Qcd (-\kk) \, \hat \kk \cdot \hat \vv,
    \label{Mon_Don}
\end{align}
where $\hat \vv$ is the unit direction of the peculiar velocity,
and the wave number is $ck=2\pi f$ due to the GW dispersion relation, 
such that we can express the response functions in terms of $f$.
The dipole response function appears due to the anisotropies induced by the
proper motion of the solar system.
It depends on the angle $\alpha$ between the orientation of the detector plane
and the peculiar velocity of the solar system.
Expressing $\kk$ in spherical coordinates, see \Eq{kvec}, we can
perform the integral over the directions on the sky,
\begin{equation} 
    \Monpm (k)= \frac{1}{\pi} \int_0^{2\pi} \dd \phi \int_0^\pi
    e_{ab}^\pm (\hat \kk)
    e_{cd}^\pm (-\hat \kk) \Qab (\kk)
    {\cal Q}_{O'}^{cd,*} (\kk) \sin \theta \, \dd \theta,
    \label{Mon}
\end{equation}
where we have used $\Qcd (-\kk)={\cal Q}_{O'}^{cd,*} (\kk)$
\cite{Bartolo:2018qqn}.
The product 
$e_{ab}^\pm (\hat \kk) e_{cd}^\pm (-\hat \kk)$ can be written as
\cite{Domcke:2019zls}
\begin{equation}
    e_{ab}^\pm  (\hat \kk) e_{cd}^\pm  (-\hat \kk) = \frac{1}{4}
    \left(\delta_{ac} - \hat k_a \hat k_c \mp i \varepsilon_{ace}
    \hat k^e\right)\left(\delta_{bd} - \hat k_b \hat k_d
    \mp i \varepsilon_{bde} \hat k^e\right),
\end{equation}
where $\delta_{ab}$ is the Kronecker delta and
$\varepsilon_{abc}$ is the Levi-Civita tensor.
It can be shown that
${\cal M}_{OO'}^+(f)={\cal M}_{OO'}^-(f)$, 
so the monopole responses to the $+$ and $-$ modes are the same,
and hence, polarization of the GW signals cannot be detected unless we consider
the dipole response or combine more than one GW detector to break the coplanarity,
as we do in \App{App_LISA_Taiji}.
The responses of the LISA channels $A$ and $E$
are the same,
i.e., 
${\cal M}_{EE}(f)={\cal M}_{AA}(f)$,
and ${\cal M}_{TT}(f)$
is much smaller in the low frequency regime, being insensitive to 
gravitational wave signals.
For this reason, the $T$ channel is known as the ``Sagnac'' or null 
channel and it is used to identify and subtract noisy signals 
\cite{Adams:2010vc}.
The response of the combined $A$ and $E$ channels is
${\cal M}_{AE} (f)=0$ \cite{Domcke:2019zls}.
Note that ${\cal M}_{AA} (f)$ corresponds to the LISA geometrical factor 
$\tilde {\cal R}^A (f)$, commonly defined
in the LISA community; see, e.g., refs.~%
\cite{Cornish:2018dyw,Caprini:2019pxz,Flauger:2020qyi,Orlando:2020oko}, 
which can be well-fit for LISA and Taiji
by
\begin{equation}
    \tilde {\cal R} (f)=
    \frac{3}{10} \left[1 + 0.6\left(2\pi fL/ c\right)^2\right]^{-1}.
\end{equation}
The monopole and dipole response functions of LISA and Taiji are shown in 
\Fig{LISA_response}.
The dipole response function can be expressed as a function of the
angle $\alpha$ between the normal of the detector and the
velocity of the proper motion $\hat \vv$.
Since we integrate over all directions of $\hat \kk$ in the sky,
we can define the vector $\vv$ in the frame of reference in
which we have defined the detector (in the $xz$-plane), such that the normal
$\hat \nn$ is in the $y$-direction, and due to the symmetry of
the integration over $\hat \kk$, we can neglect the projection of
$\vv$ on the detector plane, and write
$\vv = v(0, \cos \alpha, 0)$, or
$\hat \kk \cdot \hat \vv=\cos \alpha \sin \phi
\sin \theta$.
This allows us to compute $\Donpm (k, \alpha)$ as a function of
the angle $\alpha$ \cite{Domcke:2019zls},
\begin{equation}
    \Donpm (k, \alpha) =\frac{i}{\pi} \cos \alpha \int_0^{2\pi}
    \sin \phi \, \dd \phi \int_0^\pi e_{ab}^\pm (\hat \kk)
    e_{cd}^\pm (-\hat \kk) \Qab (\kk) {\cal Q}_{O'}^{cd,*} (\kk)
    \sin^2 \theta \, \dd \theta.
    \label{Don}
\end{equation}
It can be shown that ${\cal D}_{OO'}^+(k, \alpha)=-{\cal D}_{OO'}^-(k, \alpha)$
for $O$, $O'=A$, $E$, $T$, and ${\cal D}_{AA}^\pm(k, \alpha)=
{\cal D}_{EE}^\pm(k, \alpha)=0$
\cite{Domcke:2019zls}.
Hence, the relevant contributions from the dipole response function are 
${\cal D}_{AE}^+(k,\alpha)={\cal D}_{EA}^+(k,\alpha)=
-{\cal D}_{AE}^-(k, \alpha)=-{\cal D}_{EA}^-(k, \alpha)$,
and we can write ${\cal D}_{AE}^\pm(k,\alpha)=\pm {\cal D}_{AE} (k) \cos \alpha$.

At the present time $t_0$, the observed GW energy density,
defined in \Eq{EEGW}, by the detector is \cite{Maggiore:1999vm}
\begin{align}
    \OmGW (t_0) = &\,
    \frac{1}{12 H_0^2} \bra{\dot{h}_{ij}^{\rm phys} (\xx, t)
    \dot{h}_{ij}^{\rm phys} (\xx, t)} =
    \frac{\pi^2 f^2}{3 H_0^2} \bra{h_{ij}^{\rm phys} (\xx, t)
    h_{ij}^{\rm phys} (\xx, t)} \nonumber \\
    = &\, \frac{2 \pi^2 f^2}{3 H_0^2} \bra{h_+^2 (\xx, t) +
    h_-^2 (\xx, t)},
\end{align}
where $H_0 = 100\, h_0 \km \s^{-1} \Mpc^{-1}$ and $h_0$ takes into account
the uncertainties of the Hubble rate at the present time \cite{Maggiore:1999vm}.
We have used the relation $\bra{\dot{h}_{ij}^{\rm phys}
\dot{h}_{ij}^{\rm phys}} = c^2 k^2 \bra{h_{ij}^{\rm phys}
    h_{ij}^{\rm phys}}$ and the GW dispersion relation $ck = 2 \pi f$.
The resulting GW spectrum $\OmGW(f)=\OmGW^+ (f) + \OmGW^- (f)$ is
\begin{equation}
    \OmGW^\pm (f) = \frac{2 \pi^2 f^3}{3 H_0^2} S_h^\pm (f),
    \label{Sh_OmGW}
\end{equation}
analogous to \Eq{OmGW_k}, defined such that $\OmGW(t_0) =
\int \OmGW(f)\, \dd \ln f$.

The time delays induced by the GW background in the interferometers
lead to the signal function $S_{OO'} (f)$ of the channels
$O$ and $O'$,
which corresponds to the Fourier transform of the two-point
correlation function of the signals; see \Eq{sig_f}, obtained by transforming $t\rightarrow f$ and
$t'\rightarrow f'$, and then setting $f=f'$. 
The signal function, expressed in terms of the GW energy density
polarization spectra $\OmGW^\pm (f)$;
see \Eq{Sh_OmGW}, is \cite{Domcke:2019zls,Ellis:2020uid}
\begin{align}
     S_{OO'} (f) = & \, \frac{3H_0^2}{8\pi^2 f^3}
    \sum_{\lambda=+,-} \Biggl[ \Mon(f) \,
    \OmGW^\lambda (f) \nonumber \\
    &  - 4i \frac{v}{c} \Don(f, \alpha) \left(\OmGW^\lambda (f)
    - \frac{f}{4} \frac{\dd \OmGW^\lambda (f)}{\dd f}
    \right) + {\cal O} \left(\frac{v^2}{c^2}\right) \Biggr],
    \label{signal_f}
\end{align}
where the $\Monpm (f)$ and $\Donpm (f, \alpha)$ are the monopole and the dipole 
response functions in frequency space, obtained from \Eqs{Mon}{Don},
using the dispersion relation $ck = 2\pi f$.
The signal functions of the LISA $A$ and $E$ channels are
\begin{equation}
    S_{AA} (f) = S_{EE} (f) = \frac{3H_0^2}{8\pi^2
    f^3} {\cal M}_{AA} (f)\,
    \OmGW(f),
    \label{SAA}
\end{equation}
since ${\cal D}_{AA}^\pm (f, \alpha)={\cal D}_{EE}^\pm (f, \alpha)=0$ and
${\cal M}_{AA}^\pm (f)={\cal M}_{EE}^\pm (f) = {\cal M}_{AA} (f)$
\cite{Domcke:2019zls}; see \Fig{LISA_response}.
The signal functions obtained correlating the $A$ and $E$ channels are
\begin{equation}
    S_{AE} (f) = S_{EA} (f) = - 4i \frac{3H_0^2}{8 \pi^2 f^3}
    \frac{v}{c} {\cal D}_{AE} (f) \cos \alpha 
    \left(\XiGW(f) - \frac{f}{4} \frac{\dd \XiGW(f)}{\dd f}\right),
    \label{SAE}
\end{equation}
where we have used the properties ${\cal M}_{AE}^\pm (f)={\cal M}_{EA}^\pm (f)=0$
and ${\cal D}_{AE}^+(f, \alpha)=-{\cal D}_{AE}^-(f, \alpha)={\cal D}_{AE} (f) \cos \alpha$
\cite{Domcke:2019zls};
see \Fig{LISA_response}.
The helical GW spectrum is $\XiGW(f)=\OmGW^+(f)-\OmGW^-(f)$.

\subsection{Signal-to-noise ratio and power law sensitivity}
\label{interferometry_SignalToNoise}

The detectability of a GW signal is studied in terms of its signal-to-noise ratio (SNR).
The SNR of the detector to a GW background
combining two channels $O$ and $O'$ is
\cite{Bartolo:2018qqn,Domcke:2019zls,Ellis:2020uid}
\begin{equation}
    {\rm SNR}_{OO'} =\sqrt{\int_0^T \dd t \int_{-\infty}^\infty
    \dd f \frac{S_{OO'}^* (f) S_{OO'} (f)}
    {P^O_n(f) P^{O'}_n (f)}}=
    \sqrt{2 \int_0^T \dd t \int_{0}^\infty
    \dd f \frac{S_{OO'}^* (f) S_{OO'} (f)}
    {P^O_n(f) P^{O'}_n (f)}},
    \label{SNROO}
\end{equation}
where $T$ is the duration of the observation and $P_n^O (f)$ is the noise power 
spectral density (PSD) of the channel $O$. 
The noise PSD $P_n (f)$ of the LISA interferometer
channels $X$, $Y$, and $Z$
is based on the results from LISA Pathfinder;
see ref.~\cite{LISA_docs}, and
refs.~\cite{Cornish:2018dyw,Caprini:2019pxz} for its explicit derivation,
\begin{equation}
    P_n (f) = P_{\rm oms}(f) + \left[3 + \cos\left(\frac{4\pi fL}{c}\right)\right] P_{\rm acc} (f),
\end{equation}
where $P_{\rm oms} (f)$ is the optical metrology system noise and
$P_{\rm acc} (f)$ is the mass acceleration noise,
\begin{align}
    P_{\rm oms} (f) = P^2 & \left[\frac{\picom}{L}\right]^2 \Hz^{-1}
    \left[1 + \left(\frac{2 \mHz}{f}\right)^4\right], \\
    P_{\rm acc}(f) =  A^2 & \left[\frac{\fm}{L}\right]^2 \left[\frac{(L/c)}{\s} \right]^4
    \Hz^{-1}
    \left[1 + \left(\frac{0.4
    \mHz}{f}\right)^2\right] \nonumber \\
    \times &  \left[1 + \left(\frac{f}{8 \mHz}\right)^4
    \right] \left(\frac{c}{2\pi f L} \right)^4,
    \label{Pacc_Poms}
\end{align}
with $P=15$ and $A=3$ being the LISA noise parameters \cite{LISA_docs}.
For Taiji, these parameters are $P=8$ and $A=3$
\cite{Ruan:2020smc}.
The characteristic frequency of LISA is $f_0=c/(2 \pi L)=0.019 \Hz$ and Taiji's is $f_0=0.016 \Hz$.
The function $P_n(f)$ corresponds to the noise auto-correlation of the 
channels $X$, $Y$, and $Z$,
and the noise cross-correlation spectra of two different channels
$XY$, $XZ$, and $YZ$,
are \cite{Flauger:2020qyi}
\begin{equation}
    P_n^{\rm cross} (f) = -\frac{1}{2} \cos\left(\frac{2\pi fL}{c}\right)
    \left[4 P_{\rm acc} (f) + P_{\rm oms} (f)\right].
\end{equation}
Using the noise correlations of the interferometer channels,
we can compute the noise PSD of the LISA $A$, $E$, and $T$ channels,
\begin{equation}
    P_n^A(f) = P_n^E (f) = \twothird \left[ P_n (f) - P_n^{\rm cross} (f)\right],
    \quad 
    P_n^T(f) = \onethird \left[P_n (f) + 2P_n^{\rm cross} (f)\right].
    \label{PnA}
\end{equation}

\begin{figure}
    \centering
    \includegraphics[width=.49\textwidth]{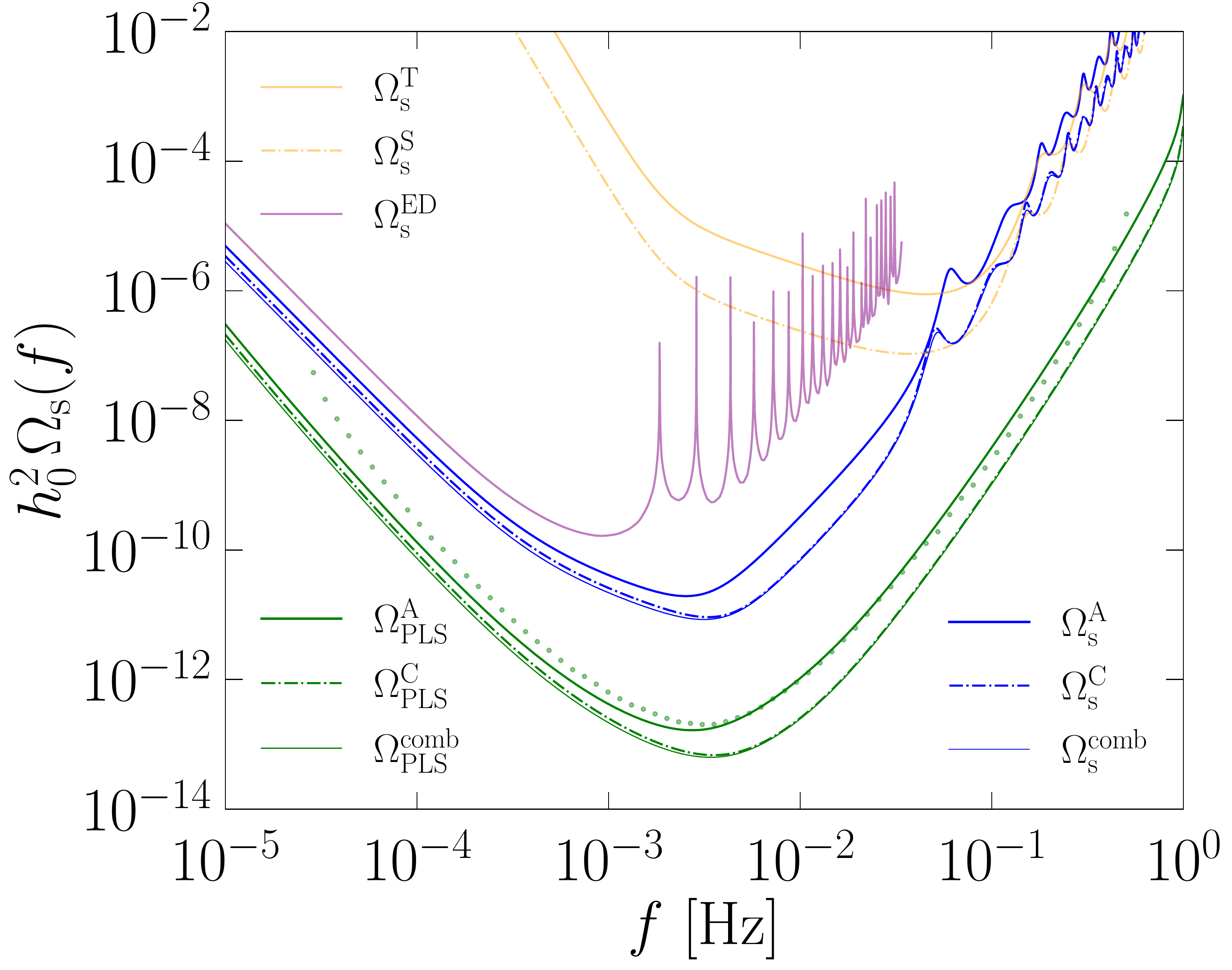}
    \includegraphics[width=.49\textwidth]{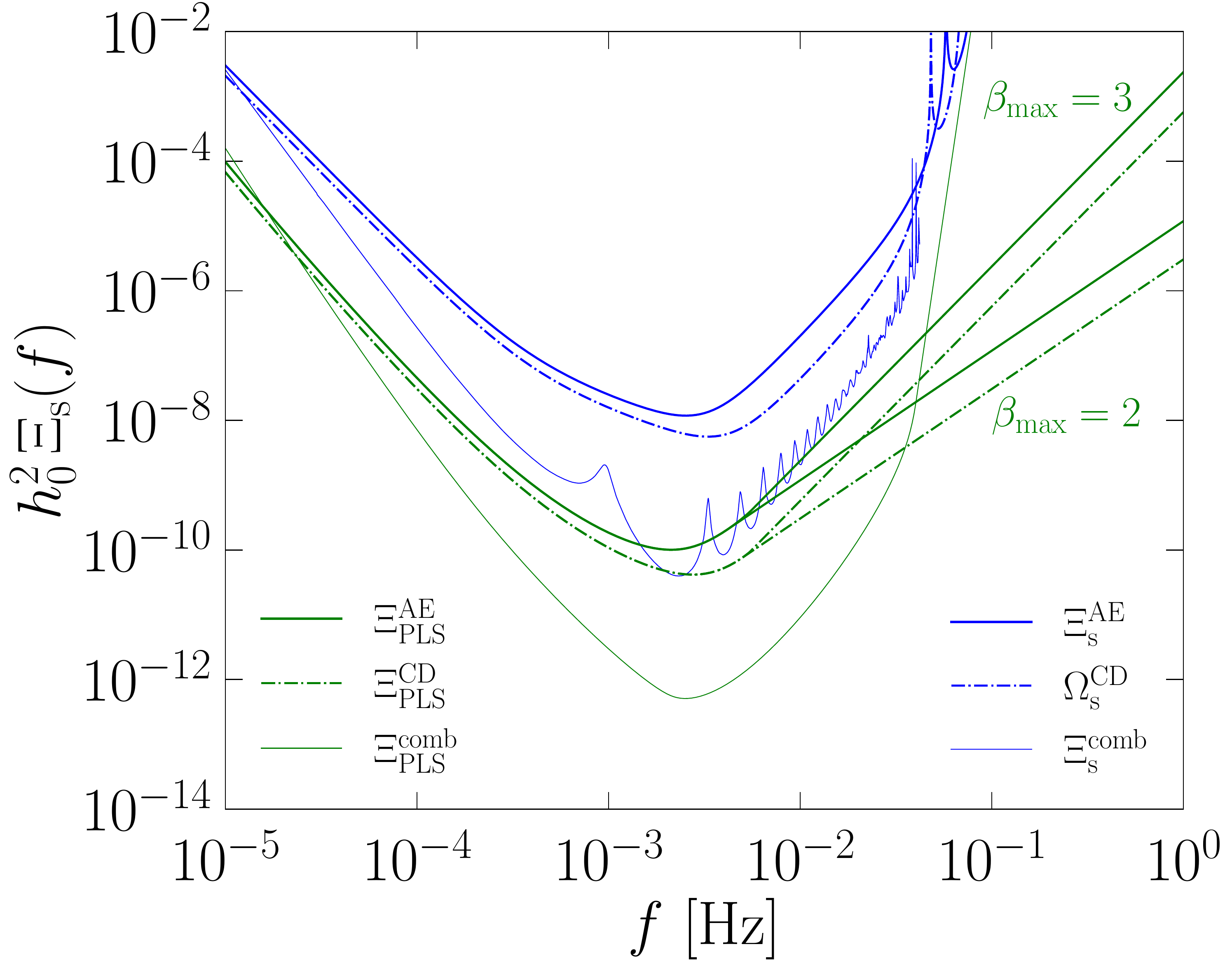}
    \caption{Sensitivities and PLS to GW energy density $\Omega_{\rm s} (f)$
    (left panel) and helicity $\Xi_{\rm s} (f)$ (right panel) of LISA (`A'), Taiji (`C'),
    and the combined LISA--Taiji network (`comb');
    see \App{App_LISA_Taiji}.
    The PLS are computed for a
    $\SNR=10$ and $T=4$ years of duration of the mission.
    The polarization PLS are
    computed limiting the maximum slope
    of the power law $\XiGW (f) = C_\beta^{\rm pol} f^\beta$ to $\beta_{\rm max}=2$ and $3$;
    see \Fig{Xi_PLS} for larger values of $\beta_{\rm max}$.
    The sensitivities of the LISA $T$ and Taiji $S$ null channels
    are much larger at low frequencies than the other channels,
    so they are insensitive
    to GW signals (left panel).
    The cross-correlations between LISA and Taiji channels yield a monopole sensitivity to
    a polarized GW background; see $\Xi_{\rm s}^{\rm comb} (f)$,
    which is smaller than the dipole sensitivity induced in LISA,
    $\Xi_{\rm s}^{AE} (f)$, and Taiji, $\Xi_{\rm s}^{CD} (f)$ (right panel),
    while their sensitivity to the GW energy density,
    e.g., $\Omega_{\rm s}^{ED} (f)$,
    is much larger compared to those of the self-correlations of LISA,
    $\Omega_{\rm s}^{A} (f)$, and Taiji, $\Omega_{\rm s}^{C} (f)$
    (left panel).
    The combination of the last two,
    $\Omega_{\rm s}^{\rm comb} (f)$,
    does not enhance significantly the detectability
    of a single detector.
    We show in green dots the PLS reported in ref.~\cite{Caprini:2019pxz} (left panel); see their figure~(2), with a SNR of 10 and
    $T=4 \yr$.
    Note that ref.~\cite{Caprini:2019pxz} uses the $X$ LISA channel instead of $A$.}
    \label{GW_sensitivity}
\end{figure}

We now define the sensitivities to the GW energy density signal, $\Omega_{\rm s}^A (f)$,
and to the helical GW signal, $\Xi_{\rm s}^{AE} (f)$, shown in \Fig{GW_sensitivity}, as
\begin{align}
    \Omega_{\rm s}^A (f) =  &\, \frac{8 \pi^2}{3 H_0^2} f^3 \frac{P_n^A(f)}
    {{\cal M}_{AA} (f)}, \label{OmsA} \\
    \Xi_{\rm s}^{AE} (f) = &\, \frac{4 \pi^2}{3 H_0^2} f^3
    \frac{\sqrt{P_n^A(f) P_n^E(f)}} {(v/c) |{\cal D}_{AE} (f)|} =
    \frac{4 \pi^2}{3 H_0^2} f^3 \frac{P_n^A(f)}{(v/c) |{\cal D}_{AE} (f)|},
    \label{XisAE}
\end{align}
such that the SNR to a stochastic GW background, defined in \Eq{SNROO}, is
\begin{equation}
    \SNR=\sqrt{\SNR_{AA}^2 + \SNR_{EE}^2}=
    2\sqrt{T}\left[\int_0^\infty \dd f \left(
    \frac{\OmGW(f)}{\Omega_{\rm s}^A (f)}\right)^2 \right]^{1/2}.
    \label{SNR_AA}
\end{equation}
Following ref.~\cite{Caprini:2019pxz}, we compute the power
law sensitivity (PLS) that corresponds to the power law
GW spectrum that leads to a specific SNR, taken to be 10
in the present work, as suggested in ref.~\cite{Caprini:2019pxz}.
First, we take $\OmGW(f) = C_\beta f^\beta$ and compute,
for a large range of $\beta$ (e.g., $-20$ to $20$), the value 
of $C_\beta$ that yields a SNR of 10 for a duration of the
mission of 4 years (the nominal duration of the
LISA mission) \cite{Audley:2017drz},
\begin{equation}
    C_\beta=\frac{\SNR}{2\sqrt{T}} \left[ \int \dd f 
    \frac{f^{2\beta}}{[\Omega_{\rm s}^A(f)]^2}\right]^{-1/2}.
    \label{C_beta_OmSA}
\end{equation}
Finally, we construct the PLS curve by taking at each frequency
the maximum value of the functions $C_\beta f^\beta$.
Note that the previous discussion also applies to Taiji
using $C$ and $D$ instead of $A$ and $E$
in \Eqss{OmsA}{C_beta_OmSA}.
The resulting PLS of LISA and Taiji are shown in
\Fig{GW_sensitivity}.

\begin{figure}
    \centering
    \includegraphics[width=.49\textwidth]{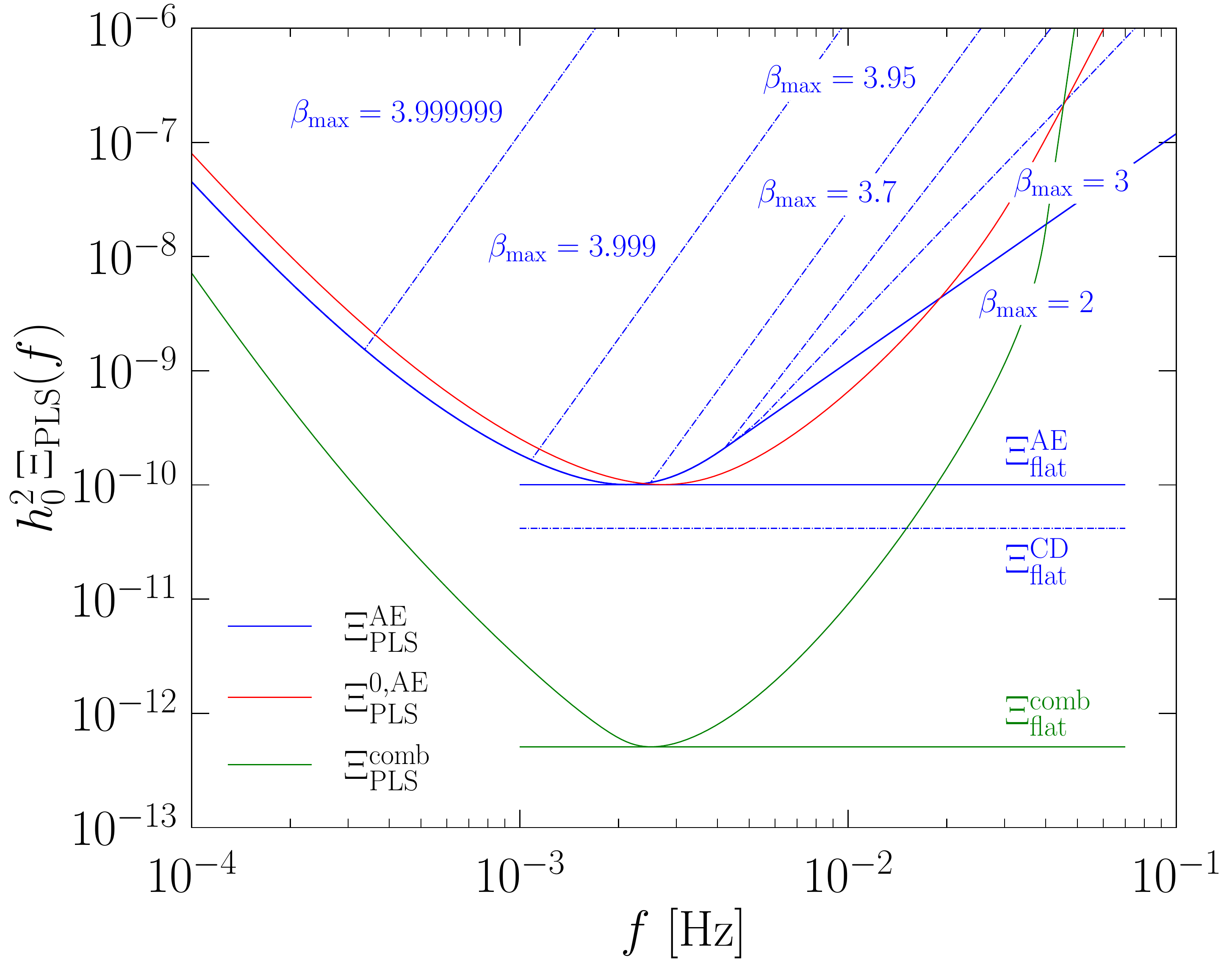}
    \includegraphics[width=.49\textwidth]{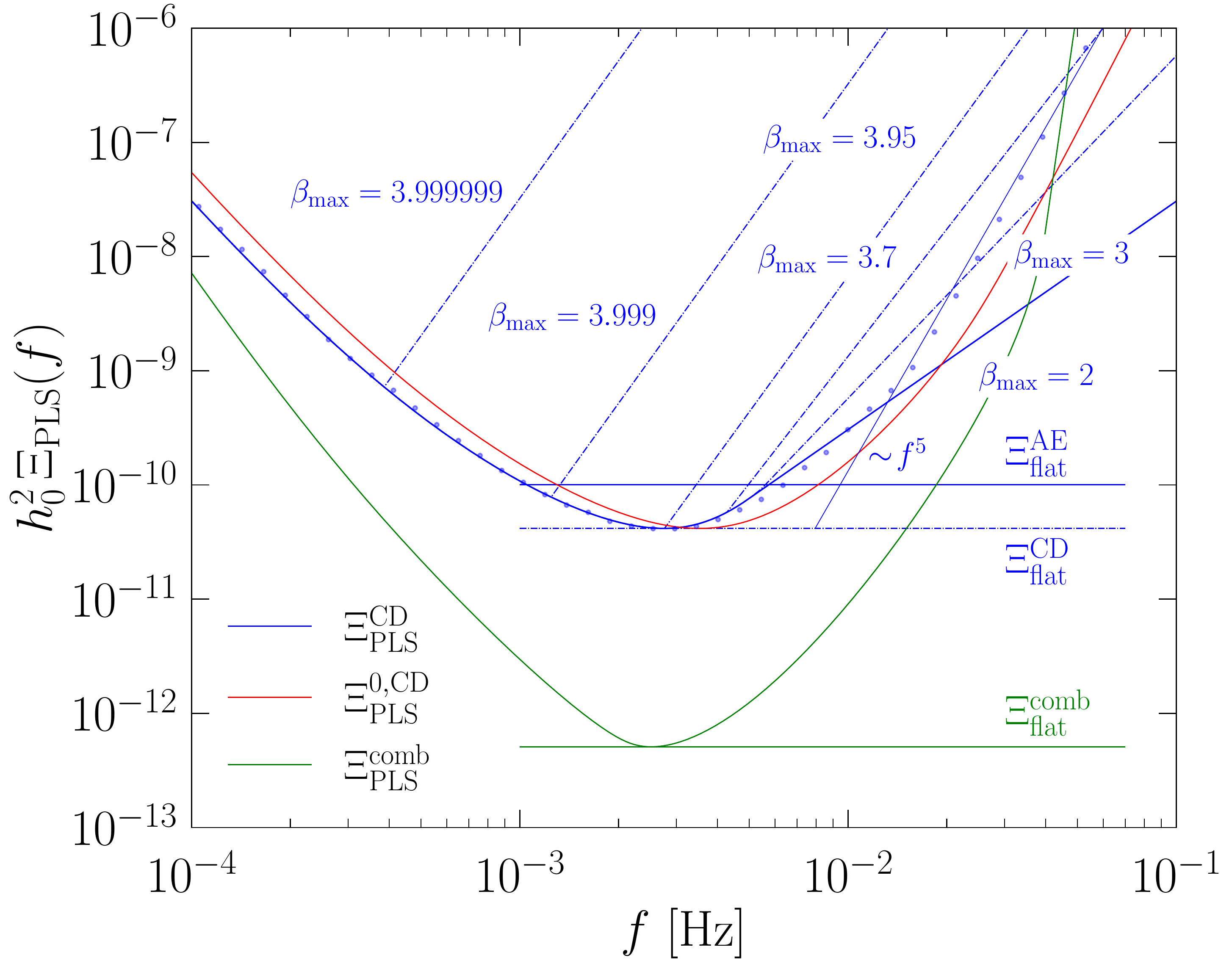}
    \caption{PLS to helical GW signals
    of LISA, $\Xi^{\rm AE}_{\rm PLS} (f)$ (left panel), and Taiji, $\Xi^{\rm CD}_{\rm PLS} (f)$ (right panel),
    computed for a $\SNR_{\rm pol}=10$ and $T=4$ years of duration of the mission, assuming power law slopes up to
    $\beta_{\rm max} = 2$, $3$, $3.7$, $3.95$,
    $3.999\equiv4 - 10^{-3}$, and $3.999 999\equiv4 - 10^{-6}$.
    The $\SNR_{\rm pol}$
    is identically 0 when $\beta=4$, such that in the limit
    $\beta \rightarrow 4$, the helical signal cannot be detected; see \Eq{SNR_pol_app}.
    The PLS that would be obtained ignoring the
    $\dd \XiGW (f)/\dd \ln f$ term, $\Xi_{\rm PLS}^{0} (f)$,
    and the PLS obtained by combining the cross-correlated
    channels of the LISA--Taiji network, $\Xi_{\rm PLS}^{\rm comb} (f)$; see \App{App_LISA_Taiji},
    are shown for comparison.
    The horizontal lines correspond to the
    flat spectra yielding a $\SNR_{\rm pol} = 10$: $h_0^2\, \Xi^{AE}_{\rm flat} = 10^{-10}$,
    $h_0^2\, \Xi^{CD}_{\rm flat} = 4.16 \times 10^{-11}$,
    and $h_0^2\, \Xi^{\rm comb}_{\rm flat} = 5.1 \times 10^{-13}$; see
    \Eqss{XiGW_LISA_dip_flat}{XiGW_LISA_Taiji_flat}.
    The helical PLS computed in ref.~\cite{Ellis:2020uid}; see
    their figure (7), is shown in blue dots (right panel),
    compared to the PLS
    obtained considering $\beta \in (-20, 2) \cup (5, 20)$,
    which shows a change of slope from $\beta = 2$ to $\beta = 5$ around $10^{-2} \Hz$.}
    \label{Xi_PLS}
\end{figure}

In the case of parity odd signals, observed via 
the induced dipole response function due to our proper motion
in the LISA $A$ and $E$
channels, the polarization
SNR of a stochastic GW
background with helical spectrum $\XiGW(f)$ is computed by integrating
$S_{AE} (f)$ in time and frequency; see \Eq{SAE},
\begin{align}
    &\,\SNR_{\rm pol}=\sqrt{\SNR_{AE}^2 + \SNR_{EA}^2}=
    \sqrt{2}\, \SNR_{AE} =\nonumber \\ &\, 4 \sqrt{T_{1 \yr}} \left[
    \int_0^{T/(1 \yr)} \cos^2 \alpha (x) \, \dd x 
    \int_0^\infty \dd f \left(\frac{\XiGW(f) - \quarter f
    \dd \XiGW(f)/\dd \ln f} {\Xi_{\rm s}^{AE} (f)}\right)^2\right]^{1/2}.
    \label{SNR_pol}
\end{align}
The integral in time of $\cos^2 \alpha (t)$ has been computed in ref.~\cite{Domcke:2019zls} assuming a circular orbit of LISA,
\begin{equation}
    \int_0^1 \cos^2 \alpha (x)\, \dd x = \frac{5 + \cos 2 (\theta_v)}{16}
    \in (0.25, 0.375),
\end{equation}
where $\theta_v$ is the angle of the peculiar velocity.
We take the minimum value of the integral, i.e., 0.25.
Introducing this value into \Eq{SNR_pol}, the resulting SNR is
\begin{equation}
    \SNR_{\rm pol}=2 \sqrt{T} \left[\int_0^\infty \dd f \left(
    \frac{\XiGW(f)-\quarter f \dd \XiGW(f)/\dd \ln f}
    {\Xi_{\rm s}^{AE} (f)}\right)^2 \right]^{1/2}.
    \label{SNR_pol_app}
\end{equation}
To obtain the PLS of a polarized GW signal, as before, we consider a power law
helical spectrum $\XiGW (f) = C_\beta^{\rm pol} f^\beta$.
We find that a special case is $\beta = 4$, for which the polarization SNR
is identically zero, and hence, such a signal is not detectable, independently
of its amplitude.
For values $\beta \neq 4$, the constant $C_\beta^{\rm pol}$ is
\begin{equation}
    C_\beta^{\rm pol}=\frac{\SNR_{\rm pol}}{2 \sqrt{T}|1-\beta/4|}
    \left[ \int \dd f \frac{f^{2\beta}}
    {[\Xi_{\rm s}^{AE}(f)]^2}\right]^{-1/2}.
    \label{Cbeta}
\end{equation}
Thus, we construct the PLS by taking the largest value of the function
$C_\beta^{\rm pol} f^\beta$ in the range of $\beta$.
Note that the same result applies to Taiji using $\Xi_{\rm s}^{CD} (f)$ in \Eqss{SNR_pol}{Cbeta}.
The resulting polarization PLS curves of LISA and Taiji are shown
in \Fig{Xi_PLS}.
As the value of $\beta$ gets close to $4$, the denominator in
\Eq{Cbeta} becomes larger, yielding large amplitudes of the PLS
at a cutoff frequency that becomes smaller as we get closer to $4$.
We compute the PLS curves for $\beta \in (-20, \beta_{\rm max})$
with $\beta_{\rm max} \in [2, 4 - 10^{-6}]$
and observe a change of slope toward $\beta_{\rm max}$
at the cutoff frequency.
For slopes $\beta \leq 3$,
the cutoff occurs at a frequency larger than the sensitivity
peak, such that the potential detectability is barely
affected by the cutoff.
However, if the helical GW signal has a slope between $3$ and $3.95$, the $\SNR_{\rm pol}$
is close to zero and the detectability 
at slightly larger frequencies than
the sensitivity peak is more challenging; see \Fig{Xi_PLS}.
At even larger slopes, between 3.95 and 4,
the cutoff frequency is below the peak sensitivity
and the $\SNR_{\rm pol}$ goes asymptotically to zero.
An analogous behavior is obtained when computing the PLS to GW signals
with slopes larger than $4$, e.g., by taking a range $\beta \in (-20, 3) \cup
(\beta_{\rm min}, 20)$ with $\beta_{\rm min} > 4$.
For slopes $\beta_{\rm min}\geq 4.2$, the change of slope occurs at frequencies
larger than $5 \times 10^{-3}$ (same as for $\beta_{\rm max}=3$).
Hence, for slopes $\beta \geq 4.2$, the potential
detectability is again unaffected by the cutoff.
The resulting PLS taking $\beta_{\rm min} = 5$ is shown in the right panel of \Fig{Xi_PLS},
compared to the PLS of ref.~\cite{Ellis:2020uid};
see their figure~7.

In the present work, we used the resulting helical PLS of LISA,
$\Xi_{\rm PLS}^{AE} (f)$, and Taiji,
$\Xi_{\rm PLS}^{CD} (f)$, obtained by taking power law spectra
with slopes in the range $\beta \in (-20, 3) \cup (4.2, 20)$,
which is valid for helical GW signals with slopes that are not
between 3 and 4.2.
The resulting PLS of LISA and Taiji are used in \Figs{XiGW_detectors}{XiGW_detectors_GWCirc}
to study the potential detectability of the polarized GW signals produced
by primordial magnetic fields, computed from numerical simulations
of MHD turbulence.

\subsection{LISA--Taiji network}
\label{App_LISA_Taiji}

We now consider the possible combination of a network
of space-based GW detectors, e.g., LISA and Taiji,
following ref.~\cite{Orlando:2020oko}.
In first place, the total SNR obtained by combining the 
self-correlations of the LISA and Taiji channels
(i.e., the correlations between two channels of the same detector)
is
\begin{align}
    \SNR= & \, \sqrt{\SNR_{AA}^2+\SNR_{EE}^2+\SNR_{CC}^2+\SNR_{DD}^2}
    =\sqrt{2(\SNR^2_{AA} + \SNR_{CC}^2)} \nonumber \\
   = & \, 
    2 \sqrt{T}\left[\int \dd f \, \OmGW^2(f) \left(\frac{1}
    {[\Omega_{\rm s}^A (f)]^2} + \frac{1}
    {[\Omega_{\rm s}^C (f)]^2} \right)\right]^{1/2} \nonumber \\
    = & \, 2 \sqrt{T}\left[\int \dd f \left(\frac{\OmGW(f)}
    {\Omega_{\rm s}^{\rm comb} (f)}\right)^2 \right]^{1/2},
\end{align}
where we have defined the GW sensitivity of the combined LISA--Taiji
network $\Omega_{\rm s}^{\rm comb} (f)$ (shown in \Fig{GW_sensitivity}),
\begin{equation}
    \Omega_{\rm s}^\text{\rm comb} (f)=\left(\frac{1}
    {[\Omega_{\rm s}^A (f)]^2} + \frac{1}
    {[\Omega_{\rm s}^C (f)]^2} \right)^{-1/2}=
    \frac{\Omega_{\rm s}^A (f)\, \Omega_{\rm s}^C (f)}
    {\sqrt{[\Omega_{\rm s}^A (f)]^2+
    [\Omega_{\rm s}^C (f)]^2}}.
    \label{Omega_s_LISATAiji}
\end{equation}
We again construct the PLS of the combined LISA--Taiji network
$\Xi_{\rm PLS}^{\rm comb} (f)$;
see \Fig{GW_sensitivity}, and
see that the improvement is very small compared to the single
detector (i.e., Taiji) PLS.

\begin{figure}
    \centering
    \includegraphics[width=.75\textwidth]{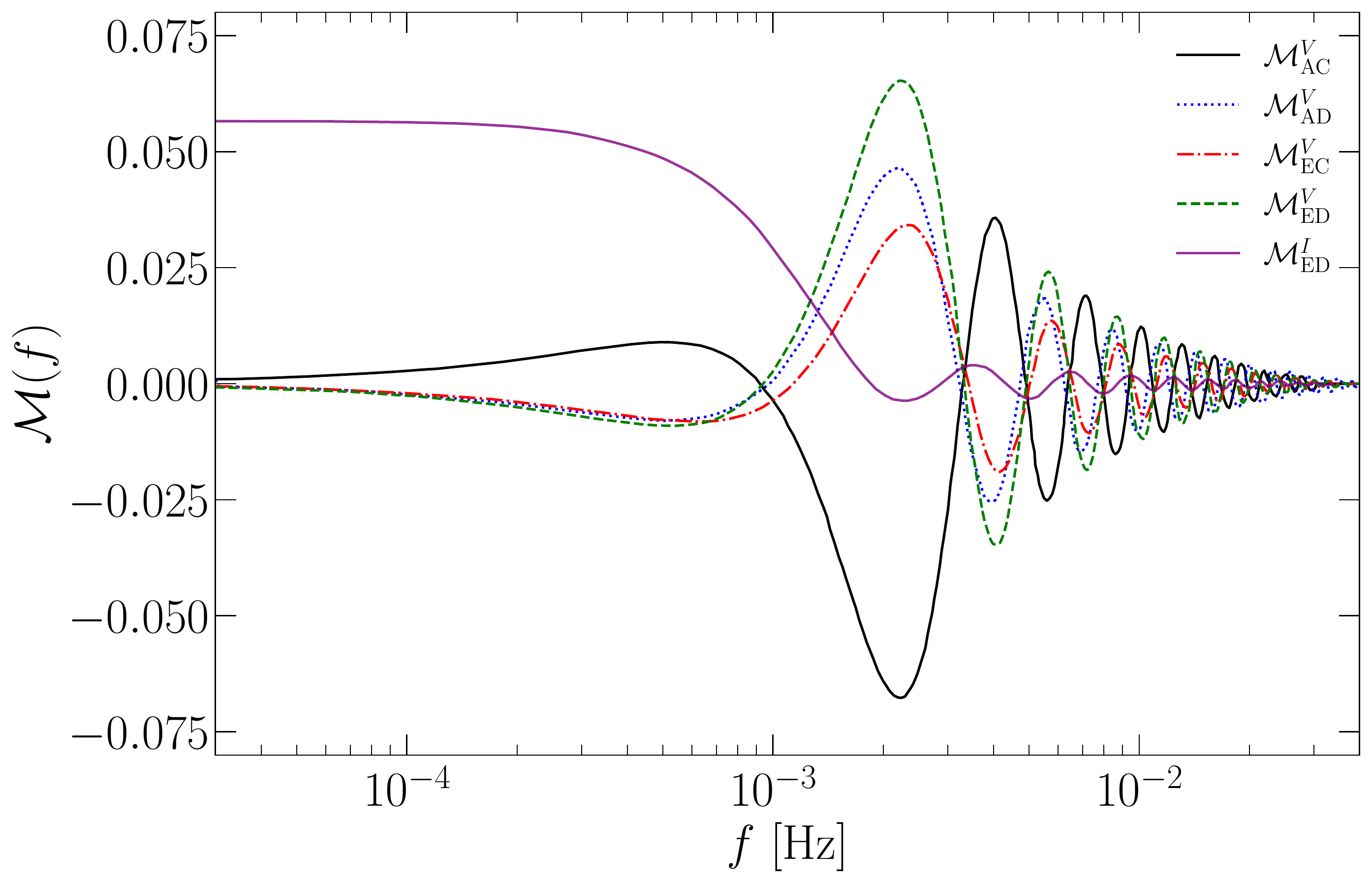}
    \caption{Helical response functions ${\cal M}^V_{OO'}(f)$ of cross-correlated channels of LISA and Taiji, with
    $OO'=AE$, $AD$, $EC$, and $ED$, and symmetric response function
    ${\cal M}^I_{ED}(f)$ \cite{Orlando:2020oko}, shown
    for comparison.}
    \label{R_cross}
\end{figure}

We now consider the cross-correlation response functions by
combining different channels of LISA and Taiji.
Similar to ref.~\cite{Orlando:2020oko}, we define the $I$ and $V$ 
monopole response functions (for the $I$ and $V$ Stokes parameters)
as
\begin{equation}
    {\cal M}_{OO'}^I(f)=\half\left({\cal M}_{OO'}^+(f) + {\cal M}_{OO'}^-(f)
    \right), \quad
    {\cal M}_{OO'}^V(f)=\half\left({\cal M}_{OO'}^+(f) - {\cal M}_{OO'}^-(f)
    \right),
\end{equation}
with $O$ and $O'=A$, $E$, $T$, $C$, $D$, or $S$, such that
${\cal M}_{OO'}^I(f)$
and ${\cal M}_{OO'}^V(f)$ contribute to the response functions
to the GW energy density spectrum $\OmGW(f)=\OmGW^+(f)+\OmGW^-(f)$ and
to the GW helicity spectrum $\XiGW(f)=\OmGW^+(f)-\OmGW^-(f)$, respectively.
The contributions to the energy density sensitivity of the 
cross-correlated symmetric responses ${\cal M}_{OO'}^I (f)$
are negligible, since their response functions are approximately zero at 
frequencies above $10^{-3}$,
and smaller than the self-correlated responses ${\cal M}_{OO'}^V (f)$,
shown in \Fig{R_cross} with ${\cal M}_{ED}^I(f)$ for comparison. 
The energy density sensitivity of the cross-correlated channels $OO'=AE$,
$AD$, $EC$, and $ED$; see \Eq{OmsA}, is
\begin{equation}
    \Omega_{\rm s}^{OO'}(f)=
    \frac{8 \pi^2}{3 H_0^2} f^3 \frac{\sqrt{P_n^O(f) P_n^{O'}(f)}}
    {{\cal M}_{OO'}^I (f)},
\end{equation}
which is shown in \Fig{GW_sensitivity} with $OO' = ED$ for comparison,
where we can see that its sensitivity
is much larger than the auto-correlation sensitivities
$\Omega_{\rm s}^{A} (f)$ and $\Omega_{\rm s}^C (f)$
so we can omit their effect on the SNR and the PLS.
Hence, the combination of LISA and Taiji does not significantly
improve the detectability of a stochastic GW background signal
$\OmGW(f)$.

We now consider the sensitivity of the combined channels to a polarized GW
background signal $\XiGW(f)$.
\FFig{R_cross} shows the $V$ response functions obtained by cross-correlating
LISA and Taiji channels, such that the helical sensitivity can be defined as
\begin{equation}
    \Xi_{\rm s}^{OO'} (f) = \frac{8 \pi^2}{3 H_0^2} f^3 \frac{\sqrt{P_n^O(f) P_n^{O'}(f)}}
    {{\cal M}_{OO'}^V (f)},
\end{equation}
with $OO'=AE$, $AD$, $EC$, and $ED$.
We define the combined LISA--Taiji sensitivity 
as
\begin{align}
\Xi_{\rm s}^{\rm comb} (f) = &\, \left(\frac{1}
    {[\Xi_{\rm s}^{AC} (f)]^2} + \frac{1}
    {[\Xi_{\rm s}^{AD} (f)]^2} + \frac{1}
    {[\Xi_{\rm s}^{EC} (f)]^2} + \frac{1}
    {[\Xi_{\rm s}^{ED} (f)]^2} \right)^{-1/2} \nonumber \\
    = &\, \frac{\Xi_{\rm s}^{AC} (f)\, \Xi_{\rm s}^{AD} (f) \, \Xi_{\rm s}^{EC} (f)\, \Xi_{\rm s}^{ED} (f)   }
    {\sqrt{[\Xi_{\rm s}^{AC} (f)]^2+[\Xi_{\rm s}^{AD} (f)]^2+
    [\Xi_{\rm s}^{EC} (f)]^2+[\Xi_{\rm s}^{ED} (f)]^2}},
    \label{Xi_s_comb}
\end{align}
such that the corresponding polarization $\SNR_{\rm pol}$ is
\begin{align}
\SNR_{\rm pol} = &\, \sqrt{2\left(\SNR_{AC}^2 +  \SNR_{AD}^2 + 
\SNR_{EC}^2 + \SNR_{ED}^2 \right)} \nonumber \\
= &\, 2 \sqrt{T} \left[ \int \dd f \, \left( \frac{\XiGW(f)}{\Xi_{\rm s}^{\text{comb}} (f)}\right)^2 \right]^{1/2}.
\label{SNR_pol_Taiji}
\end{align}
The resulting helical GW sensitivity and PLS,
$\Xi_{\rm s}^{\rm comb} (f)$
and $\Xi_{\rm PLS}^{\rm comb} (f)$, respectively,
of the combined LISA--Taiji network are
shown in \Fig{GW_sensitivity}.
We use the resulting PLS to study the potential detectability of polarized
GW signals produced by primordial magnetic fields in \Sec{detection_sec}; see \Figs{XiGW_detectors}{XiGW_detectors_GWCirc}.


\end{document}